\documentclass{article}

\usepackage[a4paper, total={6.8in, 10.5in}]{geometry}
\usepackage{subfig}
\usepackage{graphicx}
\usepackage{amsmath}
\usepackage{mathrsfs}
\usepackage{amssymb}
\usepackage{enumerate}
\usepackage{algorithm}
\usepackage{algpseudocode}
\usepackage{graphicx,picins}
\usepackage{authblk}

\newcommand{\tabincell}[2]{\begin{tabular}{@{}#1@{}}#2\end{tabular}}

\begin{document}

%\markboth{}{Poisson Vector Graphics (PVG) and Its Closed-Form Solver}

\title{Poisson Vector Graphics (PVG) and Its Closed-Form Solver} % title
%\Large\textbf{Poisson Vector Graphics (PVG) and Its Closed-Form Solver}

\author[1]{Fei Hou}
\author[1]{Qian Sun}
\author[1]{Zheng Fang}
\author[2]{Yong-Jin Liu}
\author[2]{Shi-Min Hu}
\author[3]{Hong Qin}
\author[4]{Aimin Hao}
\author[1]{Ying He\thanks{yhe@ntu.edu.sg}}
\affil[1]{Nanyang Technological University}
\affil[2]{Tsinghua University}
\affil[3]{Stony Brook University}
\affil[4]{Beihang University}

\date{}

\maketitle

\begin{abstract}
  This paper presents Poisson vector graphics, an extension of the popular first-order diffusion curves, for generating smooth-shaded images.
  Armed with two new types of primitives, namely Poisson curves and Poisson regions,
  PVG can easily produce photorealistic effects such as specular highlights, core shadows, translucency and halos.
  Within the PVG framework, users specify color as the Dirichlet boundary condition of diffusion curves and control tone by offsetting the Laplacian,
  where both controls are simply done by mouse click and slider dragging.
  The separation of color and tone not only follows the basic drawing principle that is widely adopted by professional artists,
  but also brings three unique features to PVG, i.e., local hue change, ease of extrema control, and permit of intersection among geometric primitives,
  making PVG an ideal authoring tool.

  To render PVG, we develop an efficient method to solve 2D Poisson's equations with piecewise constant Laplacians.
  In contrast to the conventional finite element method that computes numerical solutions only,
  our method expresses the solution using harmonic B-spline,
  whose basis functions can be constructed locally and the control coefficients are obtained by solving a small sparse linear system.
  Our closed-form solver is numerically stable and it supports random access evaluation, zooming-in of arbitrary resolution and anti-aliasing.
  Although the harmonic B-spline based solutions are approximate,
  computational results show that the relative mean error is less than 0.3\%, which cannot be distinguished by naked eyes.
\end{abstract}

\section{Introduction}
\label{sec:introduction}

  Vector graphics provides several practical benefits over traditional raster graphics,
  including sparse representation, compact storage, geometric editablity, and resolution-independence.
  Early vector graphics supports only linear or radial color gradients, diminishing their applications for photo-realistic images.
  Orzan et al.~\cite{Orzan2008} pioneered diffusion curve images (DCIs),
  which are curves with colors defined on either side.
  By diffusing these colors over the image, the final result includes sharp boundaries along the curves with smoothly shaded regions between them.
  Thanks to its compact nature and the ability of producing smoothly shaded images,
  diffusion curves quickly gain popularity in the graphics field and inspire many follow-up works,
  such as improving runtime performance and numerical stability~\cite{Jeschke2009,Sun2012,Sun2014},
  and generalization to 3D and non-Euclidean domains~\cite{DBLP:journals/tog/JeschkeCW09a,Sun2012,DBLP:journals/tog/TakayamaSNI10}.

  Recent research has been focused on extending the expressiveness with more user control.
  Since Laplacian diffusion does not natively support manipulation of the color gradient, higher-order interpolation is a possible way for gradient control.
  Using thin-plate splines (TPS),
  Finch et al.~\cite{Finch2011} extended diffusion curves to provide smooth interpolation through color constraints while omitting the diffusion curves blur operation.
  Although TPS allows more user control and is able to mimic smooth shading,
  it often produces unwanted local extremals (hereby unpredicted effects) out of the user-specified regions due to the violation of the maximal principle of harmonic equation.
  Moreover, solving a bi-Laplace's equation is more computationally expensive than solving Laplace's equation,
  and it may suffer from serious numerical issues since the system is less well-conditioned.
  To remove the undesired extremals, Jacobson et al.~\cite{Jacobson2012} proposed a non-linear optimization guided by a harmonic function.
  Their method allows the user to specify the exact locations and values of the local maxima and minima,
  which is a highly desired feature to authoring and editing.
  Lieng et al.~\cite{Lieng2015} proposed shading curves, which associate shading profiles to each side of the curve.
  These shading profiles, which can be manually manipulated, represent the color gradient out from their associated curves.
  However, the colors produced by shading curves are not as vivid as diffusion curves.
  Recently, Jeschke~\cite{Jeschke2016} proposed generalized diffusion curve images (GDCIs), which spatially blend multiple conventional DCIs.
  Thanks to more degrees of freedom provided by all DCIs,
  GDCI is able to provide a similar expressive power of color control as the TPS model and its solver is efficient and numerically stable.
  However, the blending functions are highly non-linear, making it difficult to design manually and control the extrema.
  Therefore, GDCI is often applied to image vectorization with simple editing (e.g., changing the colors or modifying the curves),
  rather than being used as an authoring tool.

  This paper aims at overcoming the above-mentioned limitations of the existing DCI framework.
  Towards this goal, we present a new type of vector graphics, called Poisson vector graphics (PVG),
  which extends DCI by allowing non-zero Laplacians.
  To make a PVG, users first sketch a set of sparse geometric primitives (e.g., points, curves and/or regions) $\{\gamma_i\}_{i=1}^{N}$.
  Then, for each primitive $\gamma_i$, specify its Laplacian of color $\bigtriangleup \gamma_i=f_i$,
  where $f_i$ is a \textit{piecewise constant} function defined on $\gamma_i$.
  The final image is obtained by solving the following Poisson equation
  \begin{equation}
  \left\{
  \begin{array}{ll}
  \Delta u(x)=f, & x\in\Omega\setminus\partial\Omega\\
  u|_{\partial\Omega}=g, & x\in\partial\Omega
  \end{array}
  \right.
  \label{eqn:poissoneqn}
  \end{equation}
  where $\Omega$ is the 2D domain and the constraint $f$ partitions $\Omega=\bigcup_{i=1}^N\Omega_i$
  so that on each subregion $\Omega_i$, $f|_{\Omega_i}=f_i$ is a constant.

  PVG is a natural extension of DCI, which are rasterized via Laplacian diffusion (i.e., solving a Laplace's equation $\Delta u=0$).
  The seemingly minor change of replacing the zero Laplacian by a piecewise constant function $f$ is indeed crucial for globally and locally controlling shading profiles,
  which are not able to achieve within the diffusion curve framework.
  Intuitively speaking, diffusion curve images are the result of diffusing the colors defined along control curves until the color field reaches an equilibrium, which is a harmonic function.
  Since a harmonic function is completely determined by the Dirichlet boundary condition, diffusion curves do not allow manipulating of color gradients.
  Although second-order diffusion curves~\cite{Finch2011}\cite{Boye2012}~\cite{Ilbery2013} support explicitly gradient control,
  they are usually difficult to use due to lack of intuitive interpretation between the manually manipulated gradients (which are \textit{vectors}) and the desired shading effect.
  In contrast, PVG is the solution of Poisson's equation, whose solution space is much larger than that of Laplace's equation,
  hereby providing users more control of the image.

\begin{figure*}[htbp]
  \centering
  \includegraphics[width=\linewidth]{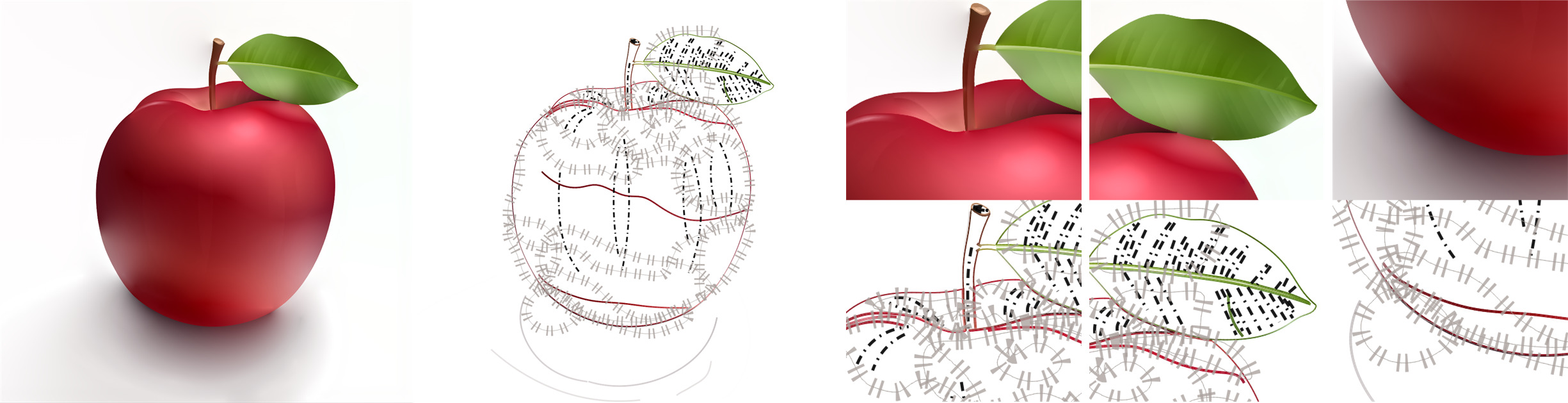}
  \caption{
  \label{fig:teaser}
  A Poisson vector graphics (PVG) consists of the popular diffusion curves (DCs)
  and two new types of primitives, called Poisson curves (PCs) and Poisson regions (PRs).
  By manipulating the Laplacian constraints associated with PCs and PRs,
  users can easily control global and local shading profiling and produce photorealistic effects such as specular highlights and core shadows,
  which cannot be achieved by using DCs only.
  Thanks to our closed-form Poisson solver, PVG also supports random access evaluation, anti-aliasing and zooming in of arbitrary resolution.
  DCs, PCs and PRs are depicted by solid lines, dashed line and loops with hatches, respectively.
  }
\end{figure*}

  Armed with two new types of primitives, namely Poisson curves and Poisson regions,
  PVG can easily produce photorealistic effects such as specular highlights, core shadows, translucency and halos (see Figure~\ref{fig:teaser}).
  Within the PVG framework, users specify color as the Dirichlet boundary condition of diffusion curves and control tone by offsetting the Laplacian,
  where both controls are simply done by mouse click and slider dragging.
  The separation of color and tone not only follows the basic drawing principle that is widely adopted by professional artists
  (e.g., see page 58 ~\cite{Hale1964}),
  but also brings three unique features to PVG, i.e., local hue change, ease of extrema control, and permit of intersection among geometric primitives,
  making PVG an ideal authoring tool.

  To render PVG, we develop an efficient method to solve 2D Poisson's equations with piecewise constant Laplacians.
  In contrast to the conventional finite element method that computes numerical solutions only,
  our method provides a closed-form solution $u(x)=\sum \lambda_i\psi_i(x)$,
  where $\{\psi_i(x)\}$ are the basis functions of harmonic B-spline~\cite{Feng2012}, and the control coefficients $\{\lambda_i\}$ are computed by solving a small sparse linear system.
  Similar to the conventional B-splines, the basis functions $\psi_i(x)$ have local support,
  therefore, evaluating the spline at a point $x$ involves only the basis functions that cover $x$.
  Our solver is numerically stable and it supports random access evaluation, zooming-in of arbitrary resolution and anti-aliasing.
  Although the harmonic B-spline based solutions are approximate,
  computational results show that the relative mean error is less than 0.3\%, which cannot be distinguished by naked eyes.

\section{Related Work}
\label{sec:relatedwork}

  This section briefly reviews the related work on various solvers for rendering diffusion curves and gradient domain image editing.

\subsection{Diffusion Curve Solvers}
\label{subsec:dcandvariants}

  To rasterize a DCI, one needs to solve a Laplace's equation defined on the entire image plane.
  Since a direct solver is expensive, Orzan et al.~\cite{Orzan2008} designed a multigrid solver,
  which uses a coarse version of the domain to efficiently solve for the low frequency components of the solution,
  and a fine version of the domain to refine the high frequency components.
  Although being fast, this solver suffers from aliasing and flickering artifacts due to the rasterization of the curves over a discrete multi-scale pixel grid.
  Jeschke et al.~\cite{Jeschke2009} used finite differences with variable step size to accelerate the convergence rate of Jacobi iterations,
  which guarantees the convergence to the right solution.
  Their method also supports zooming-in of arbitrary resolution by solving Laplace equation only for the region of interest.
  However, since the boundary condition is obtained from the coarse domain, which is not accurate enough,
  numerical issues may occur near the boundary.
  To overcome these limitations, Boy{\'e} et al.~\cite{Boye2012} developed a finite element method (FEM) based biharmonioc equation solver,
  which dynamically converts a DCI into a high-order mesh-based representation that is automatically adapted to the complexity of the input curves.
  Such an intermediate triangulation, hidden from the user, is updated only when diffusion curves are edited.
  Moreover, their solver directly deals with gradient constraints.
  Their solver is ideal for (bi-)Laplace equations, but it is unclear whether it can be directly extended to Poisson equations.

  Motivated by the parallel between diffusion curve rendering and final gathering in global illumination,
  Bowers et al.~\cite{Bowers2011} developed a stochastic ray tracing method that allows trivial parallelism by using shaders
  and provides a unified treatment of diffusion curves with classic vector and raster graphics.
  However, it densely computes values even in smooth regions and sacrifices support for instancing and layering.
  Later, Prevost et al.~\cite{Prevost2015} improved the ray tracing method by using an intermediate triangular representation with cubic patches to synthesize smooth images faithful to the per-pixel solution.

  Another family of methods for rasterizing DCIs is to use boundary element method (BEM),
  which rephrases Laplace's equation as a boundary integral along control curves.
  Sun et al.~\cite{Sun2012} formulated the solution as a sum of Green's functions of Laplacian.
  Thanks to the closed-form formula, this approach is quite fast and enables integrating the solution over any rectangular region, which allows anti-aliasing.
  However, it requires pre-calculating the weights of the Green's function kernels, which depends on normal derivatives along the control curves.
  As a result, it can only take a DCI with fixed geometry and color constraints as input, and is not suitable for interactive authoring.
  Ilbery et al.~\cite{Ilbery2013} proposed a BEM based solver for rendering TPS vector graphics in a line-by-line manner.
  Sun et al.~\cite{Sun2014} presented a fast multipole representation for random-access evaluation of DCIs.
  Their method is able to achieve real-time performance for rasterization and texture-mapping DCIs of up to millions of curves.

  Most DCI solvers require GPU acceleration to achieve real-time performance.
  Aiming at efficiently rendering DCIs on devices with only a CPU, Pang et al.~\cite{Pang2012} developed a mesh-based approach,
  which sets the diffusion curves as constraints in triangulation and employs mean value coordinate-based interpolant to estimate vertex colors.
  Their algorithm supports random access evaluation, but the produced DCI is only an approximation and it may suffer from the aliasing issue.

\subsection{Poisson Solvers for Gradient Domain Image Editing}
\label{subsec:poissonsolver}
  Solving 2D Poisson's equations plays an important role in various image processing tasks,
  such as composition~\cite{Perez2003,Agarwala2007},
  alpha matting~\cite{Sun:2004:PM:1015706.1015721}, colorization~\cite{Levin:2004:CUO:1015706.1015780},
  filtering and relighting~\cite{Bhat:2010:GGO:1731047.1731048}.
  Fast Fourier transformation (FFT) is a popular method to solve Poisson's equation in rectangular domains due to its high performance.
  For irregular domains, one often adopts the finite element method (FEM),
  which subdivides the domains into smaller parts and solves a large sparse linear system.

  Poisson image editing~\cite{Perez2003} enables seamless cloning by matching the gradients of the source and the target images, which is formulated as a Poisson's equation.
  However, directly solving such an equation is computationally expensive.
  Agarwala~\cite{Agarwala2007} improved the scalability by using quadtrees to substantially reduce both memory and computational requirements.
  McCann and Pollard~\cite{McCann2008} presented a multi-grid Poisson solver on GPUs,
  with which they can achieve real-time interactive performance.
  Taking the source image as the solution of a Poisson's equation under a different boundary condition,
  Farbman et al.~\cite{Farbman2009} converted the Poisson problem to a Laplace problem,
  which can be solved efficiently using mean value coordinates~\cite{Floater2003}.

  In contrast to the above solvers for general Poisson's equations, our solver is for a special type of Poisson's equation,
  where the function $f$ is piecewise constant.
  We derive the closed-form solution using harmonic B-spline basis functions~\cite{Feng2012}.

\section{Poisson Vector Graphics}
\label{sec:pvg}
  Poisson vector graphics extends the DC framework by adding two new geometric primitives, namely Poisson curves and Poisson regions.
  The former is to model color discontinuity across curves, while the latter is to design smooth shading within the user specified regions.
  Mathematically speaking, PVG solves a Poisson's equation with piecewise constant Laplacians $f$.
  As the \textit{first} order Laplacian, PVG takes DCI as a special case with $f\equiv 0$.
  Extending the zero Laplacian to a piecewise constant function $f$ brings three unique advantages.
  First, users can explicitly control the local and/or global shading profiling via manipulating $f$ (which is a scalar for each color channel).
  Second, users can easily control the extrema, which are either on the curves (for PC and DC) or inside a region (for PR).
  Third, PVG allows intersection among the geometric primitives.

  Although a PVG can have an arbitrary number of PCs and PRs, it must contain at least one diffusion curve, serving as the boundary condition $g$.
  In the following, we detail Poisson curves and Poisson regions.

\subsection{Poisson Curves}
\label{subsec:laplaciancurves}

\begin{figure}
  \centering
  \includegraphics[width=1.5in]{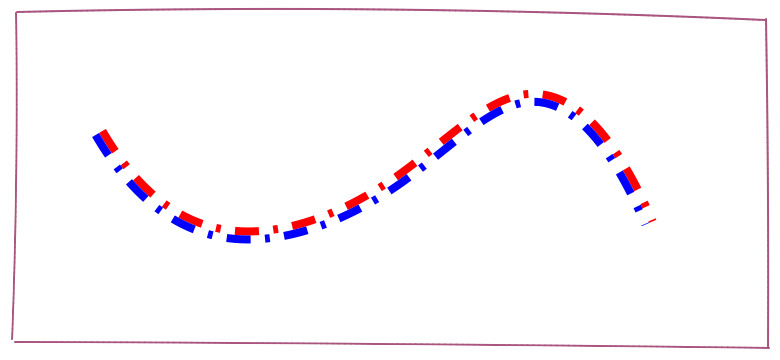}
  \includegraphics[width=1.5in]{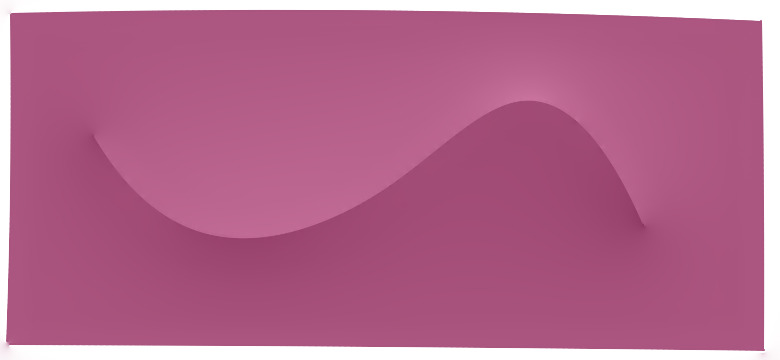}
  \includegraphics[width=1.5in]{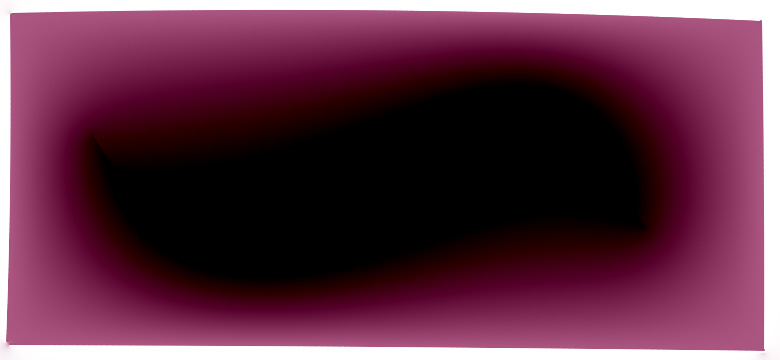}\\
  \makebox[1.5in]{(a) Poisson curve}
  \makebox[1.5in]{(b) $f_{+}=41=-f_{-}$}
  \makebox[1.5in]{(c) $f_+=41, f_-=-45$}
  \caption{A Poisson curve is a two-sided curve created by cubic B-splines,
  where each side is associated with a Laplacian constraint and the two Laplacians add up to zero.
  The zero sum is a necessary condition for local shading control.
  If it is not satisfied, unwanted artifacts occur.
  The rectangular boundary in (a) is a diffusion curve, which specifies the boundary condition of the Poisson's equation.
  }
  \label{fig:laplacian}
  \end{figure}

  Similar to diffusion curves, a Poisson curve $\gamma$ is two-sided, denoted by $\gamma^+$ and $\gamma^-$, and can be either open or closed.
  We associate each side a Laplacian value, denoted by $f_+$ and $f_-$ respectively, such that $f_++f_-=0$ (see Figure~\ref{fig:laplacian}).
  The requirement for zero sum is for \textit{local} shading control.
  To explain this, consider a region $D\supset\gamma$ that bounds the diffusion of $\gamma$.
  The divergence theorem $\iint_D\Delta udA=\oint_{\partial D}\nabla u\cdot\mathbf{n}dl$
  relates the double integral of $\Delta u$ to the line integral of $\nabla u$.
  Since the color function $u$ remains unchanged on $\partial D$,
  the line integral of $\nabla u$ is a constant, implying that the double integral of Laplacian is also a constant.
  Based on this observation, we set $(\Delta u)|_{\gamma^+}+(\Delta u)|_{\gamma^-}=0$ (see Figure \ref{fig:laplacian}).
  Although the zero-sum is only a \textit{necessary} condition to ensure local shading control,
  it works pretty well in our experiments.

\begin{figure*}
  \centering
  \includegraphics[width=1in]{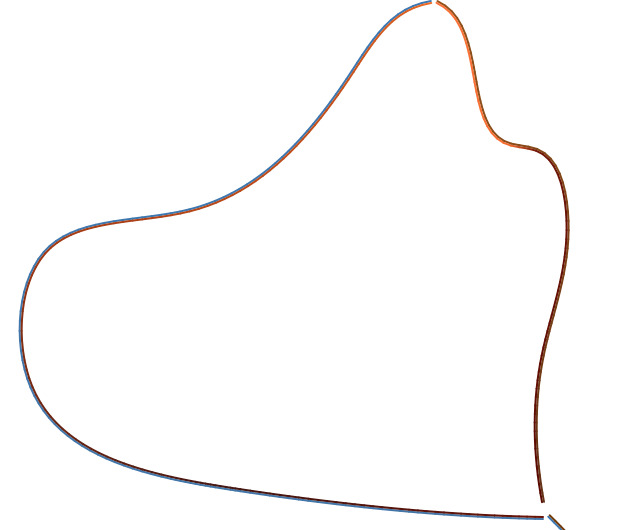}
  \hspace{2pt}
  \includegraphics[width=1in]{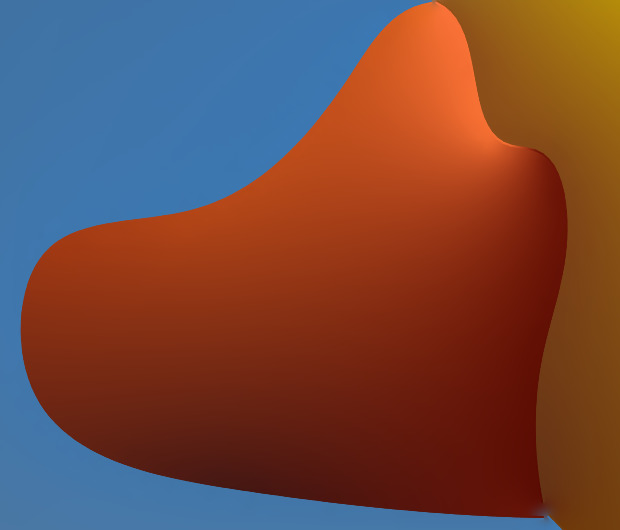}
  \hspace{2pt}
  \includegraphics[width=1in]{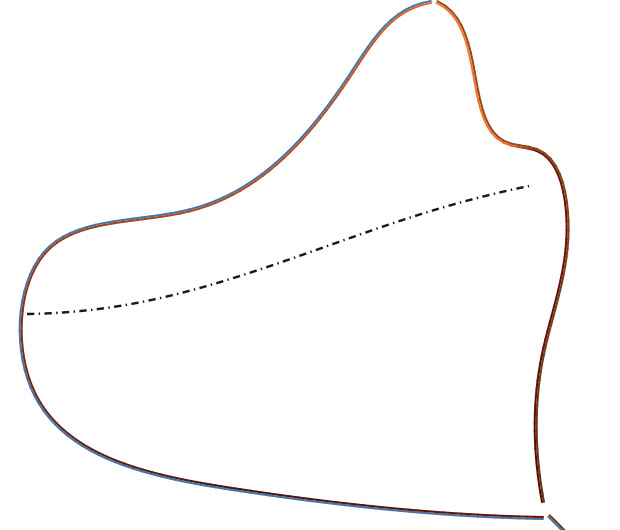}
  \hspace{2pt}
  \includegraphics[width=1in]{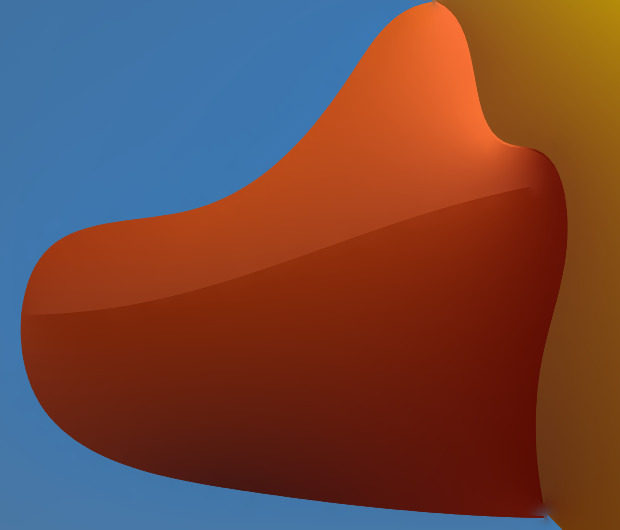}
  \hspace{2pt}
  \includegraphics[width=1in]{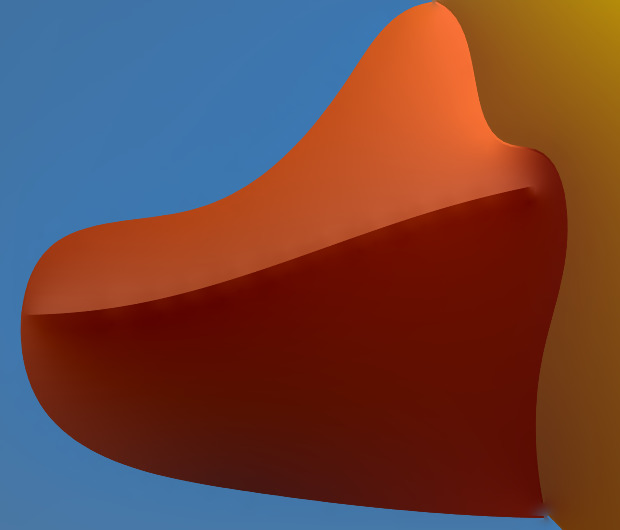}
  \hspace{2pt}
  \includegraphics[width=1in]{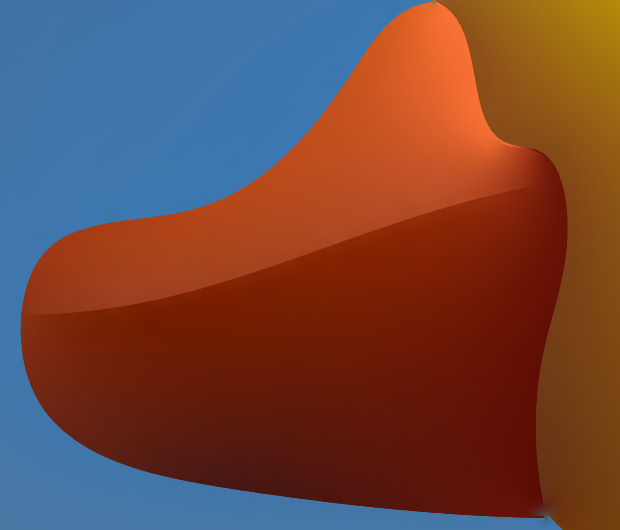}\\
  \makebox[1in]{(a) DC} \hspace{2pt}\makebox[1in]{(b) DC image} \hspace{2pt}\makebox[1in]{(c) DC+PC}\hspace{2pt}
  \makebox[1in]{(d) $f_+=19$}\hspace{2pt} \makebox[1in]{(e) $f_+=82$}\hspace{2pt} \makebox[1in]{(f) Spatially varying $f$}
  \caption{Poisson curves are used to model color discontinuity in a bounded region.
  (a)-(b): A diffusion curve is used to create the beak of the Rubber Duck.
  (c)-(d): We can create a sharp edge using a Poisson curve $\gamma$ (dashed line).
  Increasing the Laplacian constraint $|f|$ makes a stronger edge.
  (f) We can also use spatially varying constraint to define sharp edges with varying strength.
  }
  \label{fig:laplacian_pixel_curve}
  \end{figure*}

  Note that the zero-sum condition is hidden to users, who simply drag over a slider to specify $f_+$.
  As long as $f_+\neq 0$, the Poisson curve corresponds to a sharp edge.
  The larger the value $|f_+|$, the stronger the sharp feature.
  Since we model Poisson curves as cubic splines,
  users can also specify spatially varying constraints $(\Delta u)|_\gamma$ by setting weights on the control points.
  See Figure~\ref{fig:laplacian_pixel_curve}.

\subsection{Poisson Regions}
\label{sec:laplacian_region}

  We develop Poisson regions to produce photorealistic effects, such as specular highlights, core shadows, translucency and halos.
  Observe that the Laplacian of specular highlight is a ``bell'' shaped curve (see Figure~\ref{fig:laplacian_region_axis}).
  To discretize such a function, we decompose a Poisson region into two disjoint sub-regions $D_1$ and $D_2$,
  where the outer part $D_1$ is relatively thin so that it preserves the geometry of $\partial D$ well.
  We assign strictly monotonic constraints $f_i$ to $D_i$, $i=1,2$.
  Similar to Poisson curves, we also require a vanishing sum $\sum_{i=1}^2\iint_{D_i}f_idA=0$ for local shading control.
  As Figure~\ref{fig:laplacian_region_axis} shows, $f_i$s with decreasing values form a $U$-shaped curve, which simulate the Laplacian of specular reflection,
  while $f_i$s with increasing values are bell-shaped, which simulate the Laplacian of core shadows.
  To simulate halo, we also provide additional control that offsets $f_i$s (see Figure~\ref{fig:laplacian_region_axis}).

  To determine the the sub-regions $D_1$ and $D_2$, we take the boundary $\partial D$ as the source and compute the Euclidean distance transform.
  Let $d_{max}$ be the maximal distance to the boundary.
  We then define $D_1=\{x|d(x) \leq 0.05d_{max},~x\in D\}$ and $D_2=D\setminus D_1$
  (see Figure~\ref{fig:laplacian_region_axis}(a)).
  To produce halos in a Poisson region, we also allow user to add an increment $\delta_i$ to $f_i$ (see Figure~\ref{fig:halo}).

\begin{figure}
  \centering
  \includegraphics[width=2.5in]{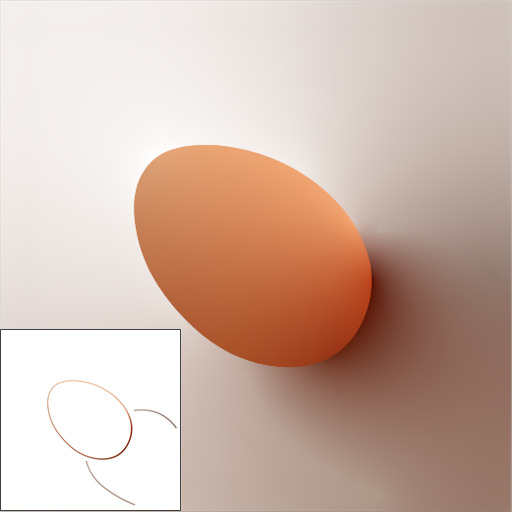}
  \includegraphics[width=2.5in]{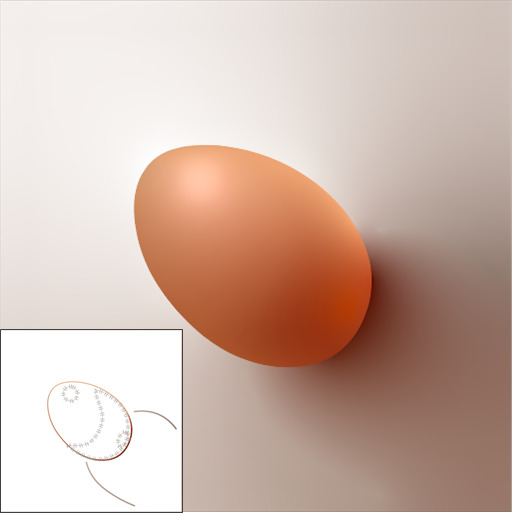}\\
  \makebox[2.5in]{DC} \makebox[2.5in]{DC+PR}
  \caption{
  Poisson regions (loops with hatches) can simulate smooth shading effects, such as specular highlights and core shadows,
  which are usually difficult to obtain using DCs only.}
  \label{fig:egg}
  \end{figure}

\begin{figure}
  \centering
  \includegraphics[width=5in]{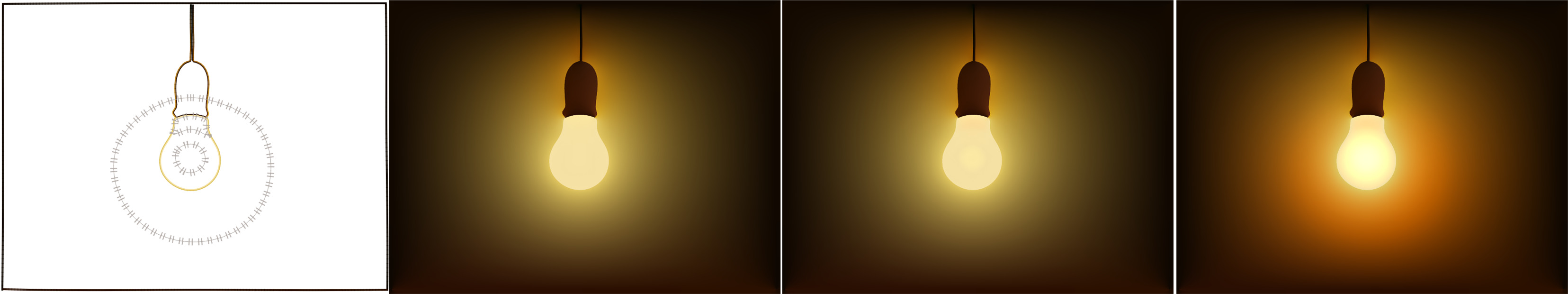}\\
  \makebox[1.25in]{(a) }
  \makebox[1.25in]{(b) }
  \makebox[1.25in]{(c) }
  \makebox[1.25in]{(d) }
  \caption{We can produce halos in a Poisson region by adding each Laplacian constraint $f_i$ an increment $\delta_i$,
  where $f_i+\delta_i$, $i=1,2,3$, are strictly monotonic.
  (a) There are two Poisson regions (loops with hatches) in this PVG.
  (b) Rendering without PR, i.e., $f_i=0$, $\delta_i=0$.
  (c) Rendering without increments, i.e., $f_i\neq 0$, $\delta_i=0$.
  (d) Rendering with increments i.e., $f_i\neq 0$, $\delta_i\neq 0$.
  }
  \label{fig:halo}
  \end{figure}

\begin{figure}
  \centering
  \includegraphics[width=1.1in]{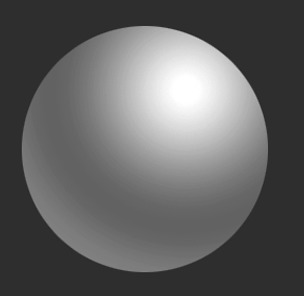}
  \includegraphics[width=1.5in]{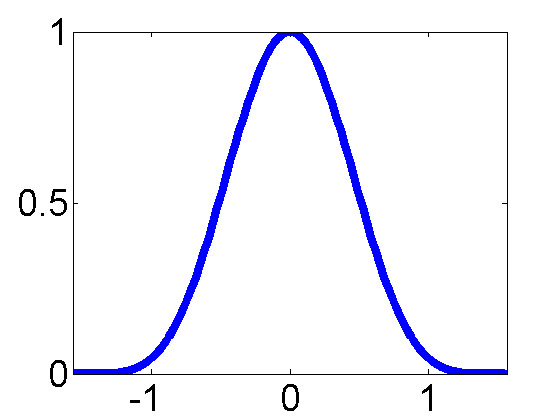}
  \includegraphics[width=1.5in]{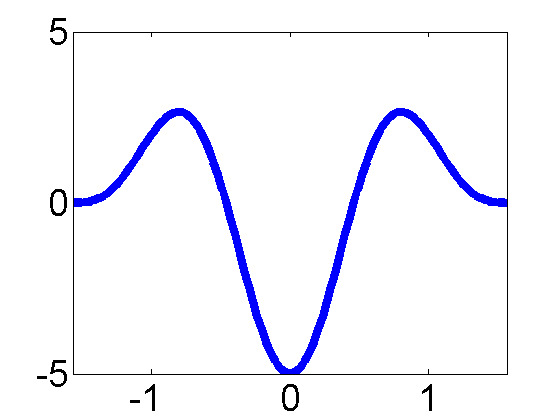}\\
  \makebox[1.1in]{(a)}
  \makebox[1.5in]{(b) $\cos^5\theta$}
  \makebox[1.5in]{(c) $\Delta\cos^5\theta$}\\
  \includegraphics[width=1.1in]{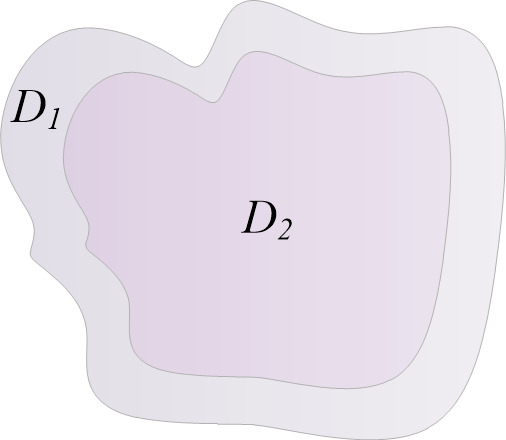}
  \includegraphics[width=1.5in]{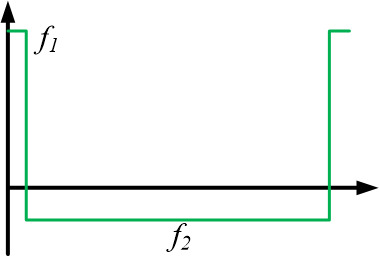}
  \includegraphics[width=1.5in]{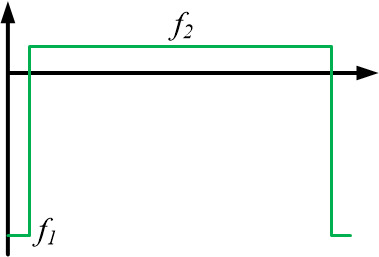}\\
  \makebox[1.1in]{(d)}\makebox[1.5in]{(e)}\makebox[1.5in]{(f)}\\
  \caption{
  (a) A shiny sphere rendered using the Phong illumination model.
  (b) The specular highlight is $\cos^k\theta$, where $\theta$ is the angle between the viewing vector and the reflection vector,
  and the exponent $k$ is the shininess constant.
  (c) The Laplacian of $\cos^k\theta$ is $U$-shaped.
  (d) A Poisson region $R$ consists of two disjoint sub-regions $D=D_1\cup D_2$,
  each of which is associated with a Laplacian constraint $f_i$.
  (e) The strictly decreasing $f_i$s are to approximate the $U$-curve of $\Delta\cos^k\theta$, simulating specular highlights.
  (f) Similarly, the strictly increasing $f_i$s are to model the Laplacian of core shadows.
  }
  \label{fig:laplacian_region_axis}
  \end{figure}

  There are 2 differences between PC and PR:
  First, a PC is a double-sided curve, which can be either open or closed, and a PR is region whose boundary is closed.
  Second, PC is mainly designed for modeling color discontinuity,
  whereas PR is to produce smooth shadings, such as specular highlights and core shadows (see Figure~\ref{fig:egg}).

  \subsection{Features}
  \label{subsec:features}

  PVG has three unique features which are favorable for authoring.

  \textbf{Local shading control.}
  Although both PCs and DCs are able to produce color discontinuity across the curves, they are fundamentally different.
  Using DCs, users explicitly specify the boundary condition (i.e., colors) on both sides of a curve.
  To change tone either locally or globally, users have to re-assign the colors for all the curves involved, which is tedious and non-intuitive.
  In contrast, PVG separates colors and tones in that the boundary condition of PC (i.e., Laplacian of colors) is a \textit{relative} value
  and the color of a PC is determined by the DC enclosing it.
  As a result, changing tone does not require any modification of the boundary condition of PCs.
  See Figure~\ref{fig:blood_seeker_change} for an example.
  Poisson regions also enable local control, since the area integral of the Laplacian on a PR is zero.

  \textbf{Ease of extrema control.}
  Diffusion curves produce harmonic color functions, whose extrema are always on the boundaries.
  Unfortunately, such a nice property is not available to biharmonic functions.
  As pointed out in~\cite{Boye2012}\cite{Jacobson2012}, it is difficult to control the locations of extrema in the TPS based vector graphics~\cite{Finch2011}.
  Since PVG is not a harmonic function, it also contains extrema points.
  However, users can control the locations of extremum in an easy and direct manner.
  Observe that non-zero Laplacians are assigned to the points on DCs and PCs, and the points inside PRs.
  Therefore, the extrema of a DC or PC are precisely on the curve, and the extrema of a PR $P$ are guaranteed to be inside $P$.
  See Section~\ref{sec:results} for more discussions.

  \textbf{Permit of intersection among geometric primitives.}
  In the DCI framework, diffusion curves are \textit{not} allowed to intersect,
  since the colors associated to the intersecting curves compete with each other (see Figure~\ref{fig:intersecting}(a)).
  Hence, users have to split those curves into disjoint segments with different colors.
  This extra operation becomes a serious drawback to designers, who have to deal with a large amount of short segments and their constraints.
  In contrast, all types of primitives can intersect each other except for two DCs.
  This feature not only simplifies the drawing process, but also allows users to use layers, which is a powerful technique for making complex drawing.

\section{Closed-Form Solver}
\label{sec:solver}

  To solve Equation~(\ref{eqn:poissoneqn}), we need Green's third identity,
  \begin{equation}
  \label{eqn:green_identity}
  \begin{split}
  u(x)= & \iint_\Omega G(x,y) \Delta u(y)d\sigma_y + \\
  & \oint_{\partial \Omega}\left (u(y)\frac{\partial G(x,y)}{\partial \bf n}-G(x,y)\frac{\partial u(y)}{\partial \bf n}\right )dl_y,
  \end{split}
  \end{equation}
  where $d\sigma$ and $dl$ are the surface and line elements, $\bf n$ is the outward pointing unit normal of $dl$,
  $G(x,y)$ is Green's function of the Laplace operator, i.e., $\Delta G(x,y)=\delta(x-y)$.
  To simplify the discussion, we assume the domain $\Omega$ is simply connected.
  However, the framework can be trivially extended to multiply connected domains in which Green's identity still hold.

\begin{figure*}
  \centering
  \includegraphics[width=1.25in]{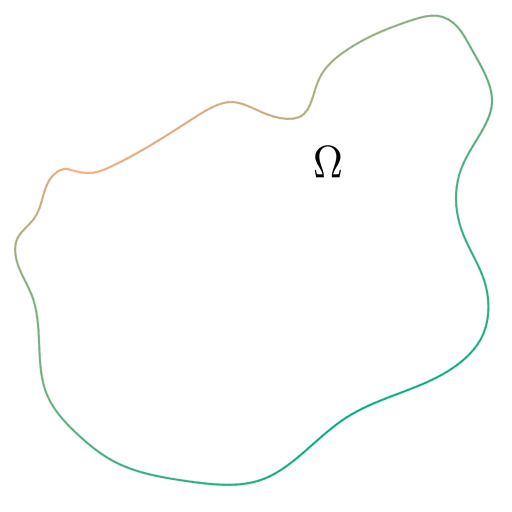}
  \includegraphics[width=1.25in]{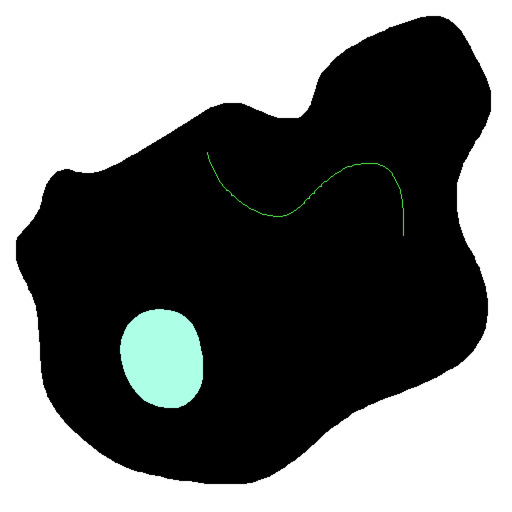}
  \includegraphics[width=1.05in]{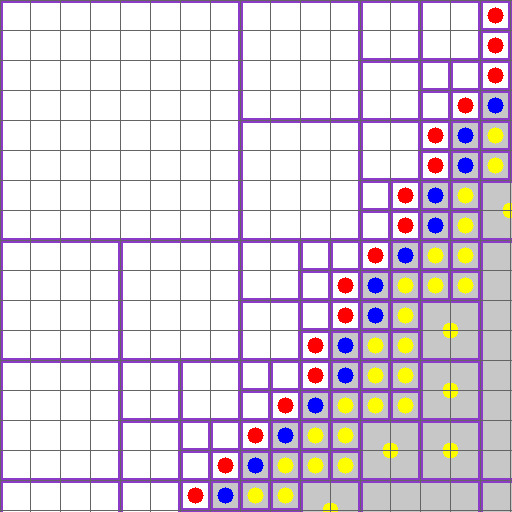}
  \hspace{5pt}
  \includegraphics[width=1.05in]{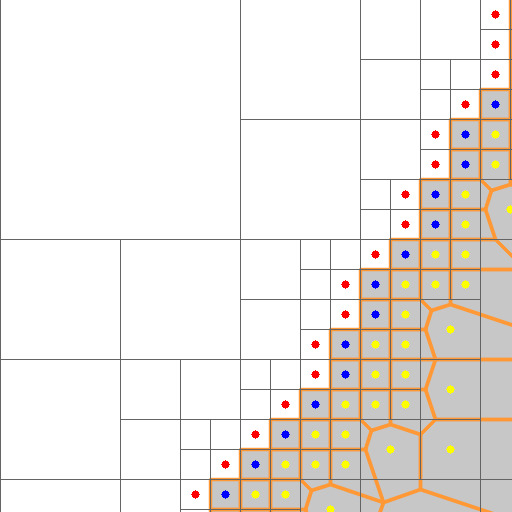}
  \includegraphics[width=1.25in]{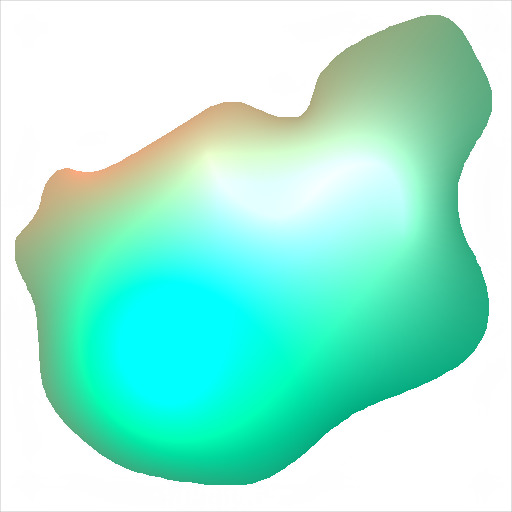}
  \makebox[1.25in]{(a) $\Omega$ and $g$}\makebox[1.25in]{(b) $f$}\makebox[1.05in]{(c) Quad-tree}\hspace{5pt}\makebox[1.05in]{(d) Voronoi diagram}\makebox[1.25in]{(e) $u$}\\
  \caption{Illustration of our Poisson solver.
  (a) shows the domain $\Omega$ and the boundary condition $g$ whose values are encoded in color.
  (b) shows the piecewise constant function $f$ that partitions $\Omega$ into three disjoint sub-regions $\Omega=\cup_{i=1}^3\Omega_i$.
  (c) We discretize $\Omega_i$ using a quad-tree, and organize its leaves into interior nodes (yellow), boundary nodes (blue) and exterior nodes (red).
  Let $n$ be the total number of interior and boundary nodes.
  (d) We construct a Voronoi diagram using the interior and boundary nodes as the generators, and clip the diagram with $\partial\Omega$.
  For a Voronoi cell $V_i$, we define a harmonic B-spline basis function $\psi_i(x)$,
  whose knot is the generator of $V_i$.
  To properly evaluate the spline on the boundary,
  we also need the squares containing the exterior nodes (the red dots in the close-up view) that are adjacent to the boundary Voronoi cells.
  (e) For a boundary Voronoi cell $V_i$, we simply set its control point $\lambda_i$ using the given boundary condition $g$.
  For internal Voronoi cells, we compute their control points by solving a sparse linear system of size $k\times k$,
  where $k$ is significantly less than the number of pixels in $\Omega$.
  Finally, the solution is given by $u(x)=\sum_{i=1}^n \lambda_i\psi_i(x)$.
  }
  \label{fig:solverpipeline}
  \end{figure*}

  The first term is a double integral, which is computationally expensive due to its \textit{global} nature.
  To improve the performance, we discretize the domain using a quad-tree and then decompose the area integral into a sum of line integral over each sub-domain.
  After reorganizing the terms, we show that the solution $u(x)$ can be written as a weighted sum of a set of locally-defined functions, called harmonic B-spline basis functions.
  The weights are the control points of the spline, and are computed by solving a small sparse linear system.
  In the following, we detail the solver using a toy model shown in Figure~\ref{fig:solverpipeline}.

  \textbf{Discretizing Domain.}
  Given a user-specified resolution, we discretize the domain $\Omega$ using a quad-tree so that each pixel $p_j\in\partial\Omega_i$
  \footnote{Note that the boundary $\partial\Omega_i$ of a sub-region $\Omega_i$ may not be part of the \textit{geometric} boundary $\partial\Omega$.} is in a square, denoted by $R_j$.
  All squares except those containing pixels of $\partial\Omega_i$ are \textit{completely} inside some subregion.
  Then we have $\Omega=\cup_{i=1}^{n}R_i$, where $n$ be the total number of such squares.
  Let us denote $A_i$ the area of the square $R_i$.
  We then define two disjoint sets, the boundary subset $\mathcal{B}=\{R_i|R_i\cap \partial\Omega\neq\emptyset,1\leq i\leq n\}$
  and the interior subset $\mathcal{I}=\{R_i| R_i\cap \partial\Omega =\emptyset,1\leq i\leq n\}$.
  See Figure~\ref{fig:solverpipeline}(c).

  \textbf{Decomposing Double Integral.}
  For each square $R_j$, $f|_{R_j}$ is a constant, since $R_j$ is either completely inside a subregion
  or a square containing only one boundary pixel of a subregion.
  So we can rewrite the first term of Eqn. (\ref{eqn:green_identity}) as
  \begin{eqnarray}
  \label{eqn:int}
  \nonumber & & \iint_\Omega G(x,y) \Delta u(y)d\sigma_y=\sum_{j=1}^n\iint_{R_j} G(x,y) \Delta u(y)d\sigma_y\\
  \nonumber & = & \sum_{j=1}^{n} \frac{\iint_{R_j}G(x,y)d\sigma_y}{A_j}\iint_{R_j}\Delta u(y)d\sigma_y\\
  & = & \sum_{j=1}^{n}\overline{G}_{R_j}(x)\oint_{\partial R_j}\frac{\partial u(y)}{\partial \bf n}dl_y,
  \end{eqnarray}
  where the last equation comes from the divergence theorem
  and $\overline{G}_{R_j}(x)\triangleq\frac{1}{A_j}\iint_{R_j}G(x,y)d\sigma_y$ is the average value of Green's function in region $R_j$.

  Consider a square $R_j\in\mathcal{B}$.
  Observe that $\partial R_j\cap\partial \Omega$ is one pixel, which is sufficiently small.
  So the line integral can be approximated by
  \begin{eqnarray}
  \nonumber &\int_{\partial R_j\cap\partial \Omega} G(x,y)\frac{\partial u(y)}{\partial \bf n}dl_y\approx \overline{G}_{R_j}(x)\int_{\partial R_j\cap\partial \Omega}\frac{\partial u(y)}{\partial \bf n}dl_y,\\
  \nonumber & \int_{\partial R_j\cap\partial \Omega} u(y)\frac{\partial G(x,y)}{\partial \bf n}dl_y \approx u(R_j)\int_{\partial R_j\cap\partial \Omega}\frac{\partial G(x,y)}{\partial \bf n}dl_y,
 \end{eqnarray}
 where $u(R_i)$ is the average color in $R_i$. We get
 \begin{eqnarray}
 \oint_{\partial \Omega} G(x,y)\frac{\partial u(y)}{\partial \bf n}\approx\sum_{R_j\in\mathcal{B}}\overline{G}_{R_j}(x)\int_{\partial R_j\cap\partial \Omega}\frac{\partial u(y)}{\partial \bf n}dl_y\label{eqn:app1}\\
 \oint_{\partial \Omega}u(y)\frac{\partial G(x,y)}{\partial \bf n}\approx\sum_{R_j\in\mathcal{B}} u(R_j)\int_{\partial R_j\cap\partial \Omega}\frac{\partial G(x,y)}{\partial \bf n}dl_y\label{eqn:app2}
 \end{eqnarray}

  Referring to Figure~\ref{fig:solverpipeline}(c), we classify the shared edges between two adjacent quad-tree cells into three groups:
  \textit{interior edge}: an edge shared by two interior (yellow) cells;
  \textit{inner boundary}: an edge shared by an interior (yellow) cell and a boundary (blue) cell, or two boundary cells;
  \textit{outer boundary}: an edge shared by an exterior (red) cell and a boundary (blue) cell.

  Substituting Equations~(\ref{eqn:int}),(\ref{eqn:app1}) and (\ref{eqn:app2}) into (\ref{eqn:green_identity}) yields
  \begin{eqnarray}
  \nonumber & & u(x)\approx \sum_{R_j\in\mathcal{B}} u(R_j)\int_{\partial R_j\cap\partial \Omega}\frac{\partial G(x,y)}{\partial \bf n}dl_y +\\
  \nonumber & &\sum_{j=1}^{n}\overline{G}_{R_j}(x) \oint_{\partial R_j} \frac{\partial u(y)}{\partial \bf n}dl_y -\sum_{R_j\in\mathcal{B}}\overline{G}_{R_j}(x)\int_{\partial R_j\cap\partial \Omega}\frac{\partial u(y)}{\partial \bf n}dl_y\\
  \nonumber &=& \sum_{j=1}^{n}\overline{G}_{R_j}(x)\int_{\partial R_j\setminus\partial \Omega} \frac{\partial u(y)}{\partial \bf n}dl_y
  + \sum_{R_j\in\mathcal{B}} u(R_j)\int_{\partial R_j\cap\partial \Omega}\frac{\partial G(x,y)}{\partial \bf n}dl_y\\
  \nonumber & = & \sum_{R_j\in\mathcal{B}}\overline{G}_{R_j}(x)\int_{\partial R_j\setminus\partial \Omega} \frac{\partial u(y)}{\partial \bf n}dl_y +\sum_{R_j\in\mathcal{I}}\overline{G}_{R_j}(x)\iint_{R_j}\Delta u(y)d\sigma_y\\
  \nonumber &&+\sum_{R_j\in\mathcal{B}}u(R_j)\int_{\partial R_j\cap\partial \Omega}\frac{\partial G(x,y)}{\partial \bf n}dl_y\\
  & = &
  \nonumber \underbrace{\sum_{R_j\in\mathcal{I}}A_jf|_{R_j}\overline{G}_{R_j}(x)}_\text{internal cells}  +
  \underbrace{\sum_{R_j\in\mathcal{B}}\overline{G}_{R_j}(x)\int_{\partial R_j\setminus\partial \Omega} \frac{\partial u(y)}{\partial \bf n}dl_y}_{\text{inner boundaries}}\\
  &+&
  \underbrace{\sum_{R_j\in\mathcal{B}}u(R_j)\int_{\partial R_j\cap\partial \Omega}\frac{\partial G(x,y)}{\partial \bf n}dl_y}_{\text{outer boundaries}}
  \label{eqn:expression}
  \end{eqnarray}

  We denote by $e_{ij}$ the common edge between two adjacent squares $R_i$ and $R_j$.
  Define $u_i\triangleq \frac{1}{A_i}\iint_{R_i}u(x)dA$ the average of $u(x)$ in square $R_i$.
  Now we discretize the sum of line integrals
  \begin{eqnarray}
  \nonumber && \overline{G}_{R_i}(x)\int_{e_{ij}} \frac{\partial u(y)}{\partial \mathbf{n}_{ij}}dl_y+\overline{G}_{R_j}(x)\int_{e_{ij}} \frac{\partial u(y)}{\partial \mathbf{n}_{ji}}dl_y\\
  \nonumber &\approx& a_{ij}(u_j-u_i)\overline{G}_{R_i}(x)+a_{ij}(u_i-u_j)\overline{G}_{R_j}(x)\\
  \nonumber &=& a_{ij}(\overline{G}_{R_j}(x)-\overline{G}_{R_i}(x))u_i+a_{ij}(\overline{G}_{R_i}(x)-\overline{G}_{R_j}(x))u_j\\
  &\approx & u(R_i)\int_{e_{ij}}\frac{\partial G(x,y)}{\partial\mathbf{n}_{ij}} dl_y + u(R_j)\int_{e_{ij}}\frac{\partial G(x,y)}{\partial\mathbf{n}_{ji}} dl_y
  \label{eqn:sumoflineintegral}
  \end{eqnarray}
  where the coefficients $a_{ij}$ is the ratio of the length $\|e_{ij}\|$ to the distance between the centers of $R_i$ and $R_j$.
  This implies the inner boundary and outer boundary terms in Eqn.~(\ref{eqn:expression}) are interchangeable in the \emph{discrete} sense.

  \textbf{Forming Basis Functions.}
  Denote by $\mathcal{N}_1(R_j)$ the set of 1-ring neighbors of $R_j$.
  The sum\\ $\sum_{R_i\in \mathcal{N}_1(R_j)}a_{ij}\left(\overline{G}_{R_i}(x)-\overline{G}_{R_j}(x)\right)$
  has a form similar to the basis functions of a harmonic B-spline~\cite{Feng2012},
  motivating us to explore the connection between Eqn.~(\ref{eqn:expression}) and harmonic B-splines.
  Note that $a_{ij}\left(\overline{G}_{R_i}(x)-\overline{G}_{R_j}(x)\right)$ is a poor approximation of the normal derivative across the common edge $e_{ij}$,
  since $R_i$ and $R_j$ may have different sizes,
  hereby the line between their centers is not perpendicular to $e_{ij}$ and it does not bisect $e_{ij}$ either.
  To obtain a high-quality discretization,
  we construct a Voronoi diagram using the interior and boundary nodes as generators and clip it with the boundary $\partial\Omega$.

  Denote by $\hat{e}_{ij}$ the Voronoi edge of two neighboring Voronoi cells $V_i$ and $V_j$.
  Then we compute the coefficient $\hat{a}_{ij}$ as the ratio of the length of Voronoi edge $\|\hat{e}_{ij}\|$ to the distance between the two generators.
  Finally, we define the basis function for Voronoi cell $V_j$ as
  \begin{equation}
  \label{eqn:basisfunction}
  \psi_j(x)=\sum_{V_i\in \mathcal{N}_1(V_j)}\hat{a}_{ij}\left(\overline{G}_{V_i}(x)-\overline{G}_{V_j}(x)\right).
  \end{equation}

  The functions $\{\psi_j\}_{j=1}^{n}$ are the basis functions of a harmonic B-spline, each of which is defined on a Voronoi cell.
  However, the line integrals in Equation~(\ref{eqn:expression}) are defined on rectangular areas $R_j$.
  Observe that there is a one-to-one correspondence between the Voronoi cells $\{V_j\}_{j=1}^n$ and the squares $\{R_j\}_{j=1}^n$,
  and $\overline{G}_{V_j}(x)\approx\overline{G}_{R_j}(x)$, since $\overline{G}$ is the average.
  Therefore, we can bridge the gap by adopting an alternative definition of $\psi_j$ as follows:
  \begin{equation}
  \label{eqn:approxbasisfunction}
  \psi_j(x)=\sum_{R_i\in \mathcal{N}_1(R_j)}\hat{a}_{ij}\left(\overline{G}_{R_i}(x)-\overline{G}_{R_j}(x)\right).
  \end{equation}

  Similar to the quad-tree edges, we classify the Voronoi edges into three disjoint sets, inner boundaries $\mathcal{E}_{ib}$,
  outer boundaries $\mathcal{E}_{ob}$ and interior edges $\mathcal{E}_{ie}$.
  Since each \textit{boundary} Voronoi cell is a quad-tree cell (see Figure~\ref{fig:solverpipeline}(d)),
  an inner (resp. outer) Voronoi boundary edge is also an inner (resp. outer) quad-tree boundary edge.

  \textbf{Computing Control Points.}
  Our goal is to express the solution $u(x)$ using a harmonic B-spline, i.e.,
  \begin{equation}
  \label{eqn:u_x}
  u(x)=\sum_{j=1}^n\lambda_j\psi_j(x),
  \end{equation} where $\lambda_j$ is the control point.

  As Equation~\ref{eqn:approxbasisfunction}) shows, each basic function $\psi_j(x)$ is computed by finite difference of Green's functions along Voronoi \textit{edges}.
  Therefore, we re-organize $\sum_j\lambda_j\psi_j(x)$ into the three terms in Equation~(\ref{eqn:expression}) by summing over Voronoi \textit{edges},
  \begin{eqnarray}
  \label{eqn:discrete_u}
  \nonumber  u(x)&=&\sum_{e_{ij}\in\mathcal{E}_{ib}\cup\mathcal{E}_{ie}}\left(\hat{a}_{ij}(\overline{G}_{R_j}-\overline{G}_{R_i})\lambda_i
  +\hat{a}_{ij}(\overline{G}_{R_i}-\overline{G}_{R_j})\lambda_j\right)\\
  \nonumber &+&\sum_{e_{ij}\in\mathcal{E}_{ob}}\hat{a}_{ij}(\overline{G}_{R_i}-\overline{G}_{R_j})\lambda_j\\
  \nonumber  &=& \sum_{e_{ij}\in\mathcal{E}_{ib}\cup\mathcal{E}_{ie}}\left(\hat{a}_{ij}(\lambda_j-\lambda_i)\overline{G}_{R_i}+\hat{a}_{ij}(\lambda_i-\lambda_j)\overline{G}_{R_j}\right)\\
  \nonumber  &+&\sum_{e_{ij}\in\mathcal{E}_{ob}}\hat{a}_{ij}(\overline{G}_{R_i}-\overline{G}_{R_j})\lambda_j \\
  \nonumber &=& \sum_{R_j\in\mathcal{I}}\left(\overline{G}_{R_j}\sum_{R_i\in \mathcal{N}_1(R_j)}\hat{a}_{ij}(\lambda_i-\lambda_j)\right)\\
  \nonumber &+&\sum_{R_j\in\mathcal{B}}\left(\overline{G}_{R_j}\sum_{e_{ij}\in\mathcal{E}_{ib}}\hat{a}_{ij}(\lambda_i-\lambda_j)\right)\\
  &+&\sum_{R_j\in\mathcal{B}}\left(\lambda_j\sum_{e_{ij}\in\mathcal{E}_{ob}}\hat{a}_{ij}(\overline{G}_{R_i}-\overline{G}_{R_j})\right).
  \end{eqnarray}

  For a boundary square $R_j\in\mathcal{B}$, we simply set $\lambda_j=g|_{R_j}$.
  To compute the control points $\lambda_j$ for internal cells, we solve the following linear system.
  Observe that for an internal square $R_j\in\mathcal{I}$, the coefficient $A_jf|_{R_j}$ is a constant.
  Define a Laplacian matrix $\mathbf{L}\in\mathbb{R}^{|\mathcal{I}|\times|\mathcal{I}\cup\mathcal{B}|}$,
  where $L_{ii}=\sum_j \hat{a}_{ij}$ and $L_{ij}=-\hat{a}_{ij}$ for $i\neq j$.
  We then partition $\bf L$ into block matrices $\mathbf{L}=[\mathbf{L}^\mathcal{I}|\mathbf{L}^\mathcal{B}]$,
  where $\mathbf{L}^\mathcal{I}\in\mathbb{R}^{|\mathcal{I}|\times|\mathcal{I}|}$ and $\mathbf{L}^\mathcal{B}\in\mathbb{R}^{|\mathcal{I}|\times|\mathcal{B}|}$.
  We also define vectors $\boldsymbol{\lambda}^\mathcal{I}\in\mathbb{R}^{|\mathcal{I}|}$,
  $\boldsymbol{\lambda}^\mathcal{B}\in\mathbb{R}^{|\mathcal{B}|}$,
  and $\mathbf{b}\in\mathbb{R}^{|\mathcal{I}|}$,
  where $\boldsymbol{\lambda}^\mathcal{I}$ is the unknowns,
  $\boldsymbol{\lambda}^\mathcal{B}$ is the given boundary conditions,
  and $b_j=A_jf|_{R_j}$.
  Then we solve the following linear system
  \begin{equation}
  \mathbf{L}^\mathcal{I}\boldsymbol{\lambda}^\mathcal{I}=\mathbf{b}-\mathbf{L}^\mathcal{B}\boldsymbol{\lambda}^\mathcal{B}.
  \label{eqn:linearsystem}
  \end{equation}
  Since the matrix $\mathbf{L}^\mathcal{I}$ is symmetric and positive definite,
  the linear system can be solved efficiently using Cholesky decomposition.

  \textbf{Remark.} Although Equation~(\ref{eqn:linearsystem}) is designed to match the first term of Equation~(\ref{eqn:discrete_u}) and the first term of Equation~(\ref{eqn:expression}),
  we show that with the computed control points $\boldsymbol\lambda$, the other two terms also match.
  Observe that Equation~(\ref{eqn:linearsystem}) is indeed Equation~(\ref{eqn:poissoneqn}) discretized at the quad-tree nodes.
  implying that for an internal cell $R_j$, the control point $\lambda_j\approx u(R_j)$.
  Also note that the coefficients $\hat{a}_{ij}$ are the weights of the discrete Laplace-Beltrami operator.
  Given an interior boundary edge $e_{ij}\in\mathcal{E}_{ib}$,
  we can verify that
  \begin{displaymath}
  \hat{a}_{ij}(\lambda_i-\lambda_j)=\hat{a}_{ij}(u(R_i)-u(R_j))=\int_{\partial R_j\cap\partial R_i}\frac{\partial u}{\partial\mathbf{n}}.
  \end{displaymath}
  Similarly, for an outer boundary edge $e_{ij}\in\mathcal{E}_{ob}$,
  we have
  \begin{displaymath}\hat{a}_{ij}(\overline{G}_{R_i}-\overline{G}_{R_j})
  = \int_{\partial R_j\cap\partial R_i}\frac{\partial G}{\partial\mathbf{n}}.
  \end{displaymath}
  Therefore, the second and third terms of Equation~(\ref{eqn:discrete_u}) match those of Equation~(\ref{eqn:expression}).

  \section{Rendering}
  \label{sec:rendering}

  \subsection{Algorithm}
  \label{subsec:renderingalgorithm}

  Given a set of PVG primitives in a domain $D$ and the user-specified resolution $l\times m$,
  we first discretize $D$ into an $l$-by-$m$ image $I$.
  Then we partition $D$ into a set of disjoint sub-domains $D=\cup_i D_i$,
  where each $D_i$ is bordered by a closed, double-sided diffusion curve.
  Since color diffusion cannot exceed the boundary $\partial D_i$ of each sub-domain $D_i$, we can render them in parallel.
  For each $D_i$, we rasterize the PVG primitives and their constraints that are inside $D_i$.
  After discretizing a Poisson region, we apply the flood-fill algorithm to fill in the region.

  \begin{algorithm}[htbp]
  \begin{algorithmic}
  \Require A set of PVG primitives in a domain $D$ and their constraints (color or Laplacian of color);
  $l\times m$: image resolution
  \Ensure PVG image
  \State Discretize the domain $D$ into an $l$-by-$m$ image $I$ and partition $D$ into a set of disjoint sub-domains, $D=\cup_{i} D_i$,
  where each $D_i$ is bordered by a closed diffusion curve
  \For {each sub-domain $D_i$}
  \State Rasterize the PVG primitives and their constraints and partition $D_i$ using a quad-tree
  \State Construct the Voronoi diagram using the leaf nodes of the quad-tree
  \State Construct the harmonic B-spline basis function $\{\psi_j\}$
  \State Compute the control points $\{\lambda_j\}$
  \For {every pixel $x\in D_i$}
    \State $I(x)=\sum_{j}\lambda_j\psi_{j}(x)$
  \EndFor
  \EndFor
  \end{algorithmic}
  \caption{Rendering PVG image}
  \label{alg:pvgrendering}
  \end{algorithm}

  Next we construct a quad-tree to partition $D_i$.
  We split a cell $c$ of the quad-tree is further split if the Laplacian constraint in $c$ is not a constant,
  it contains pixels on $\partial D_i$, or it is on the image boundary.
  The last two conditions are to ensure the leaf nodes are able to accurately capture the geometry of boundary $\partial D_i$.

  Taking the leaf nodes of the quad-tree as the generators, we compute a Voronoi diagram $\{V_j\}$.
  For each Voronoi cell, we construct a harmonic B-spline basis function $\psi_j$
  and compute its control point $\lambda_j$ by solving the sparse linear system in Eqn.~(\ref{eqn:linearsystem}).
  Note that the dimension of the linear system is equal to the number of leaf nodes of the quad-tree, which is significantly smaller than the number of pixels in $D_i$.

  Finally, we evaluate the color for a pixel $x\in D_i$ using the harmonic B-spline $I(x)=\sum_j\lambda_j\psi_j(x)$.
  Since the basis functions $\{\psi_j\}$ are localized, the summation applies to $x$'s \textit{local} neighbors $\mathcal{N}(x)$ only.

  PVG can handle anti-aliasing easily.
  Recall that diffusion curves and Poisson curves are double sided.
  For each pixel on a diffusion curve, we simply set its color using the boundary condition $g$.
  For each Poisson curve, we adopt the standard curve-drawing algorithm to re-draw both sides of the PC.
  Due to ``soft'' boundaries, Poisson regions do not have aliasing issue.

  \begin{figure}
  \centering
  \includegraphics[width=1.5in]{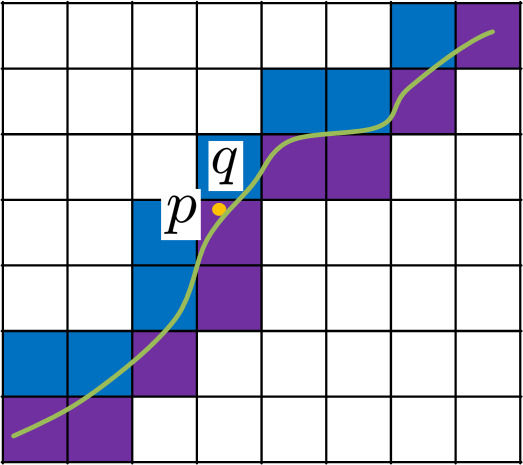}
  \includegraphics[width=1.5in]{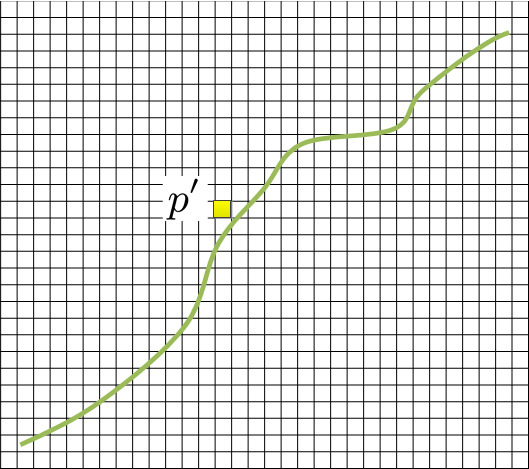}
  \includegraphics[width=1.35in]{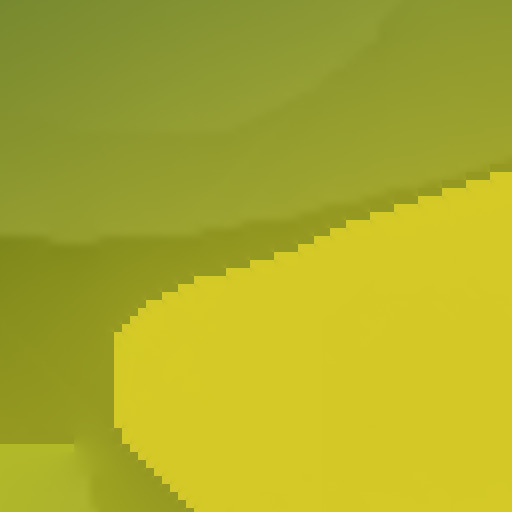}
  \includegraphics[width=1.35in]{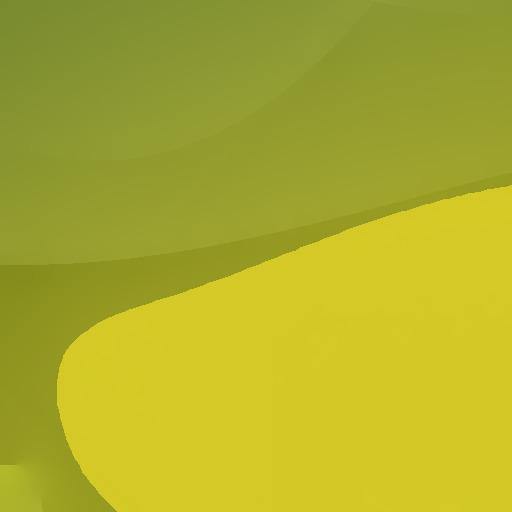}\\
  \makebox[1.5in]{(a)}\makebox[1.5in]{(b)}\makebox[1.35in]{(c)}\makebox[1.35in]{(d)}\\
  \caption{Location ambiguity and zooming-in.
  (a) In the low resolution image, the pixel $p$ is on the right side of the curve.
  (b) In the high resolution image, the sub-pixel $p'$ of $p$, is indeed on the left side.
  (c) Zigzagged boundaries occur if one doesn't address the issue of location ambiguity.
  (d) Find a pixel $q$ in the image of the original resolution so that $q$ is closest to $p'$ and it is on the same side as $p'$.
  Move $p'$ to $q$ and set $u(p')$ by $u(q)$.
  Then we smooth the color for the pixels near the boundary using Jacobi iterations [Jeschke et al. 2009].
  }
  \label{fig:infzoom}
  \end{figure}

  \subsection{Zooming-In}
  \label{subsec:zooming}

\begin{figure}[!ht]
  \centering
  \includegraphics[width=5.5in]{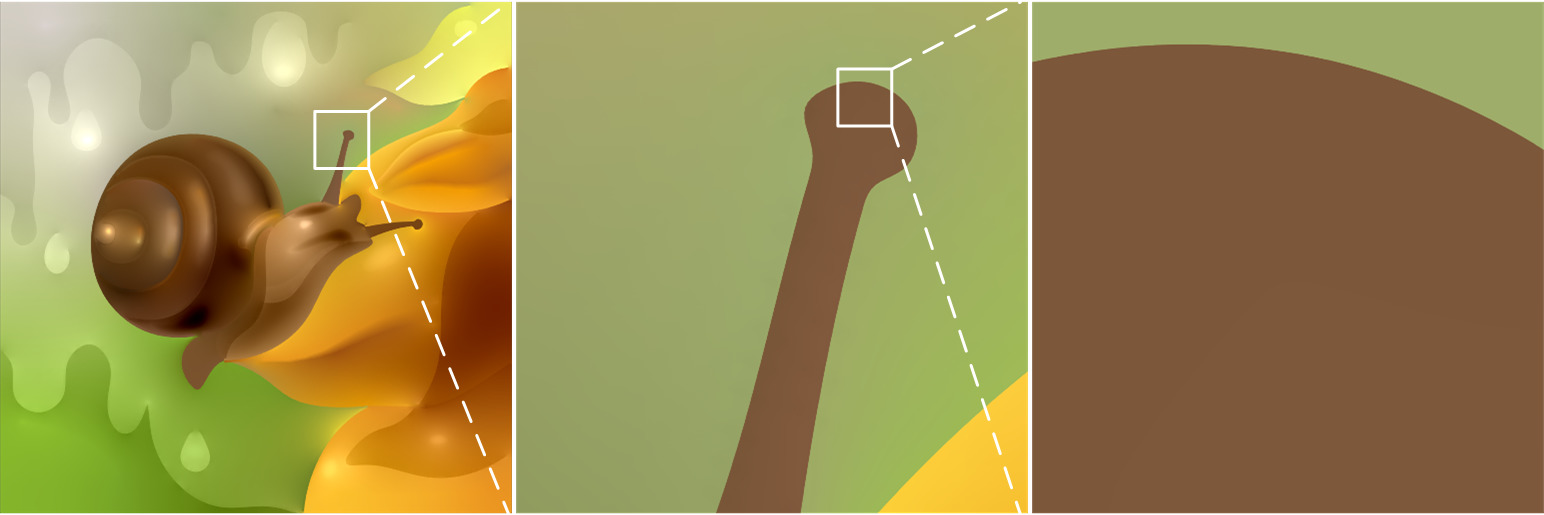}\\
  \makebox[1.9in]{$1\times$} \makebox[1.9in]{$10\times$} \makebox[1.9in]{$100\times$}
  \caption{Our closed-form solver supports random access evaluation, resolution-independent zoom-in and anti-aliasing.}
  \label{fig:zoom}
  \end{figure}

  As a kind of vector graphics, PVG is resolution independent and naturally supports zooming-in of arbitrary resolution.
  Consider an arbitrary rectangular region $\Omega'\subset\Omega$.
  Obviously, it is time consuming to re-discretize the sub-domain $\Omega'$ using a quad-tree of higher resolution and then construct a new harmonic B-spline.
  Instead, we simply keep the quad-tree, the Voronoi diagram and the harmonic B-spline constructed from the original domain $\Omega$.
  For each \textit{interior} point $p\in\mathcal{I}$, we can directly evaluate its color using Eqn.~(\ref{eqn:u_x}).
  As Figure~\ref{fig:infzoom}(a)-(c) shows, the tricky part comes from the boundary Voronoi cells $\mathcal{B}$, due to location ambiguity.
  We fix the problem by resolution-aware point location and Jacobi iteration-based smoothing (see Figure~\ref{fig:infzoom}(d)).
  Note that our zooming-in algorithm neither increases the resolution of the quad-tree nor recomputes the Voronoi diagram.
  Therefore, it is both time and space more efficient than na\"ively applying Algorithm 1 to a high resolution image.
  Computational results show that the performance of the zooming-in algorithm is independent of the zoom factor.
  See Figure~\ref{fig:zoom} for an example.

\section{Results \& User Study}
\label{sec:results}

\begin{figure}
  \centering
  \includegraphics[width=2in]{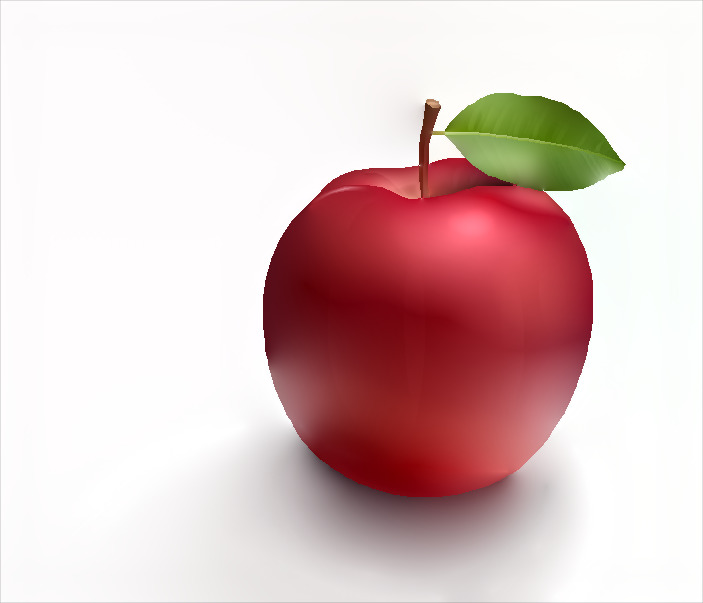}
  \includegraphics[width=2in]{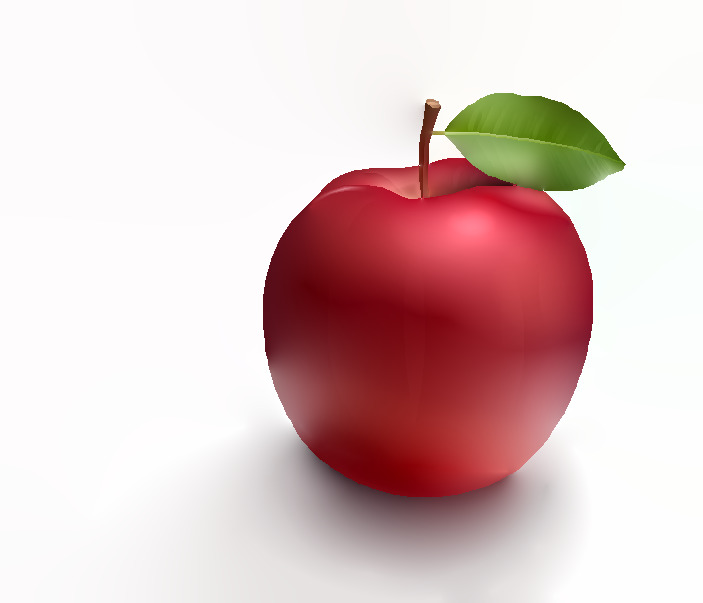}
  \includegraphics[width=2in]{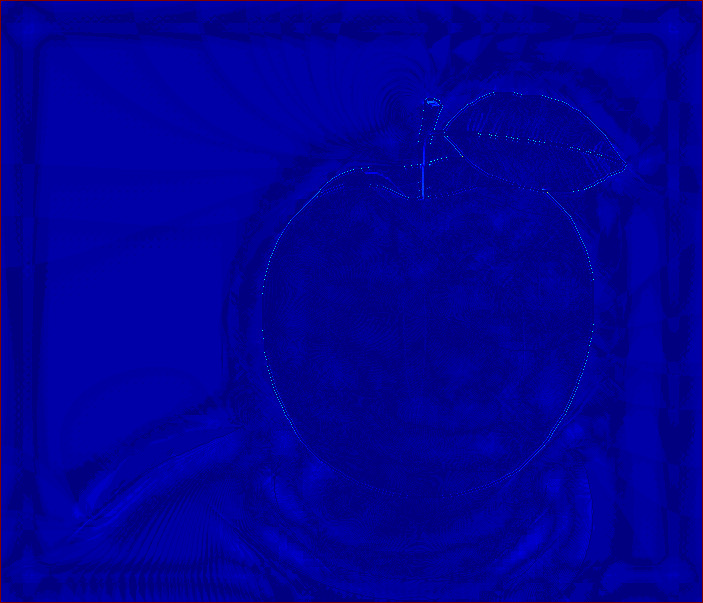}
  \includegraphics[width=0.35in]{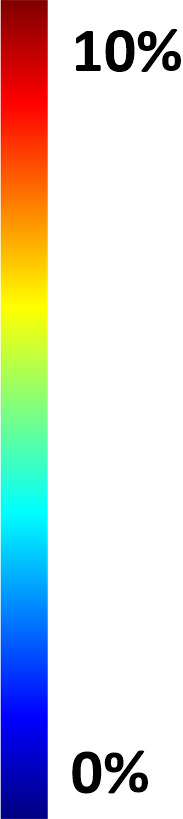}\\
  \makebox[2.0in]{Our solver} \makebox[2.0in]{Finite element method} \makebox[2.0in]{Error map}\\
  \caption{Accuracy. Although our solution is approximate, the relative mean error is only 0.301\%,
  which is equivalent to 0.77 in the scale of 255 as in the 24-bit color model.
  Such small differences are indistinguishable by naked eyes.
  }
  \label{fig:accuracy}
  \end{figure}

  We implemented our PVG solver in C++ and CUDA7.5,
  and tested it on a PC with an Intel Xeon E5-2609 v2~\@2.50GHz and an Nvidia Quadro K5000 GPU.
  Given a PVG with user-specified resolution, we first discretize the geometric primitives (represented by spline curves)
  and partition the image domain into disjoint regions.
  Then we run our solver in parallel on each individual region.
  As Table~\ref{tab:performance} reports, our solver is efficient and can render most testing PVGs in less than 1 second.
  Since the harmonic B-spline based solver provides an approximate solution, we evaluated its accuracy by comparing to the standard finite element method, which solves a large sparse linear system, whose dimension is the total number of pixels in the image.
  We observe that our solver is accurate, producing relative mean errors no more than 0.3\%.
  As Figure~\ref{fig:accuracy} shows, our solver produces PVG images with no visual difference from those of the exact solver,
  but it runs much faster.
  The weights $\hat{a}_{ii}$ are always positive and the Laplacian matrix is positive definite and diagonally dominant,
  implying that the condition number of $\bf L$ is low.
  Therefore, solving the linear system (\ref{eqn:linearsystem}) is numerically stable.

\begin{figure*}
  \centering
  \includegraphics[width=\linewidth]{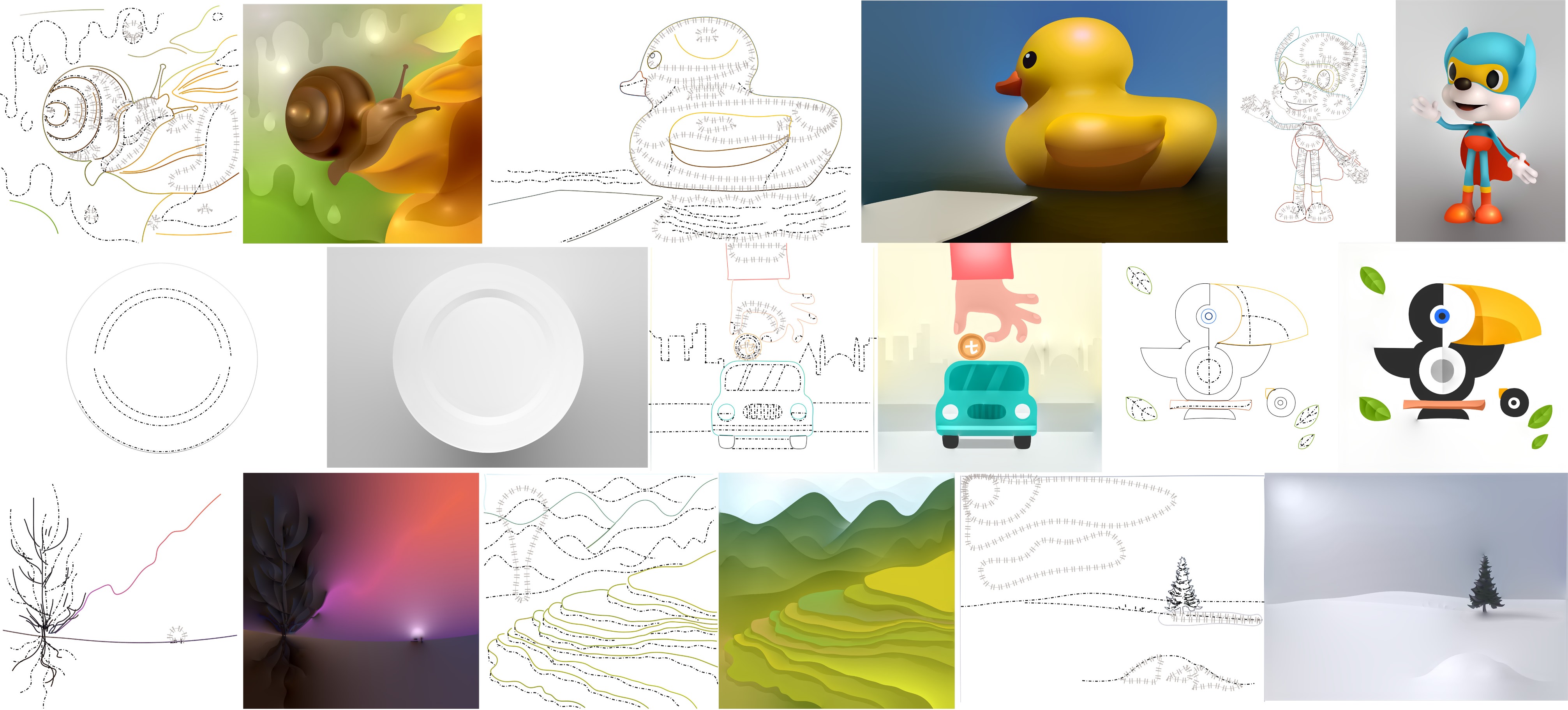}
  \caption{A gallery of Poisson vector graphics.}
  \label{fig:gallery}
  \end{figure*}

\begin{figure}
  \centering
  \includegraphics[width=5.5in]{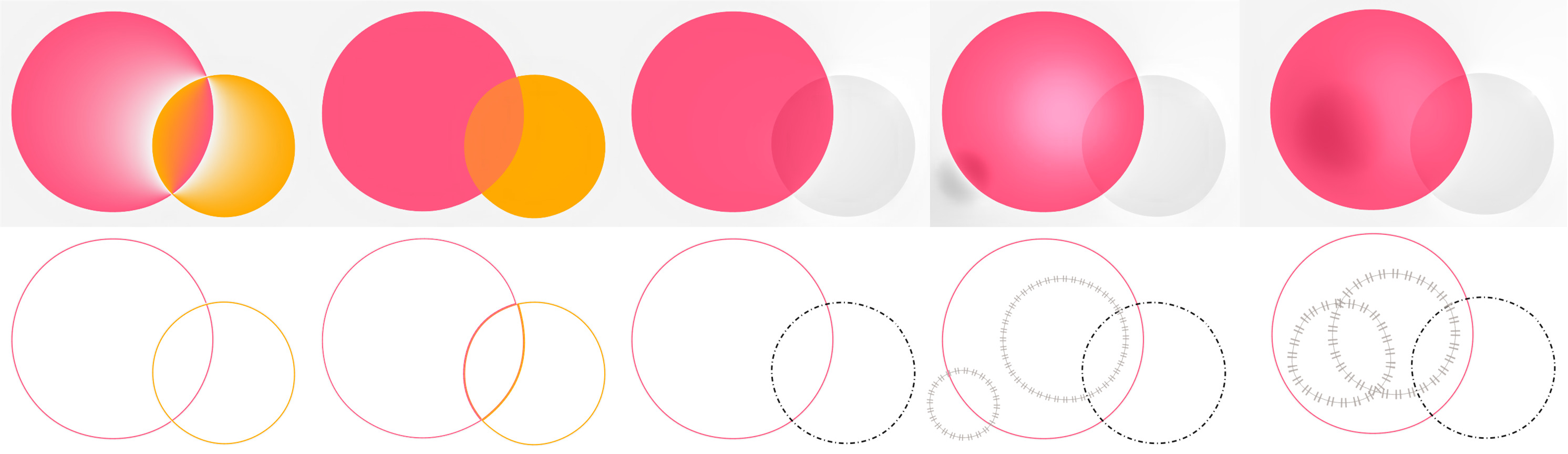}\\
  \makebox[1in]{(a)}\makebox[1in]{(b)}\makebox[1in]{(c)}\makebox[1in]{(d)}\makebox[1in]{(e)}\\
  \caption{Intersecting primitives. (a) The colors attached to two intersecting DCs compete with each other,
  leading to undesired artifacts.
  (b) Within the diffusion curve framework, users have to split the curves into disjoint segments with different colors in order to produce smooth colors.
  (c)-(e) In contrast, PVG allows all types of intersection except for DC-DC intersection.
  }
  \label{fig:intersecting}
  \end{figure}

  Figure~\ref{fig:gallery} demonstrates that PVG is able to express both cartoon-like images and photorealistic images.
  Poisson curves, as a double-sided curve with opposite Laplacian constraint,
  create high contrast (i.e., color discontinuity) across the curve,
  which is desired to model object boundaries and sharp features.
  Poisson regions, on the other hand, have controllable ``soft'' boundaries,
  are effective to produce highlights, core shadows, halos and transparency.
  In addition, we allow PRs intersecting each other and other types of primitives,
  providing more flexibility and expressive power to mimic complex shadings in photorealistic images.

  \begin{table*}[ht]
  \centering
  \caption{Statistics.
  $N_{dc}$, $N_{pc}$, $N_{pr}$, $N_s$: the number of diffusion curves, Poisson curves, Poisson regions and sub-domains, respectively;
  $T_{d}$, $T_s$: time for domain discretization and our PVG solver;
  $T$: the total running time;
  $\varepsilon$: the relative mean error.
  }
  \setlength\tabcolsep{3pt}

  \begin{tabular}{|l|c|c|c|c|c|c|c|c|c|}
  \hline
  Image & Resolution & $N_{dc}$ & $N_{pc}$ & $N_{pr}$& $N_{s}$ & $T_{d} (s)$ & $T_{s} (s)$ & $T (s)$ & $\varepsilon$\\
  \hline
  \hline
  Apple &$703\times 603$ & 38 & 54 & 16  & 30 & 0.148 &  0.414 & 0.562 & 0.301\%\\
  \hline
  Bird & $512\times 512$ & 54 & 16 & 0  & 16 & 0.066 &  0.227  & 0.293 & 0.306\%\\
  \hline
  Blood Seeker & $765\times 765$ & 85 & 71 & 24  & 38 & 0.217 & 0.93  & 1.147 & 0.260\% \\
  \hline
  City & $512\times 512$ & 35 & 76& 3  & 7 & 0.089 & 0.438  & 0.527& 0.339 \%\\
  \hline
  Dawn & $512\times 512$ & 48 & 16 & 1  & 31 & 0.064 &  0.376  & 0.440 & 0.162\%\\
  \hline
  Duck & $ 512\times 512$ & 26 & 1 & 24  & 10 & 0.066 &  0.306  & 0.372 & 0.259 \%\\
  \hline
  Egg & $512\times 512$  & 3 & 0 & 3  & 2 & 0.048 & 0.169 & 0.217 & 0.275 \% \\
  \hline
  Light bulb & $827\times 620$ & 8 & 0 & 3  & 31 & 0.099 &  0.451 & 0.550 & 0.0934\% \\
  \hline
  Macau & $800\times 600$ & 45 & 80 & 18  & 51 & 0.185 &  0.486 & 0.671 & 0.213\%\\
  \hline
  Mountains & $512\times 512$ & 15 & 29 & 1  & 1  & 0.077 &  0.708 & 0.785 & 0.254\%\\
  \hline
  Plate & $716\times 492$ & 1 & 5 & 0  & 2 & 0.069 &  0.263 & 0.332 &0.267\%\\
  \hline
  Rubber Duck & $940\times 622$ & 12 & 21 & 11  & 7 & 0.149 &  0.579 & 0.728 & 0.164\%\\
  \hline
  Snail & $512\times 512$ & 30 & 20 & 20  & 4 & 0.077 & 0.463 & 0.540 & 0.240\% \\
  \hline
  Snow & $753\times 565$ & 34 & 39 & 8  & 30 & 0.121 &  0.519 & 0.640 & 0.303\% \\
  \hline
  Superhero & $702\times 1000$ & 70 & 14 & 44  & 35 & 0.213 &  0.634 & 0.847 & 0.245\%\\
  \hline
  Teapot & $800\times 600$ & 21 & 0 & 5  & 6 & 0.093  &  0.332 & 0.425 & 0.233\% \\
  \hline
  \end{tabular}

  \label{tab:performance}
  \end{table*}

  \begin{figure}[htbp]
  \centering
  \includegraphics[width=3in]{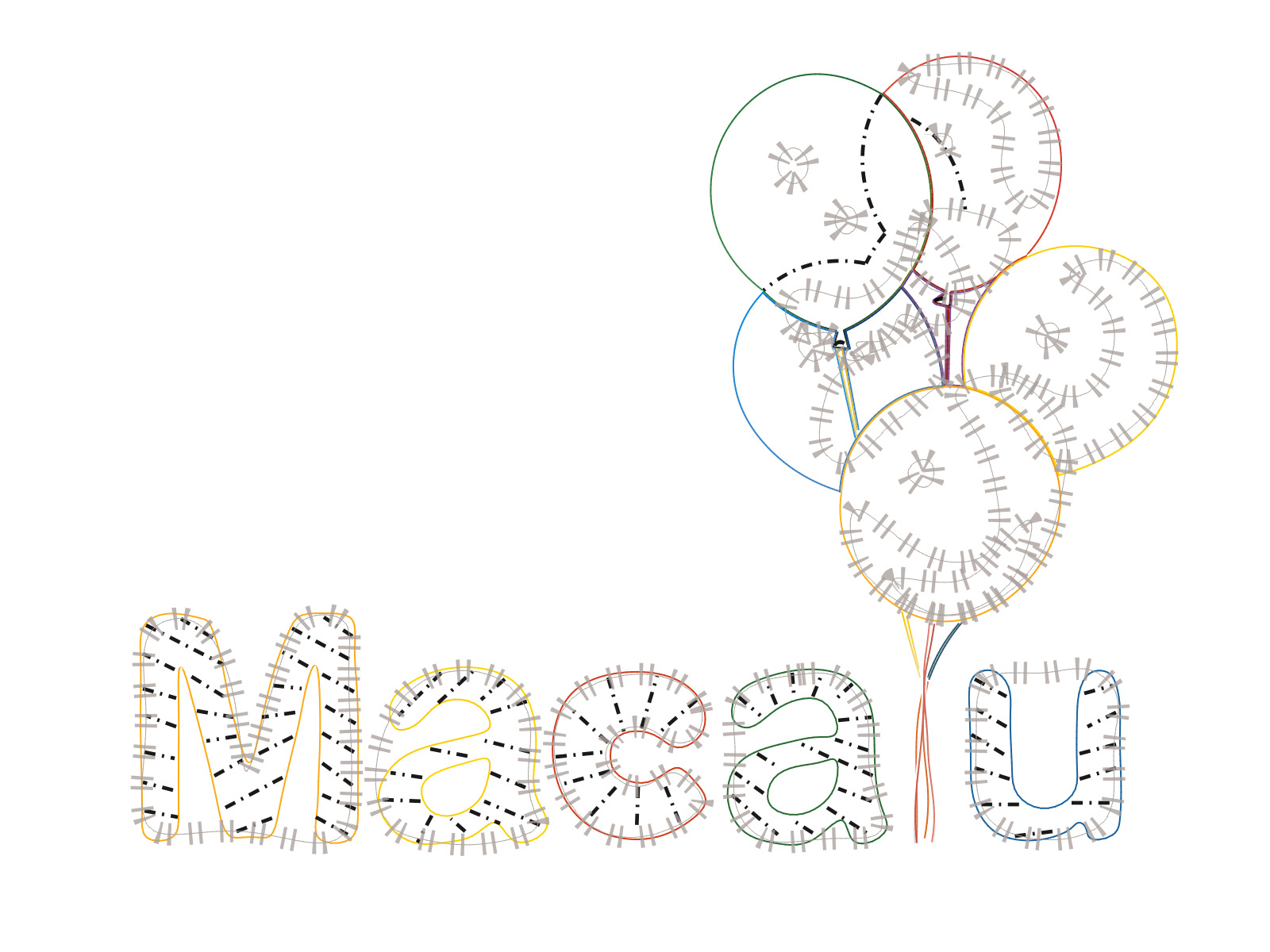}
  \includegraphics[width=3in]{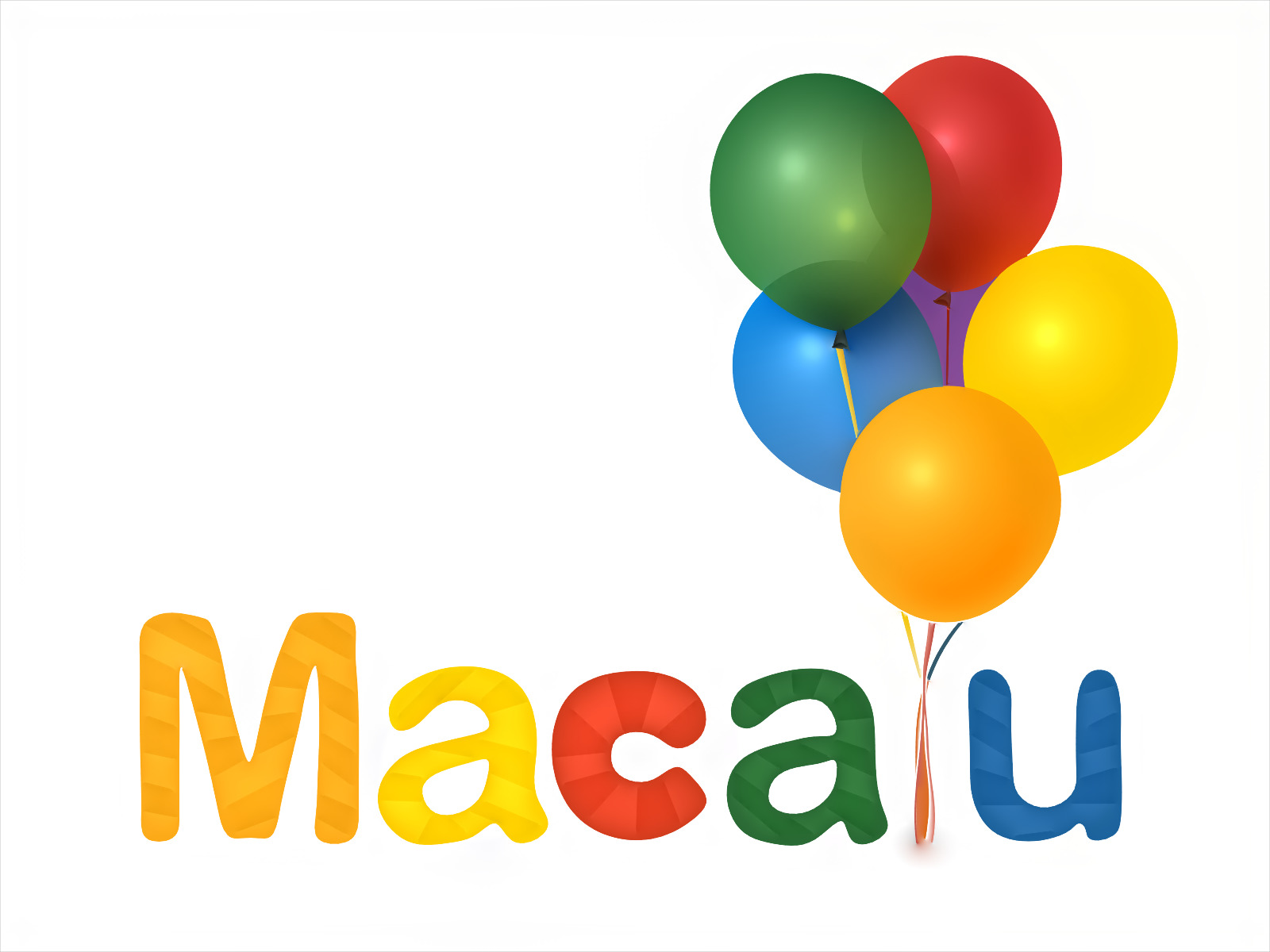}\\
  \caption{Poisson regions produce translucency on the balloons.}
  \label{fig:macau}
  \end{figure}

  We conducted a preliminary user study to explore benefits (efficiency and  usability) of PVG and compared it with TPS based vector graphics developed by Finch et al.~\cite{Finch2011}.
  We recruited 12 participants who are professional 2D artists with 9.3 years painting experience on average.
  Before going to the tasks in the experiments,
  we introduced the key concepts of DC, PR, PC and TPS to the participants,
  and gave them 20 minutes to familiarize both software.
  After that, they were asked to complete two painting tasks.
  To help the participants complete the tasks, we showed them the expected results.

\begin{figure}
  \centering
  \includegraphics[width=1.5in]{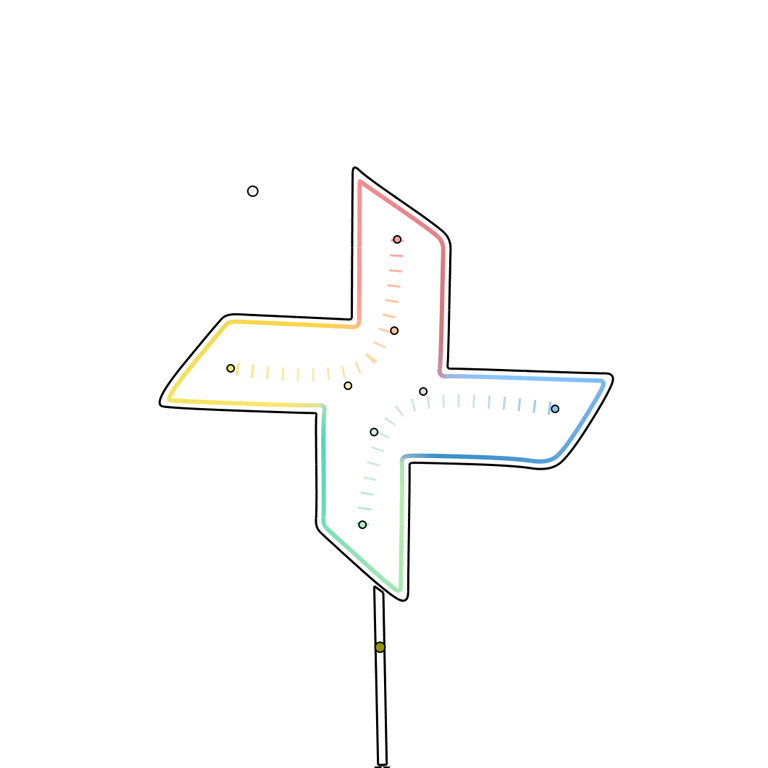}
  \includegraphics[width=1.5in]{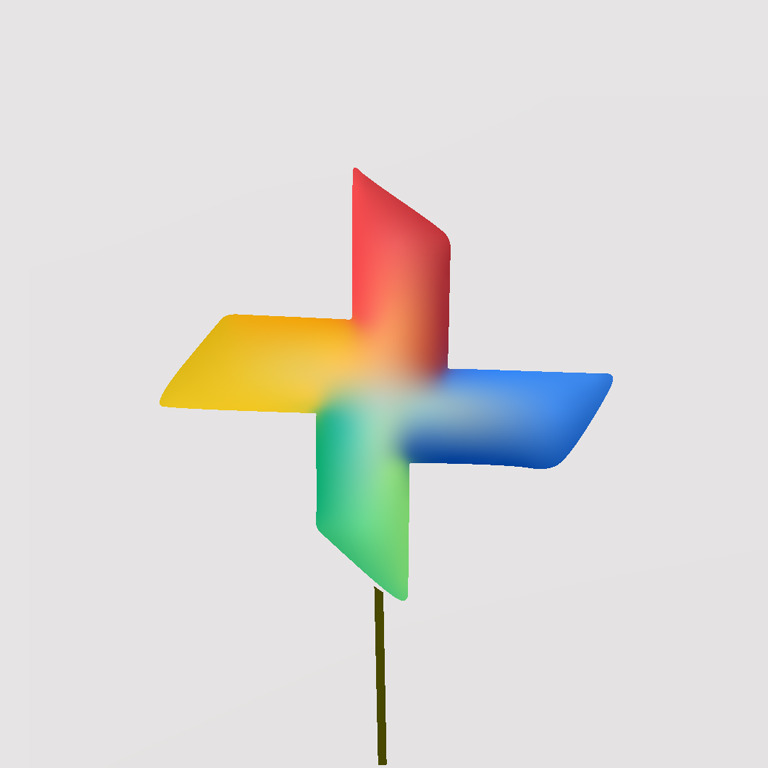}
  \includegraphics[width=1.5in]{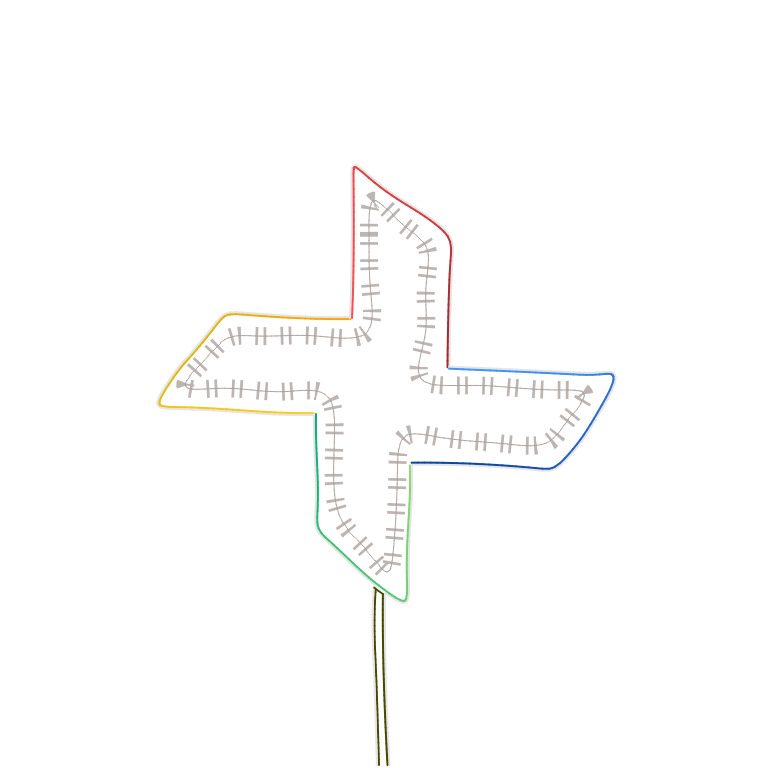}
  \includegraphics[width=1.5in]{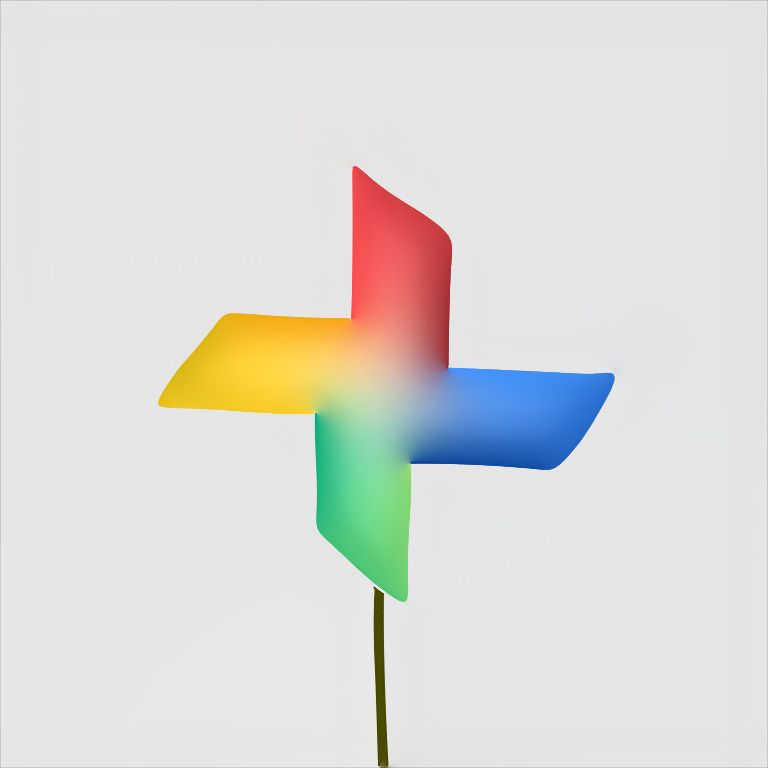}\\
  \makebox[3in]{(a) Thin-plate splines}
  \makebox[3in]{(b) PVG}\\
  \includegraphics[width=3in]{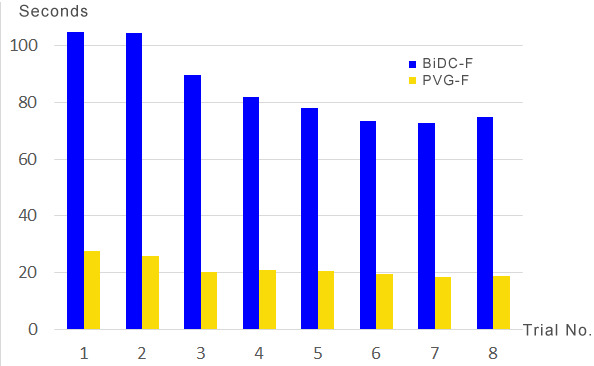}
  \includegraphics[width=3in]{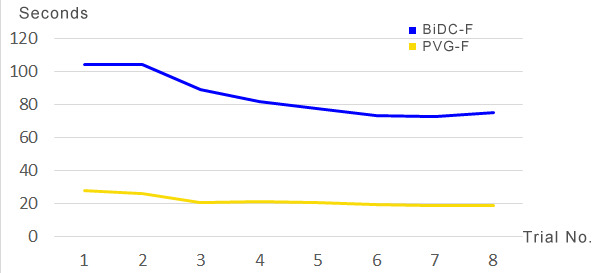}\\
  \makebox[3in]{(c) Time}\makebox[3in]{(d) Learning curve}\\
  \caption{\label{fig:windwill} Experiment \#1: Local shading control.
  }
  \end{figure}

  We designed two tasks to evaluate local shading control and permit of intersection among geometric primitives in the user study.
  In experiment \#1, the participants were given a colorful windmill and they were asked to add a highlight on it (see Figure~\ref{fig:windwill}).
  With TPS, they sketched two VS curves - the compound value and slope curve - each curve has at least 4 control points for specifying the boundary colors.
  Note that, four colors are the least number of colors to simulate the rich colors of the area we provided.
  With PVG, participants sketched one Poisson region and then specified its Laplacian constraint using slider,
  which is much easier than specifying \textit{discrete} colors.
  Also, it is non-intuitive to \textit{explicitly} specifying colors for \textit{shaded} objects with spatially-varying colors but the same tone
  As Figure~\ref{fig:windwill}(c) shows, PVG reduces roughly 75\% of drawing time as compared to DC and TPS.

\begin{figure}
  \centering
  \includegraphics[width=2.8in]{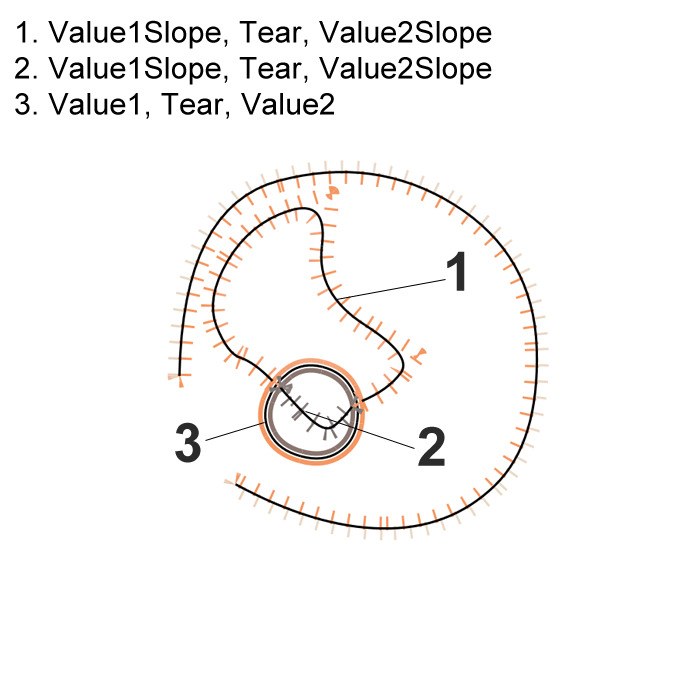}
  \includegraphics[width=2.8in]{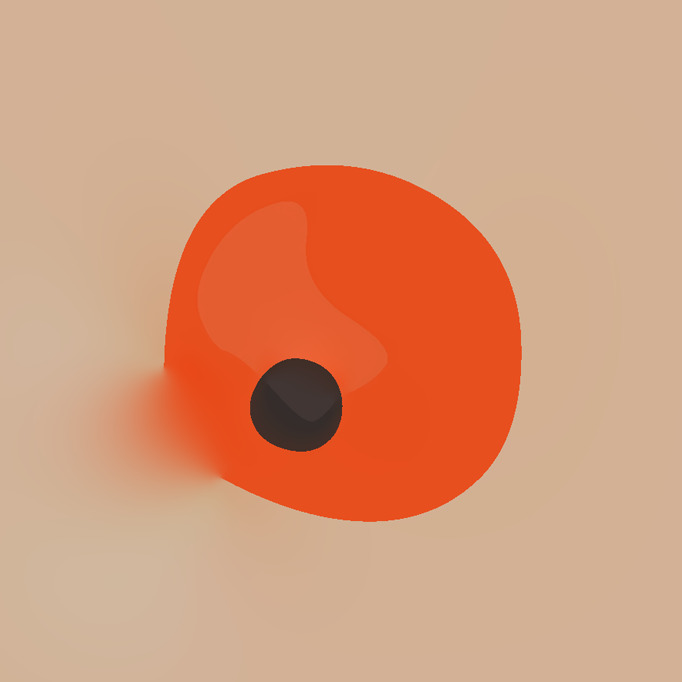}\\
  \makebox[3.20in]{(a) Thin-plate splines}\\
  \includegraphics[width=2.8in]{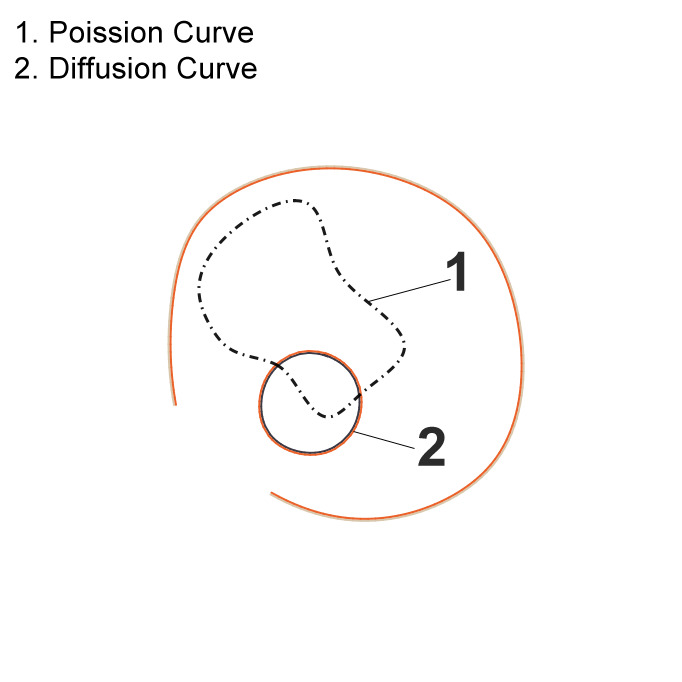}
  \includegraphics[width=2.8in]{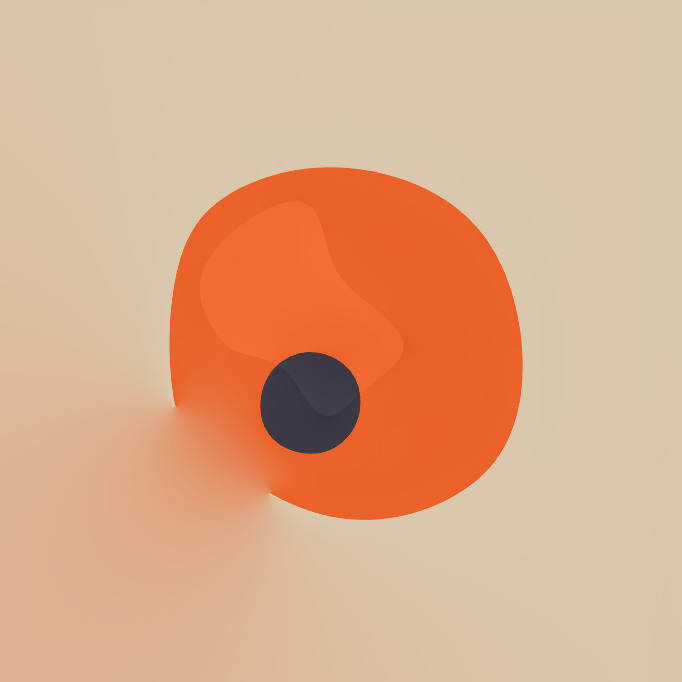}\\
  \makebox[3.2in]{(b) Poisson vector graphics}\\
  \includegraphics[width=2.8in]{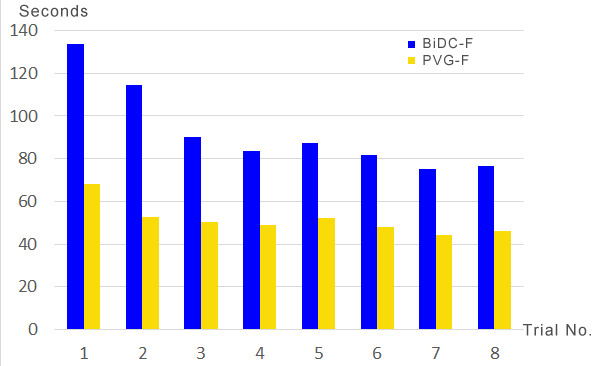}
  \includegraphics[width=2.8in]{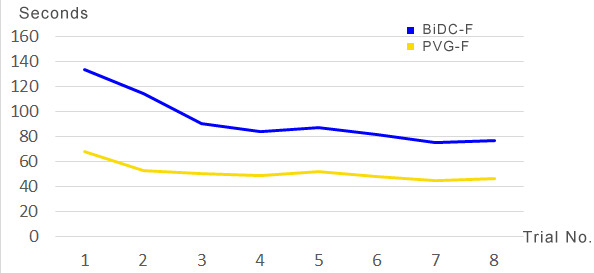}\\
  \makebox[3.2in]{(c) Time}\makebox[3.2in]{(d) Learning curve}\\
  \caption{\label{fig:bug} Experiment \#2: Handling intersecting primitives.
  DC and TPS do not allow intersecting primitives.
  Therefore, one has to partition the closed curve into two disjoint segments (labeled 1 and 2 in (a)).
  With PVG, one can draw Poisson curves/regions anywhere.
  }
  \end{figure}

  In experiment \#2, they were given a ``ladybug'' and required to add a highlight on its back (see Figure~\ref{fig:bug}).
  To simplify the task, the ladybug has only one spot.
  Since the highlight crosses the black spot,
  the participants had to partition DCs and TPS into disjoint segments and then specify their colors separately.
  With PVG, they simply sketched one PC in a single stroke and specified only one Laplacian constraint.
  As a result, PVG saves 40\% of the drawing time as compared to DC and TPS (see Fig.~\ref{fig:bug}(e)).

  To evaluate the ease of learning, we asked the participants to repeat each experiment 8 times.
  As the learning curves in Figures~\ref{fig:windwill}(d) and~\ref{fig:bug}(f) show,
  the PVG curves tend to stabilize in only 3 or 4 trials, whereas the TPS curves take longer, from trial 6 onwards, to become stable.
  implying that PVG is easier to learn compared to TPS.
  See the accompanying video for details.

  The participants commented that PVG faithfully follows the basic painting principle by separating color and tone.
  In painting theory~\cite{Staiger2016}, tones are comprised of highlight, halftone, core shadow, reflected light and cast shadow,
  which are the key factors to produce photorealistic rendering.
  With PC and PR, they can simulate various types of tones and control them in an easy and intuitive manner (e.g, experiment \#1).
  Although raster graphics naturally supports this painting style (e.g., the dodging and burning in Adobe Photoshop),
  this is the first time that they saw it in vector graphics.
  They also commented that PVG is more flexible than DC and TPS, since it allows geometric primitives intersecting each other.
  This feature enables them to use layers in complex drawings, such as experiment \#2.

\section{Comparison}
\label{sec:compare}

  Color diffusion based vector graphics can be classified according to the order of the PDE.
  First-order DCs solve Laplace's equation with Dirichlet boundary condition,
  while second-order DCs solve the bi-Laplace's equation with Neuman boundary condition.
  Table~\ref{tab:comparison} summarizes the major features of various vector graphics.
  In the following, we compare PVG with first-order and second-order DCs as well as their variants.

  \textbf{Comparison with diffusion curves.}
  Since diffusion curve images are harmonic functions, they do not support control of color gradient, which, however, is highly desired for producing photorealistic effects.
  To overcome this limitation,
  Orzan et al.~\cite{Orzan2008} blurred the harmonic color function with a spatially varying blur attribute associated to each curve.
  However, the blurring operation also brings three new problems:
  First, it does not allow explicit control over the value or position of color extrema~\cite{Finch2011}. See Figure~\ref{fig:DCIartifact} for an example.
  Second, user cannot directly specifying the blur kernel size.
  Instead, user specifies a blur attribute for each curve and the kernel size is determined by diffusion those attributes.
  Third, since the blur attributes are defined on diffusion curves, one cannot blur the regions which are far away from the curves.
  Compared with DC, PVG does not require any post-processing and users can produce photorealistic effects easily.

\begin{figure}
  \centering
  \includegraphics[width=5in]{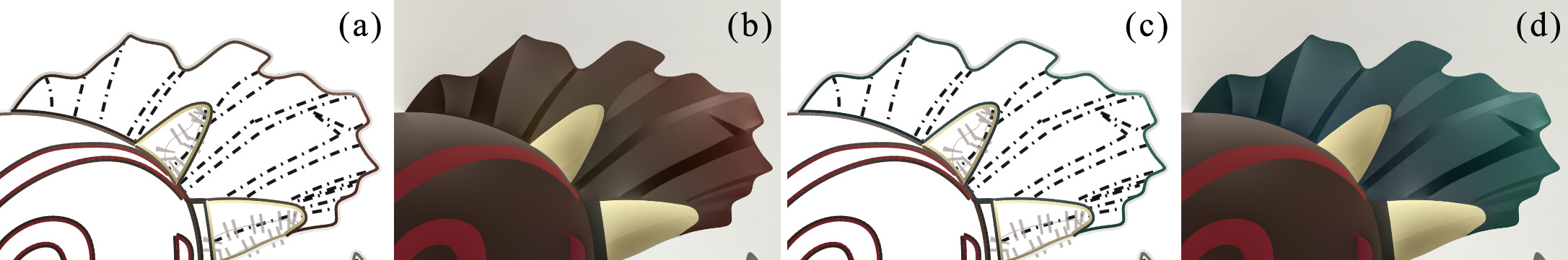}\\
  \includegraphics[width=5in]{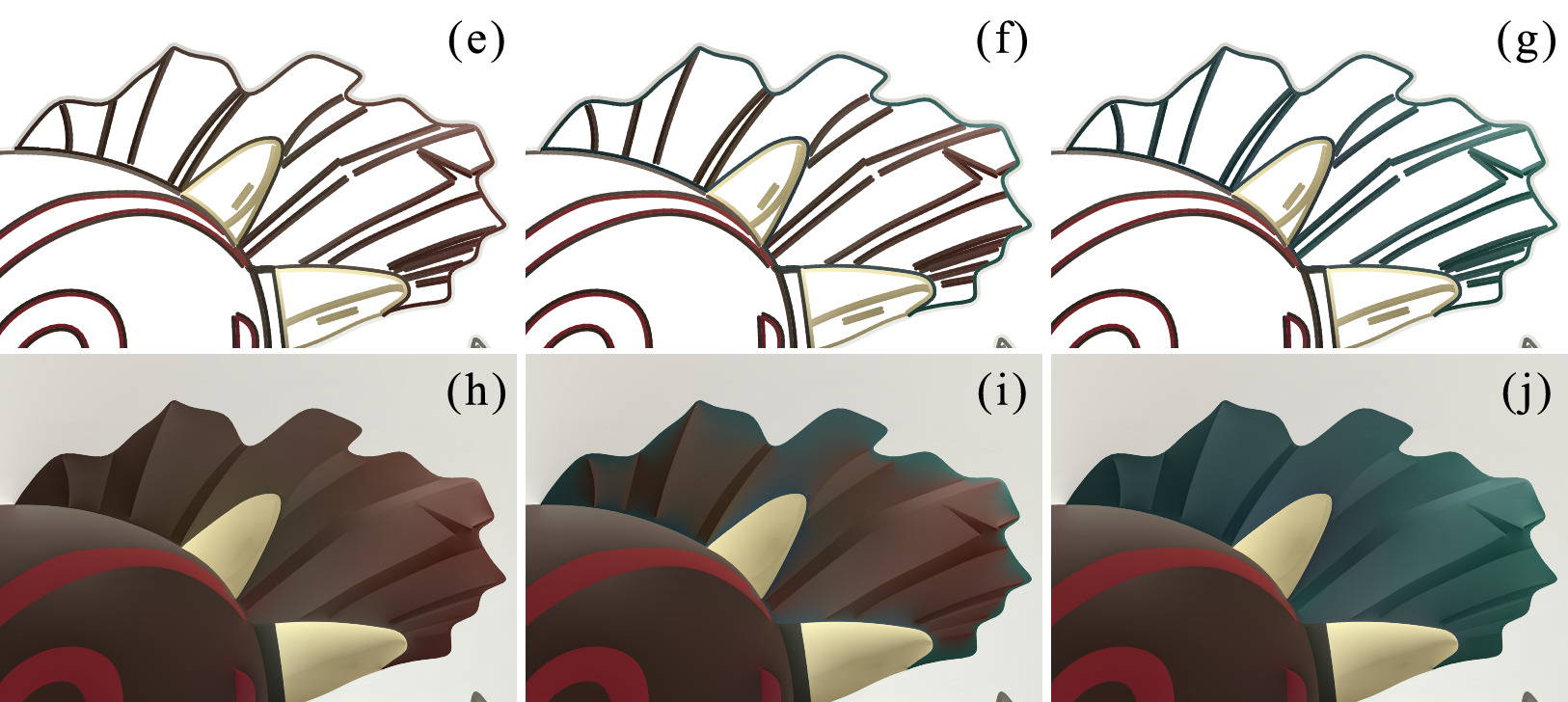}\\
  \caption{Changing hue in the blood seeker's cap,
  The region of interest $R$ is bounded by a closed diffusion curve, whose inner side color is dark brown.
  Row 1 (PVG): we simply changes the boundary color of $\partial R$ to blue,
  then the Poisson curves inside $R$ (dashed lines in (a) and (c)), still with the same Laplacian constraints,
  can automatically adjust to the new boundary condition so that the hue in $R$ are coherent with the boundary color.
  Rows 2 and 3 (DCI): replacing each Poisson curve to a diffusion curve with proper color conditions (see colored lines in (e)),
  one can generate a DCI image (see (h)) similar to the PVG image in (b).
  However, simply changing the boundary color of $\partial R$ (see (f)) in the DCI produces strange hue,
  due to the significant difference between the colors of the interior DCs and the boundary DC.
  To fix this, one has to manually adjust the boundary condition for each interior DC (see (g)),
  which is tedious and error prone.
  }
  \label{fig:blood_seeker_change}
  \end{figure}

  PVG is also superior in \textit{local} shading control.
  If two diffusion curves are close to each other, color diffusion is controlled by both curves.
  Therefore, changing the boundary condition (i.e., colors) of one diffusion curve often leads to a chain reaction to the neighboring curves.
  In contrast, PCs and PRs, constraining the Laplacians rather than colors,
  are \textit{loosely} coupled with the neighboring diffusion curves.
  Figure~\ref{fig:blood_seeker_change} shows an example that user wants to change the hue in a region of interest $\Omega$.
  For DCI, user has to manually adjust the colors for all DCs inside $\Omega$.
  With PVG, user only needs to change the color of the boundary DC $\partial\Omega$,
  then the Poisson curves inside $\Omega$, still with the same Laplacian $f$, produce colors that are coherent with the boundary color.

  Another limitation of DC is that it does not allow intersecting curves,
  since the colors associated to them compete with each other.
  In contrast, PVG allows all types of primitives except for two DCs intersecting each other.
  This feature enables the PVG users to create layers and provides them more flexibility in the design process.
  Take the Apple (Figure 1) as an example.
  Note that there are a large number of intersections among the primitives.

  \textbf{Comparison with constrained diffusion curves.}
  Bezerra et al.~\cite{Bezerra2010} proposed several techniques, such as diffusion barriers, diffusion anisotropy, and spatially varying color strength,
  to control the diffusion process.
  Their approach is able to diffuse both colors and normal maps, hereby producing interesting non-photorealistic effects.
  However, the constrained diffusion has two limitations.
  First, it is non-intuitive to specify the boundary condition for normals.
  Second, normals are only diffused \textit{within} the user-specified region (which may not be a closed diffusion curve),
  some artifacts (such as color discontinuity) on the region boundary may occur.
  With PVG, users can directly produce specular reflection using Poisson regions.
  Since the constraint is the Laplacian of colors, colors on the boundary of a PR are continuous.
  See Figure~\ref{fig:windwill}(c) for an example.

\begin{figure}
  \centering
  \includegraphics[width=2.5in]{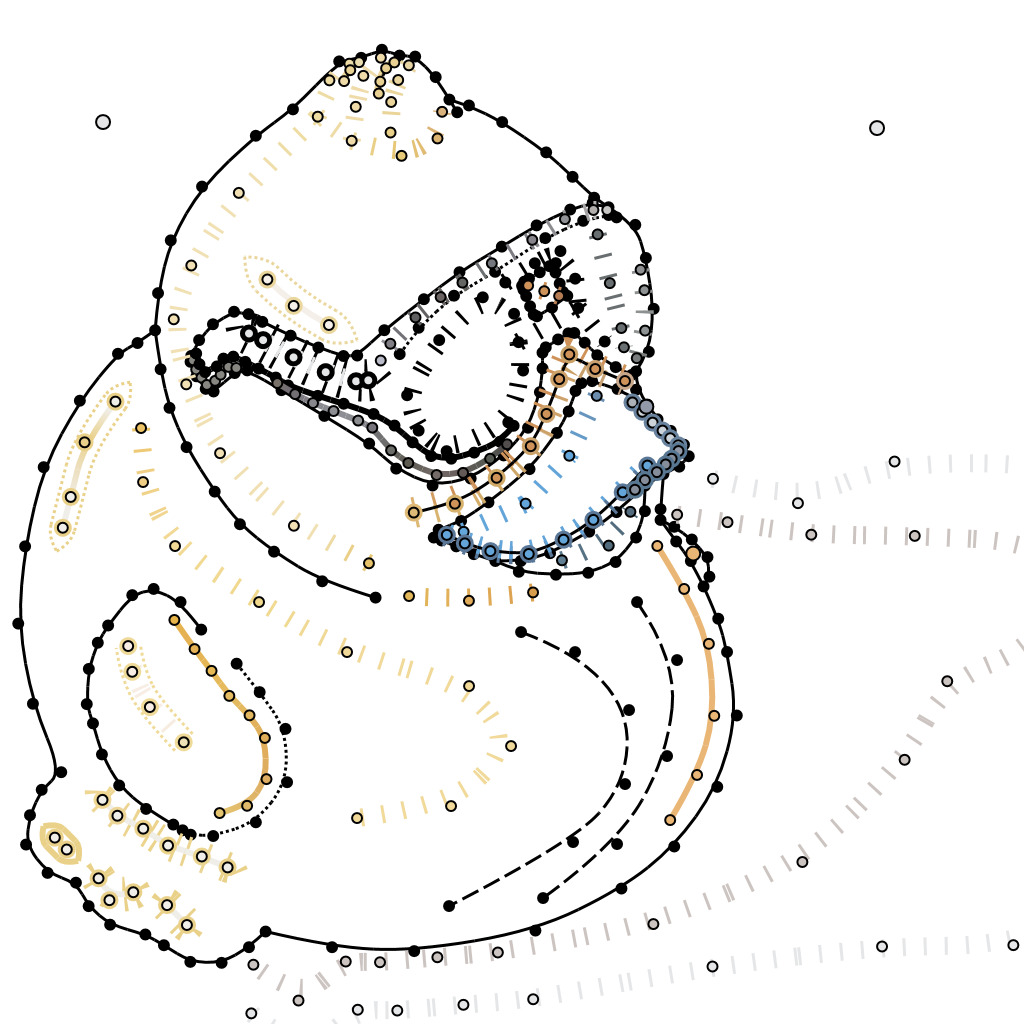}
  \includegraphics[width=2.5in]{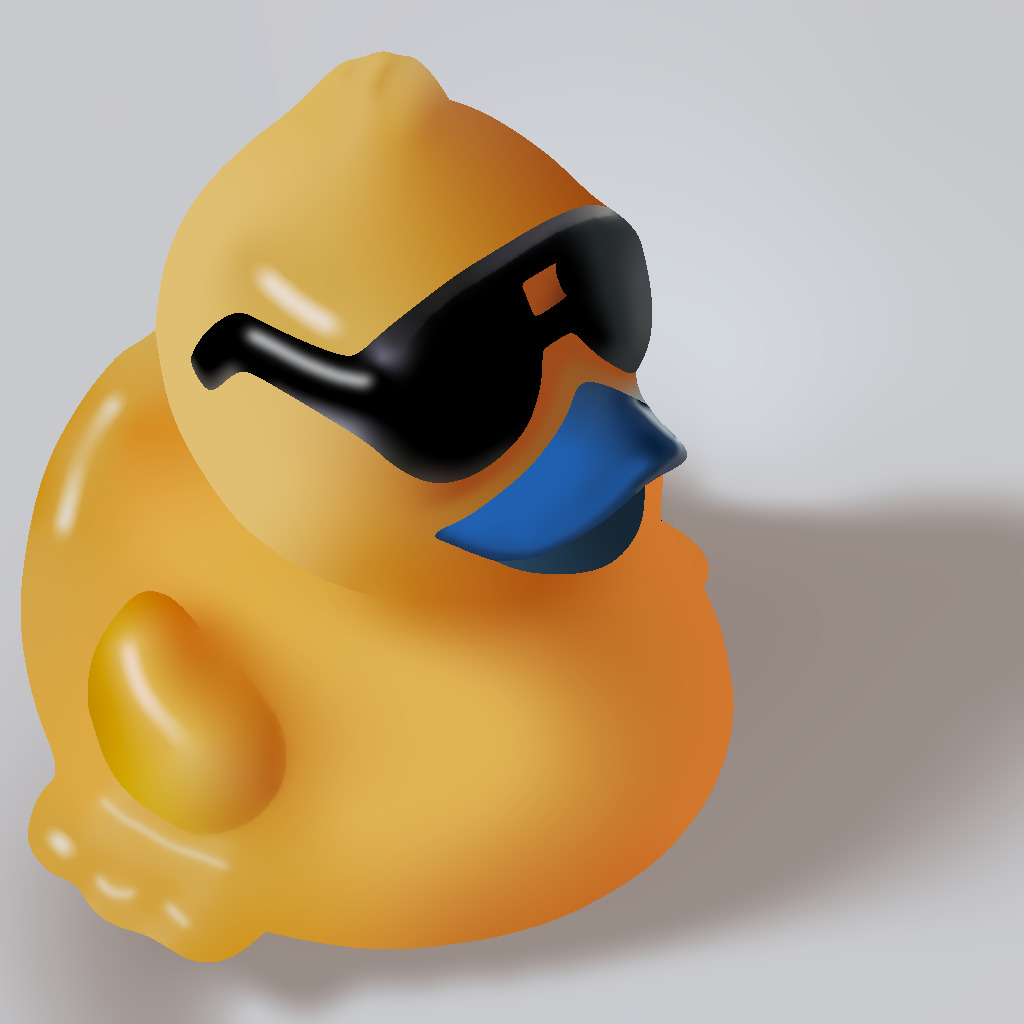}\\
  \includegraphics[width=2.5in]{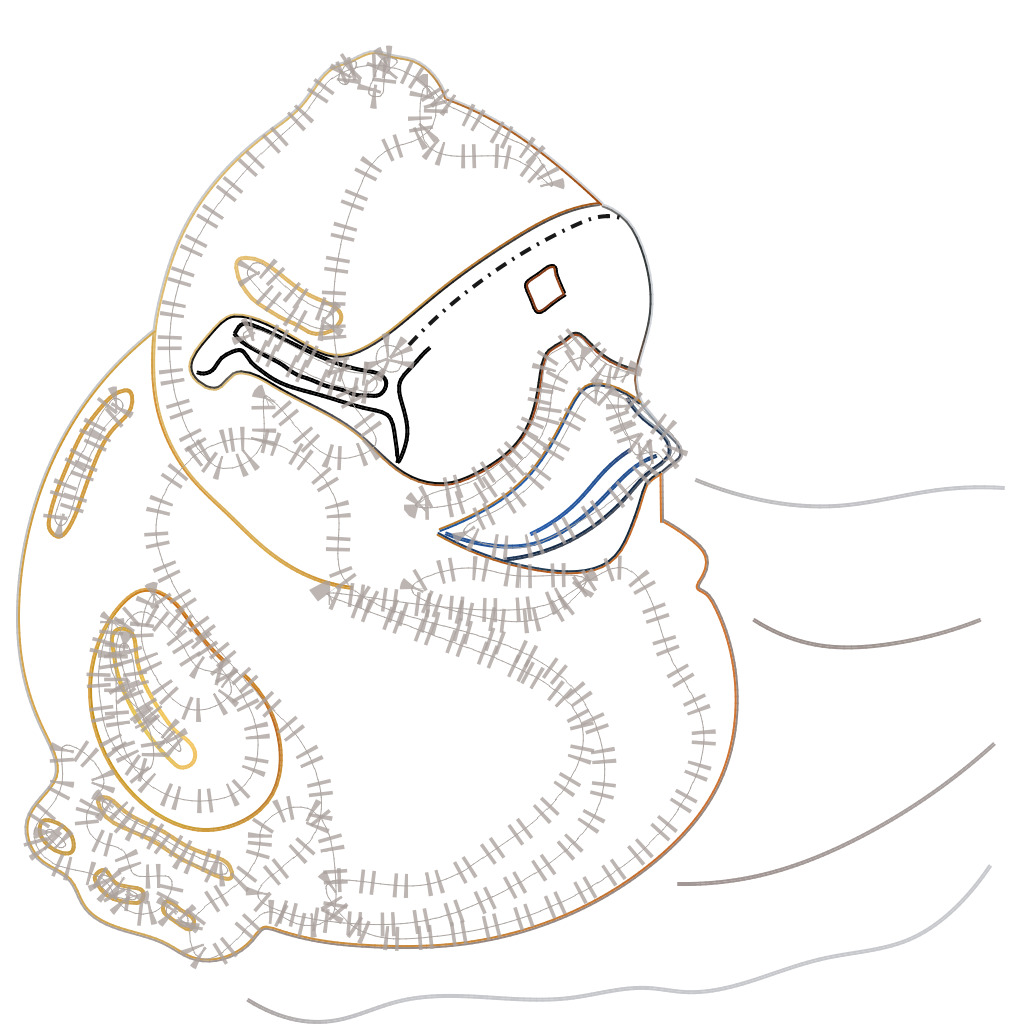}
  \includegraphics[width=2.5in]{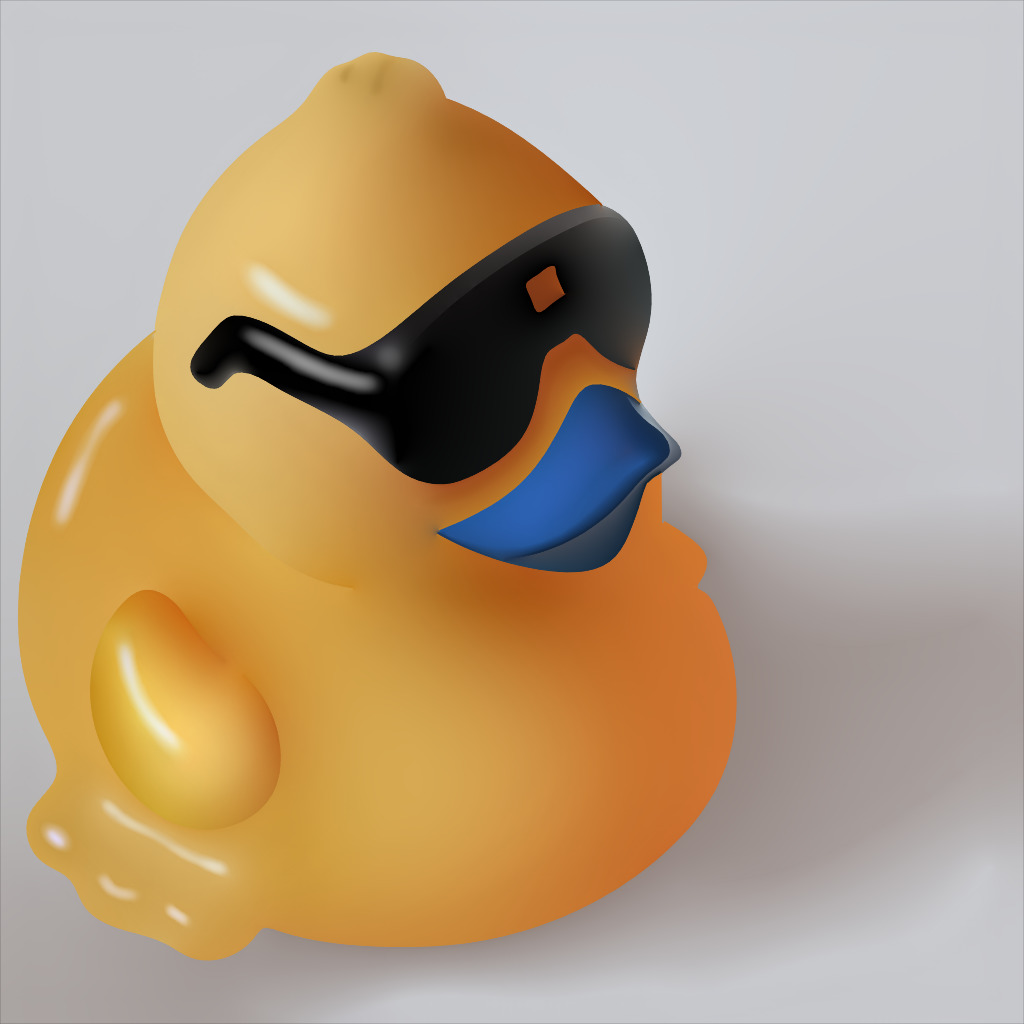}\\
  \caption{Comparison with TPS.
  Row 1: The TPS [Finch et al. 2011] consists of 53 primitives in 13 categories.
  Row 2: Our PVG, producing a similar image, contains 51 primitives in only 3 categories: 26 DCs, 1 PCs and 24 PRs.}
  \label{fig:comparewithhoppe}
  \end{figure}

\begin{figure}
  \centering
  \includegraphics[width=0.7in]{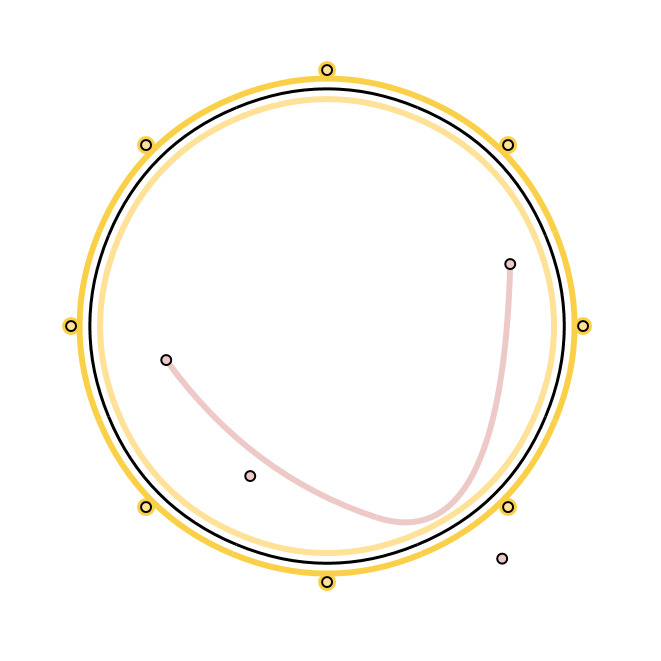}
  \includegraphics[width=0.7in]{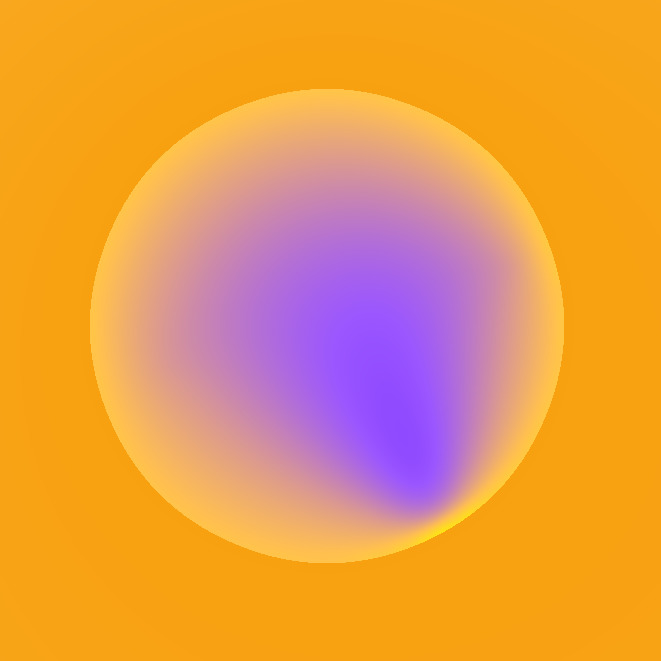}
  \includegraphics[width=1.66in]{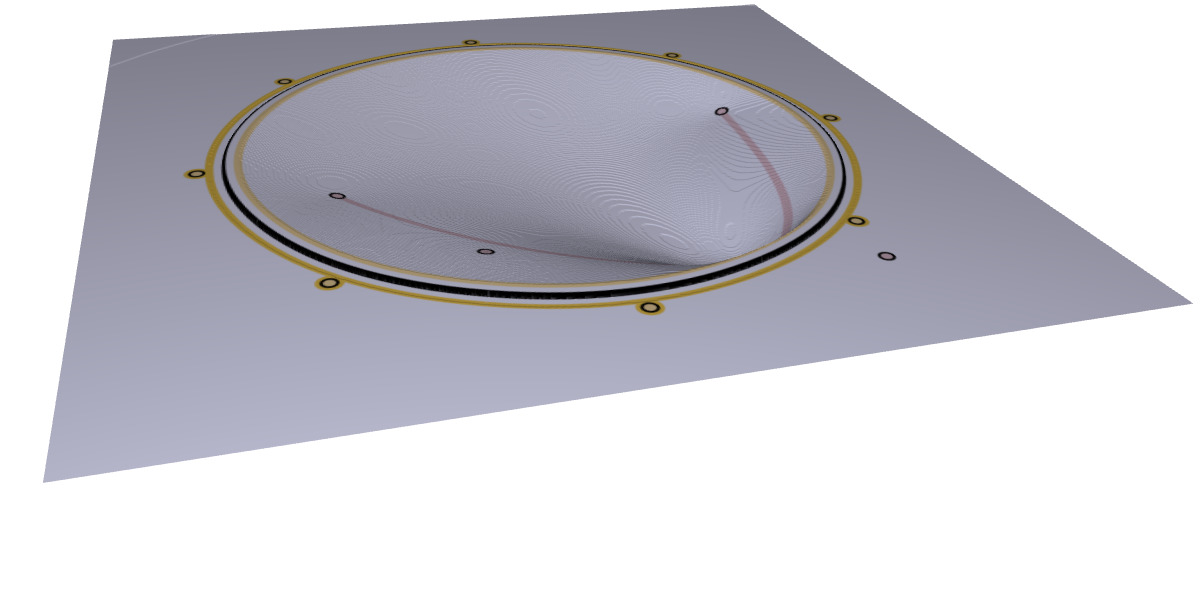}
  \includegraphics[width=1.66in]{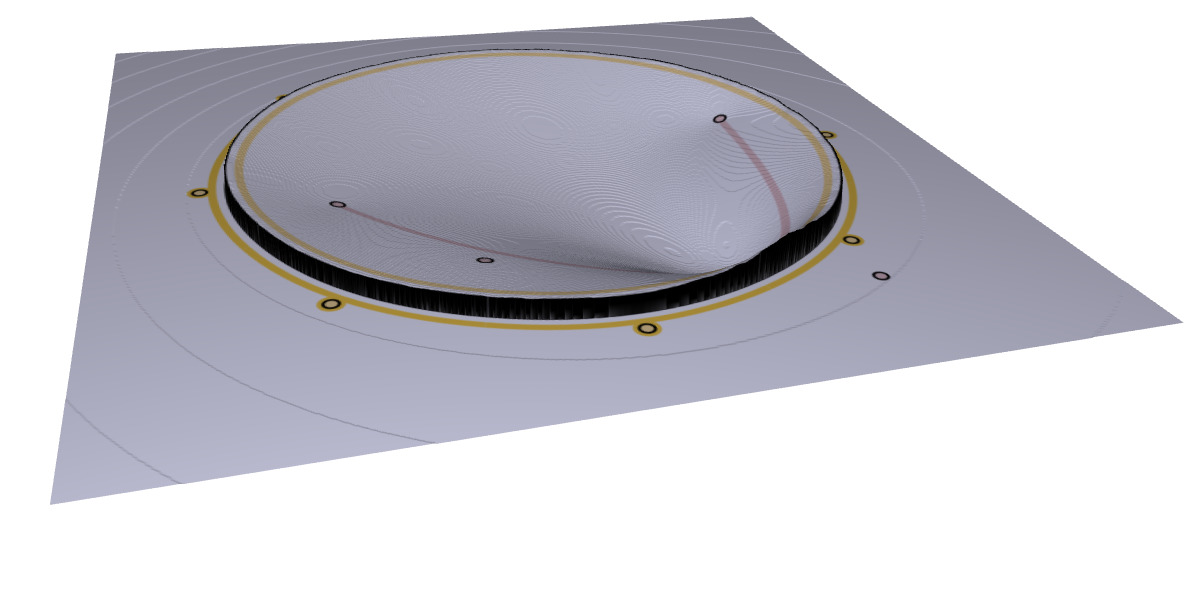}
  \includegraphics[width=1.66in]{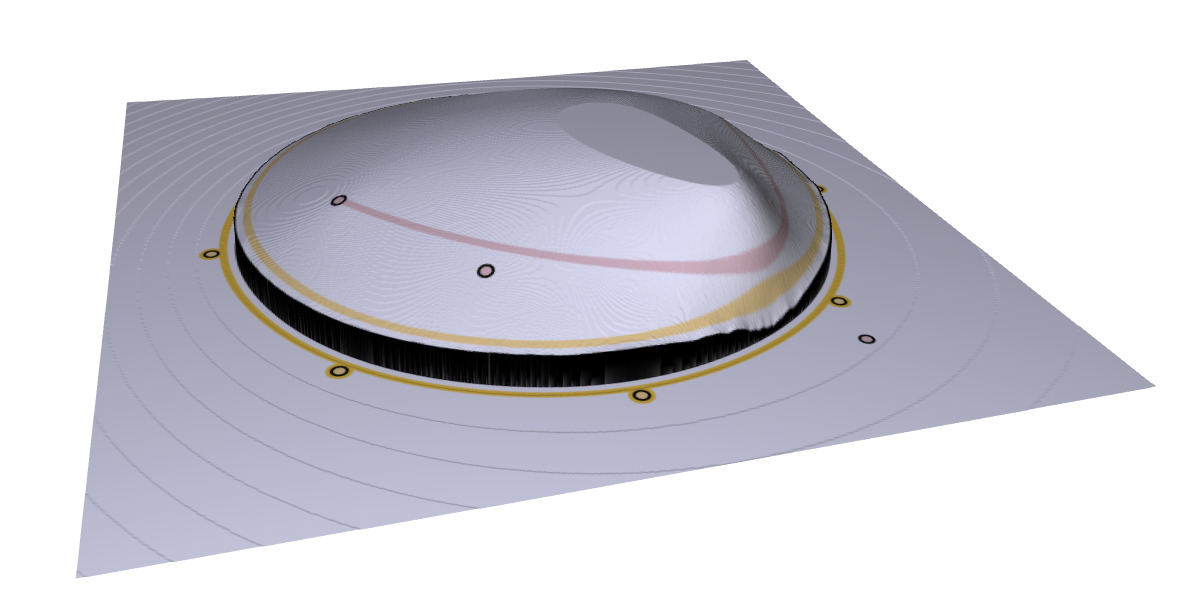}\\
  \includegraphics[width=0.7in]{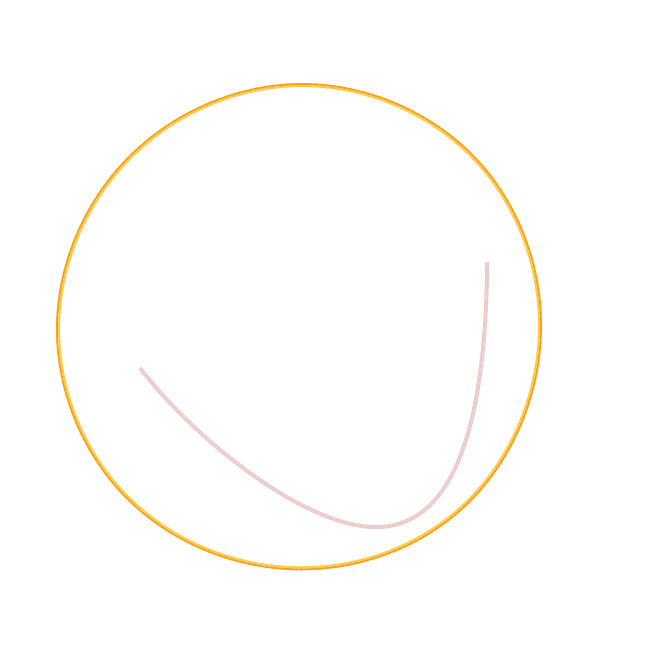}
  \includegraphics[width=0.7in]{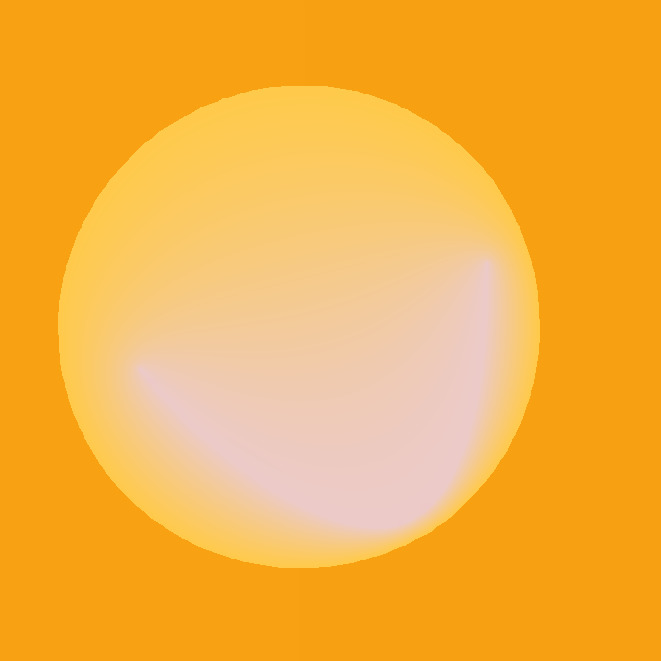}
  \includegraphics[width=1.66in]{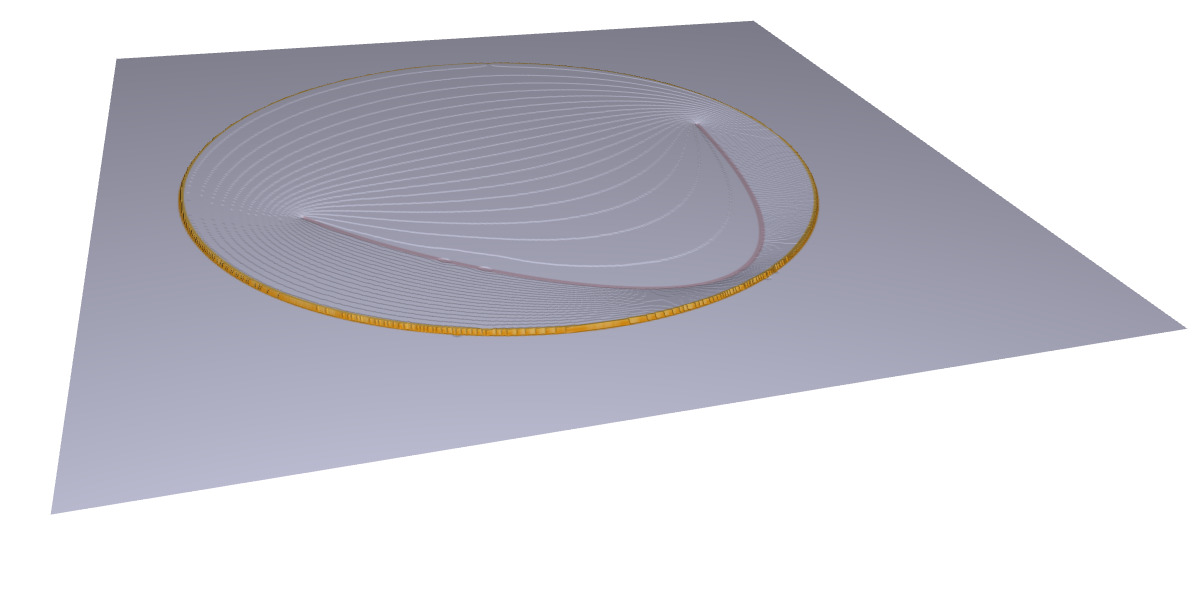}
  \includegraphics[width=1.66in]{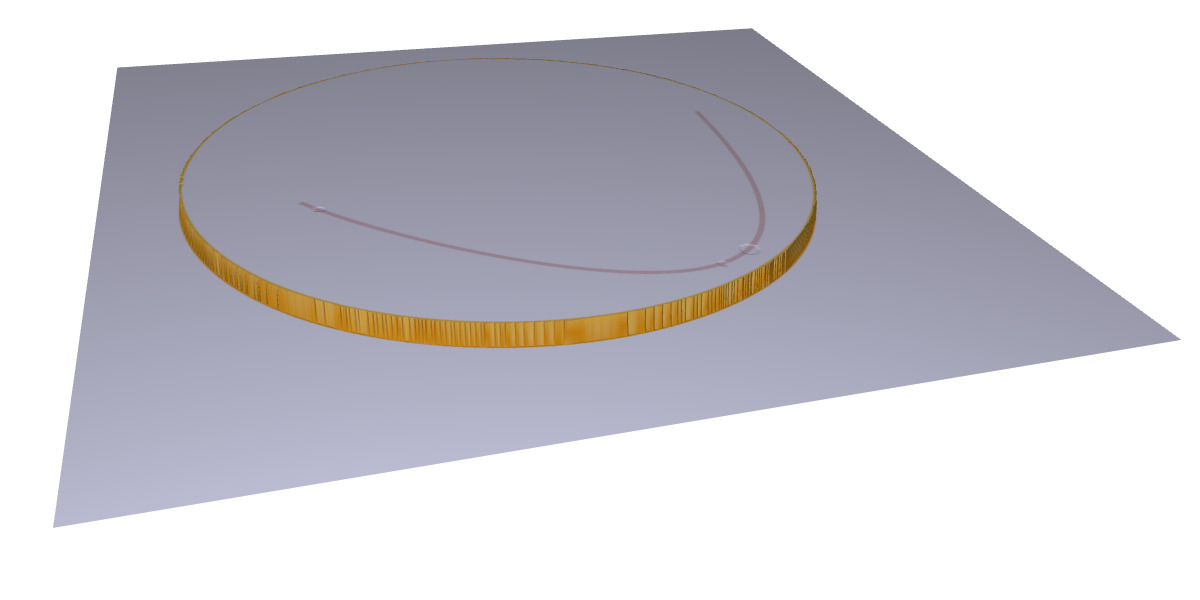}
  \includegraphics[width=1.66in]{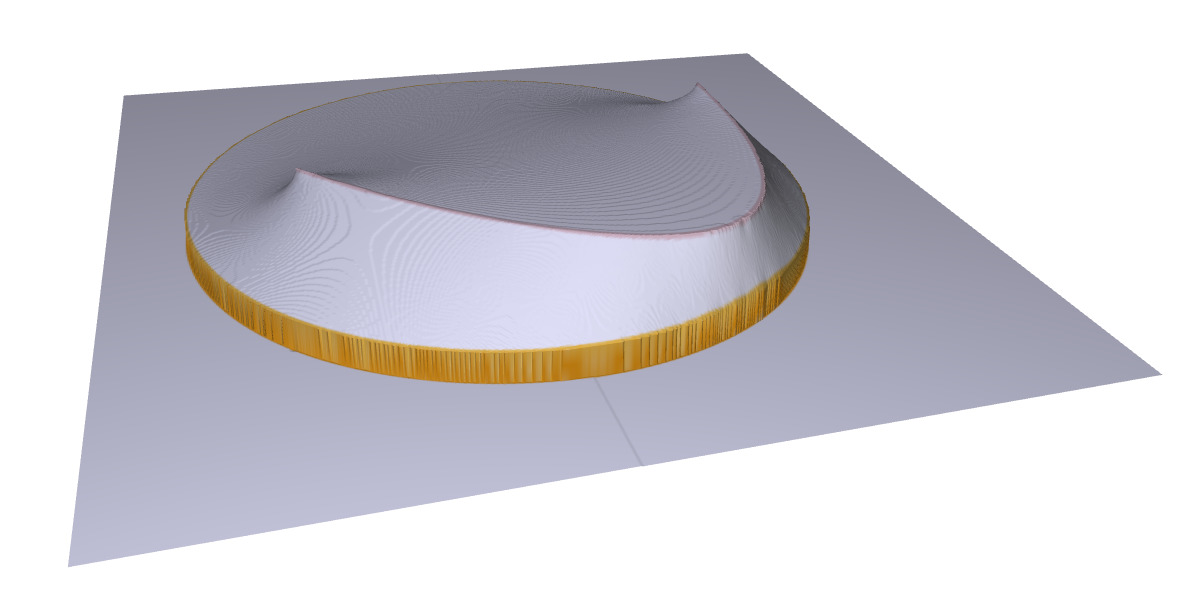}\\
  \includegraphics[width=0.7in]{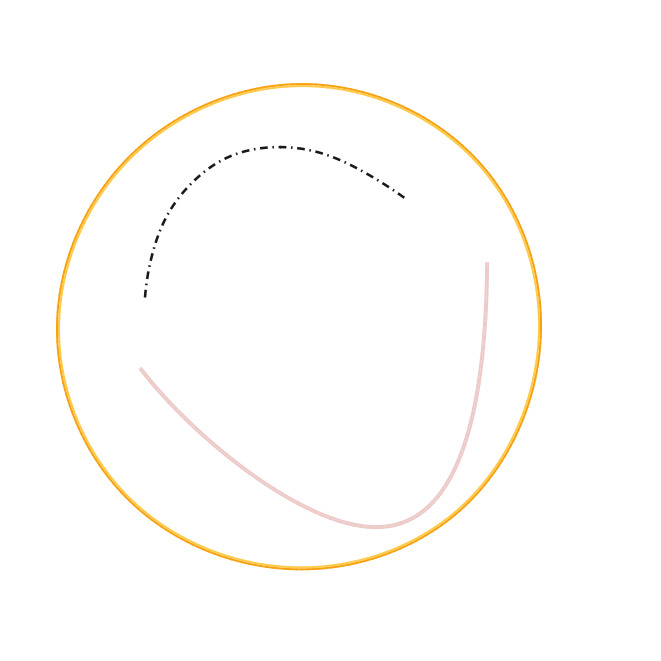}
  \includegraphics[width=0.7in]{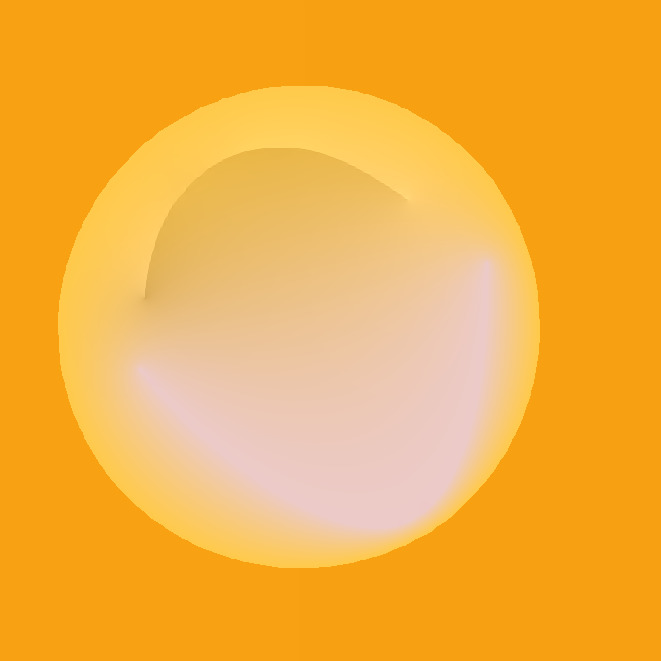}
  \includegraphics[width=1.66in]{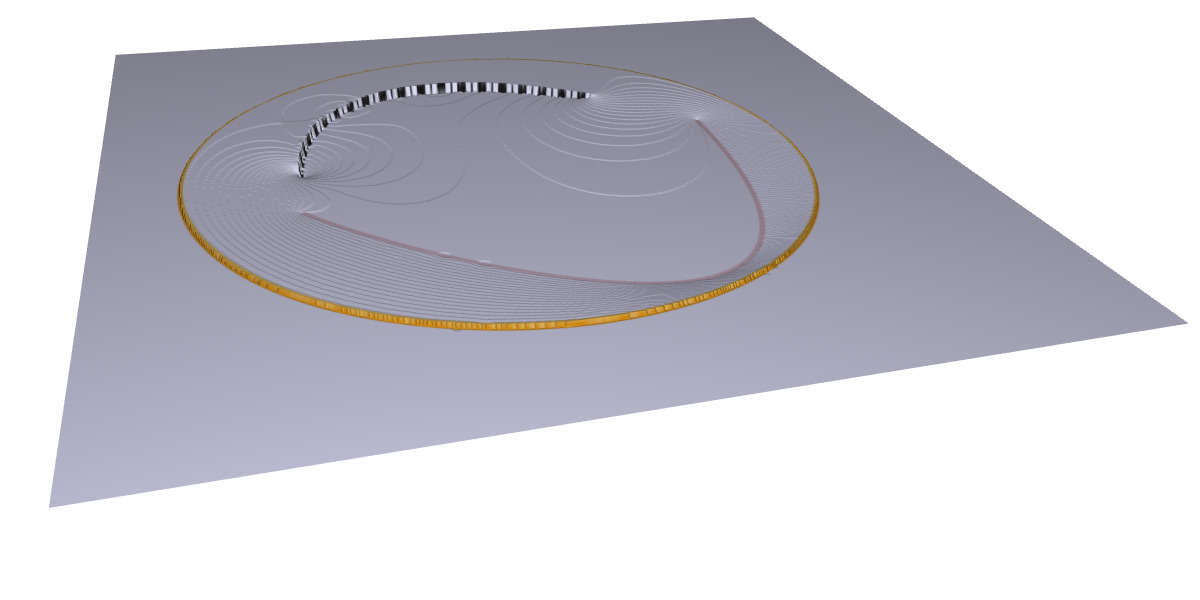}
  \includegraphics[width=1.66in]{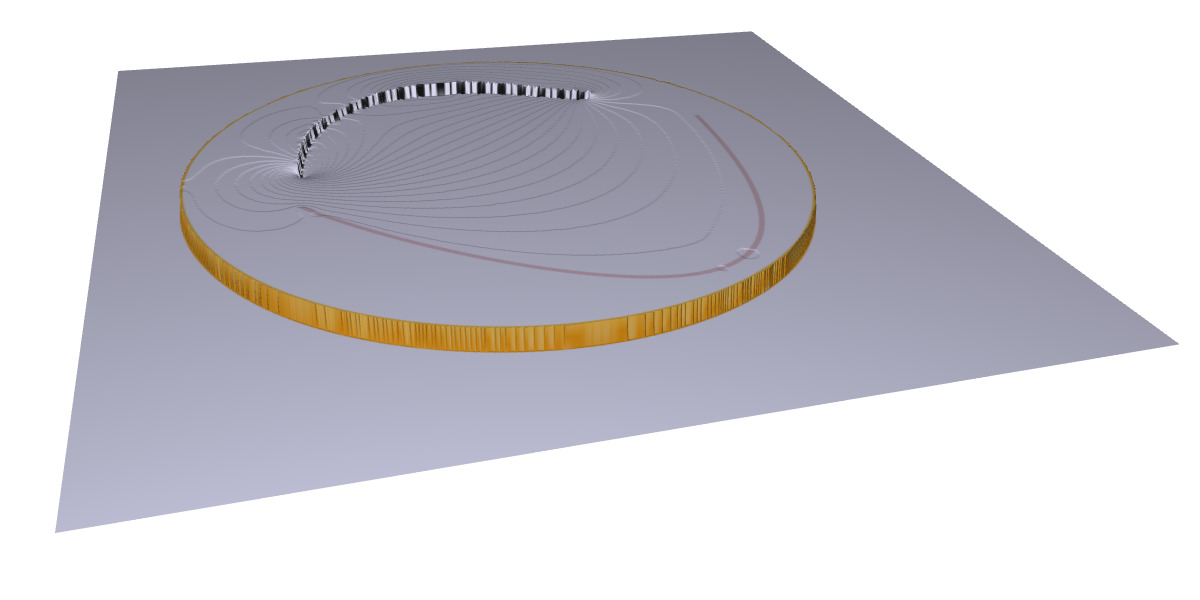}
  \includegraphics[width=1.66in]{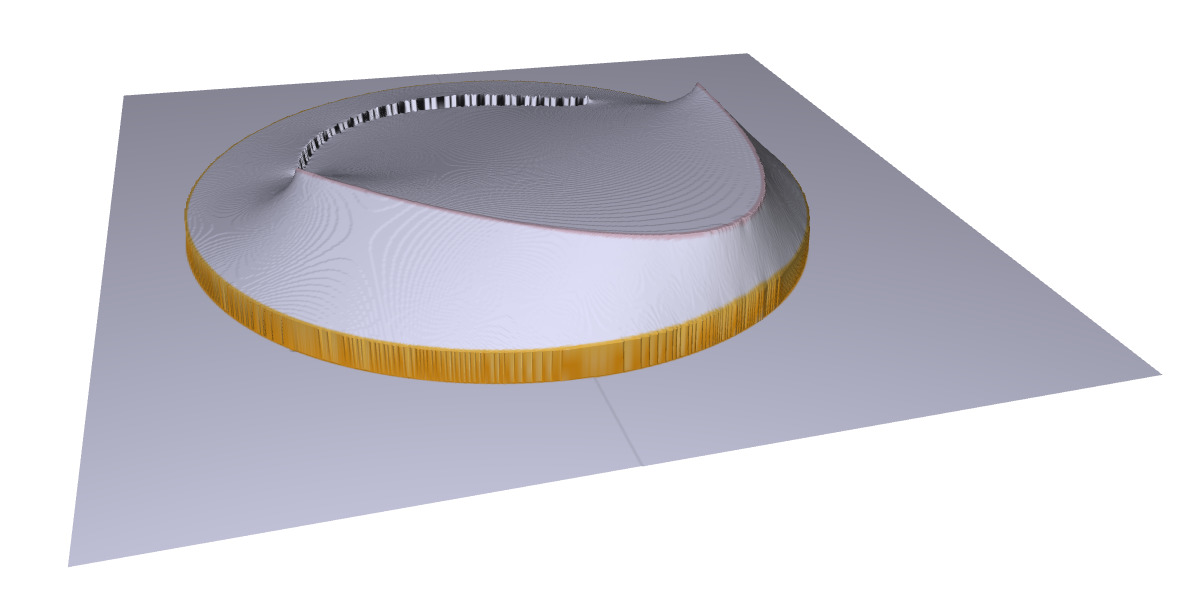}\\
  \includegraphics[width=0.7in]{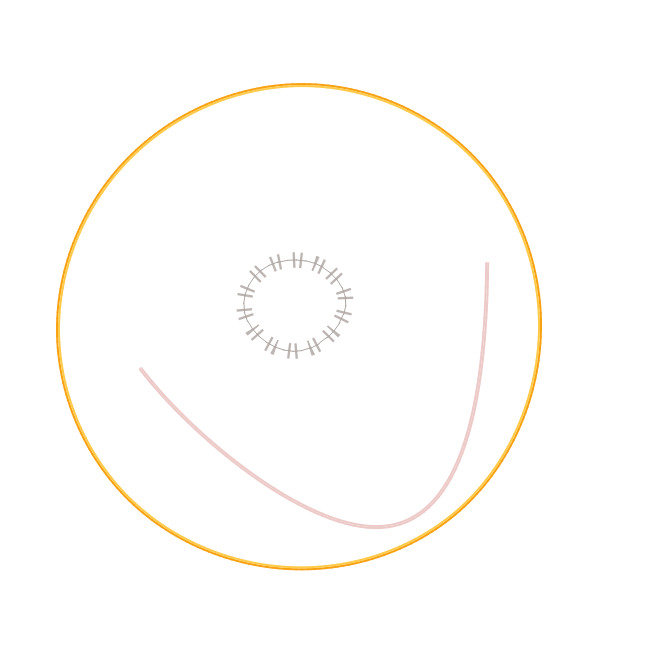}
  \includegraphics[width=0.7in]{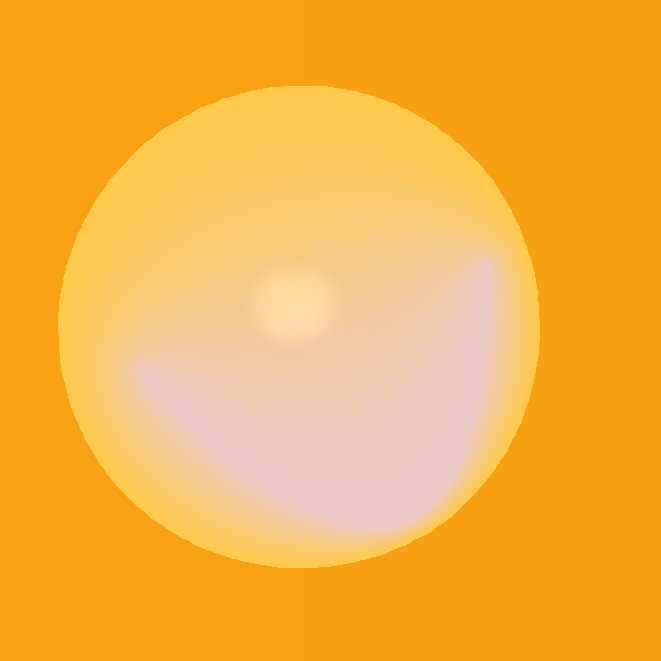}
  \includegraphics[width=1.66in]{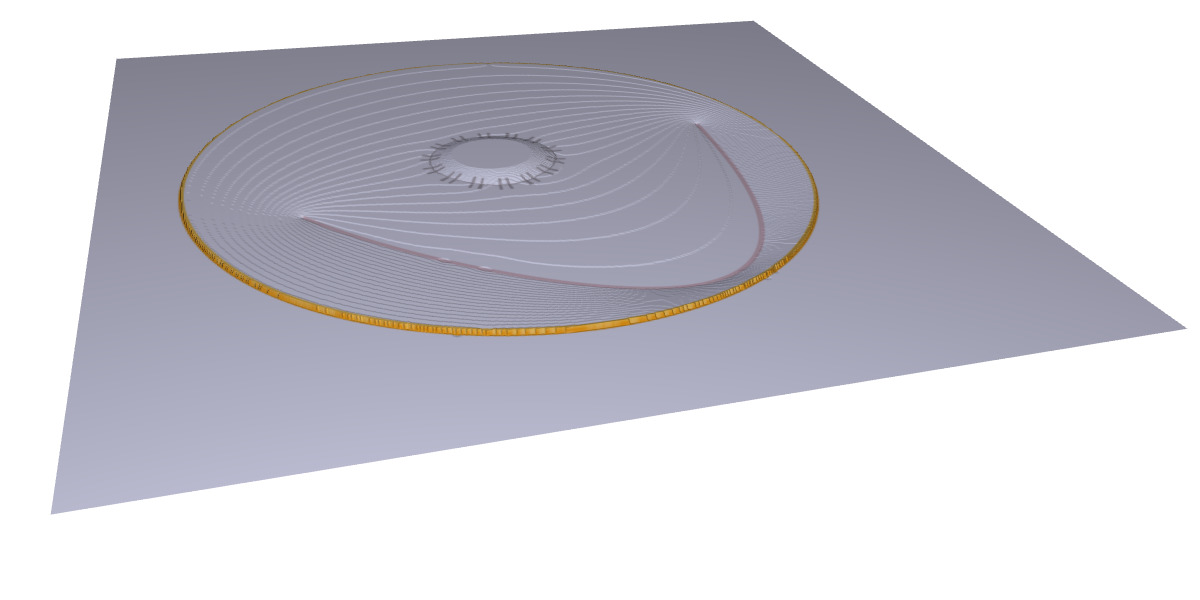}
  \includegraphics[width=1.66in]{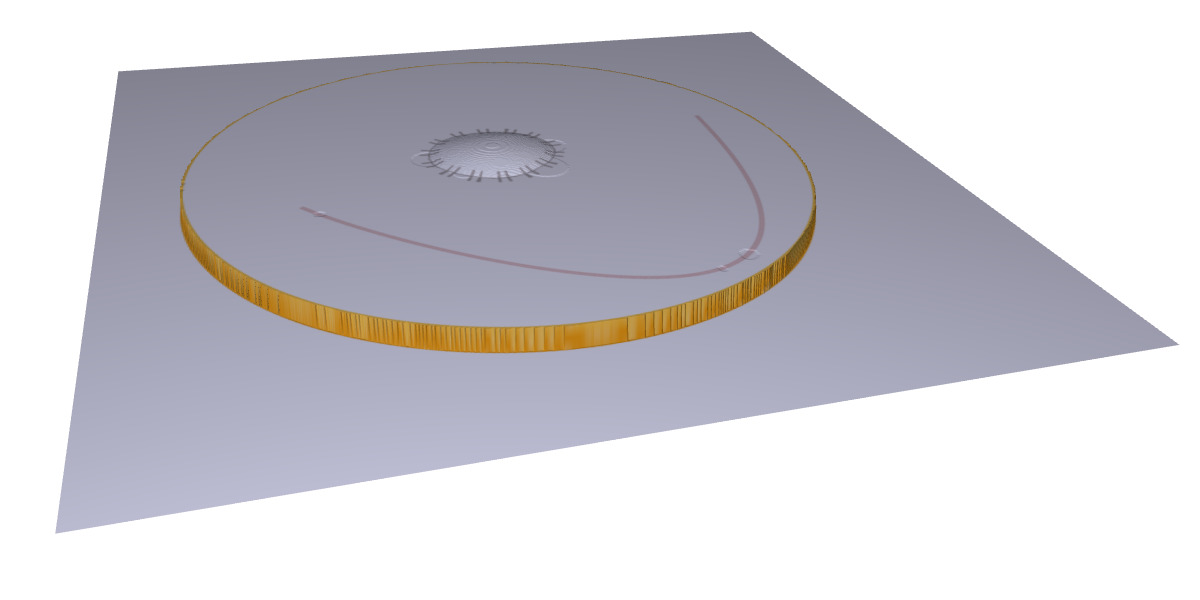}
  \includegraphics[width=1.66in]{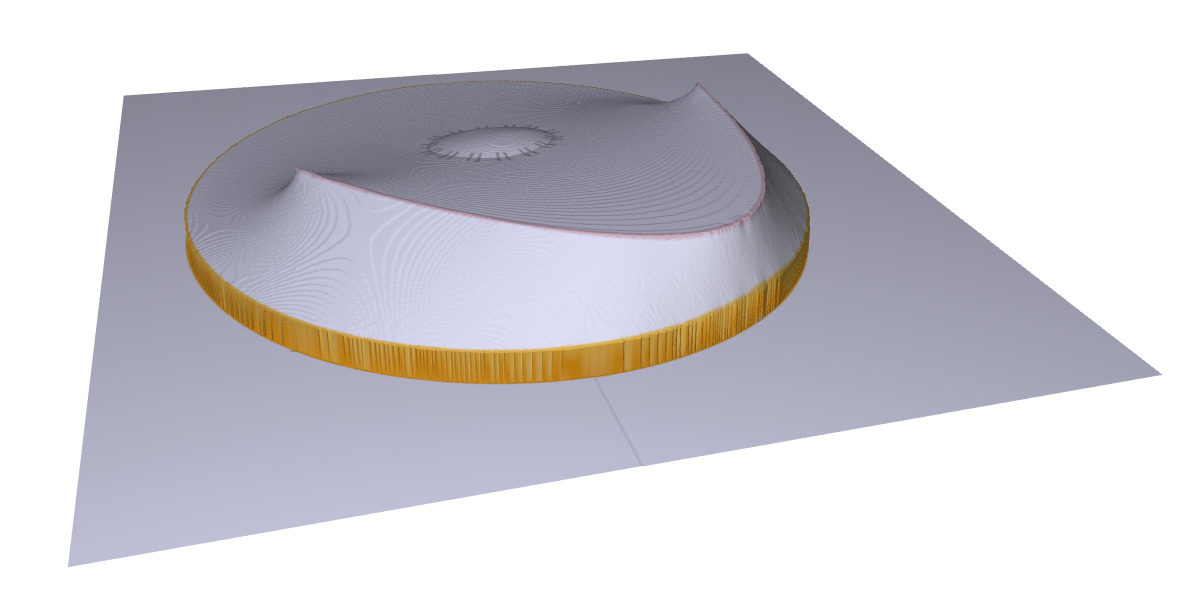}\\
  \makebox[0.8in]{}\makebox[1.86in]{Red} \makebox[1.6in]{Green}\makebox[1.6in]{Blue}\\
  \caption{First-order vs second-order diffusion.
  Row 1 shows a TPS with a value curve (the U-shaped curve) and a value-tear curve (the circle).
  The two curves are close to each other, producing undesired extrema (see the height functions for the red, green and blue channels),
  which is on neither curve.
  Rows 2-4 show one DCI and two PVGs in which the DCs/PCs/PRs are placed at the same locations.
  For DCs and PCs, the extrema are exactly on those curves.
  For PRs, the extrema are guaranteed to be inside the regions.
  }
  \label{fig:artifactshoppe}
  \end{figure}

\begin{figure}
  \centering
  \includegraphics[width=0.8in]{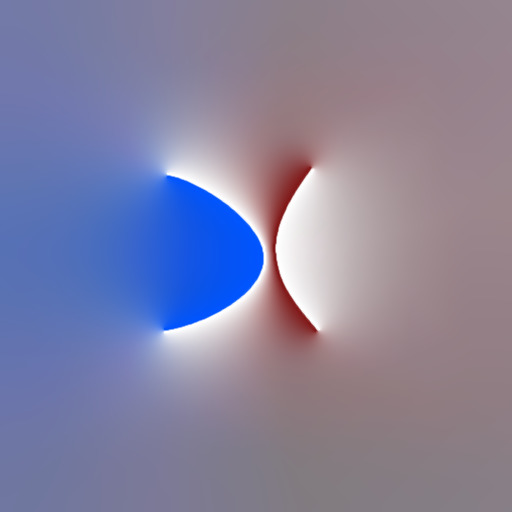}
  \includegraphics[width=1.86in]{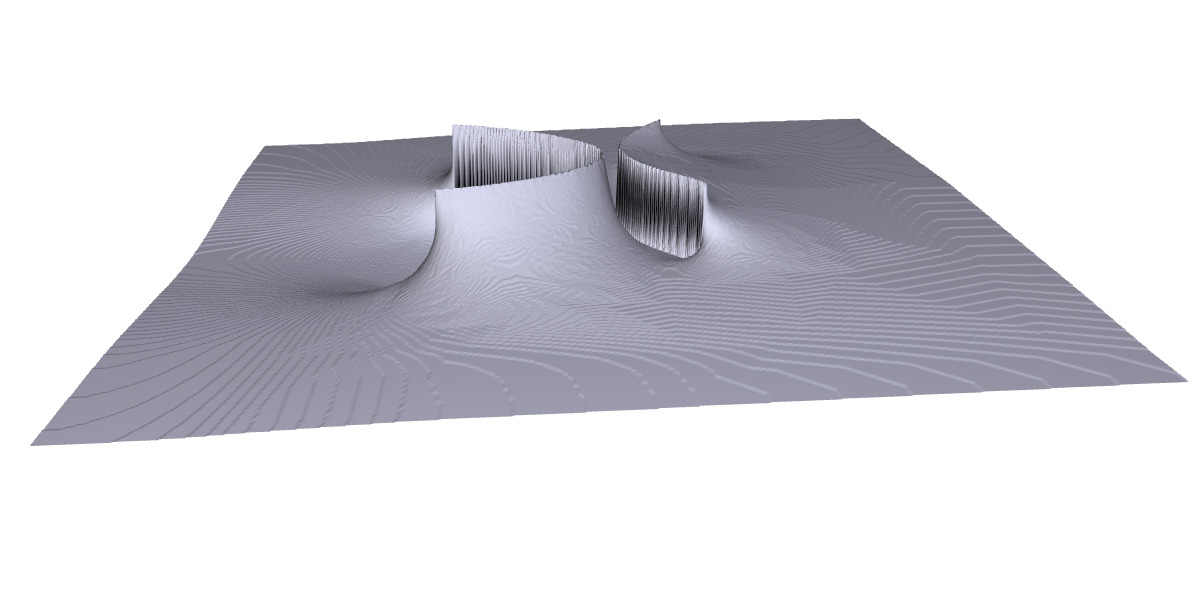}
  \includegraphics[width=1.86in]{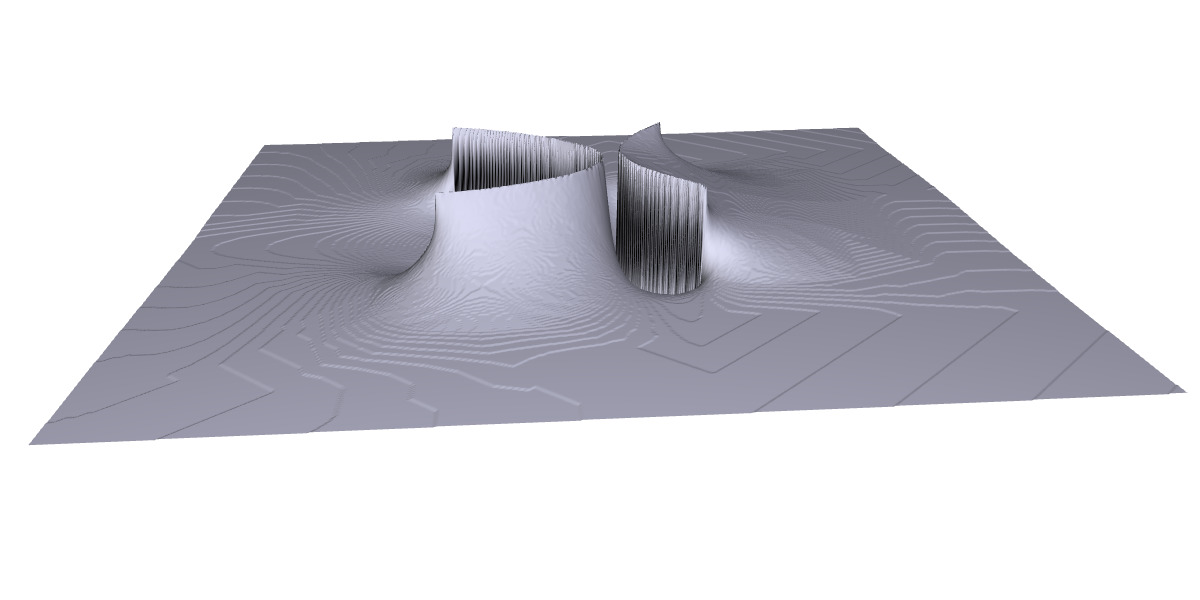}
  \includegraphics[width=1.86in]{DC_G_noblur}\\
  \includegraphics[width=0.8in]{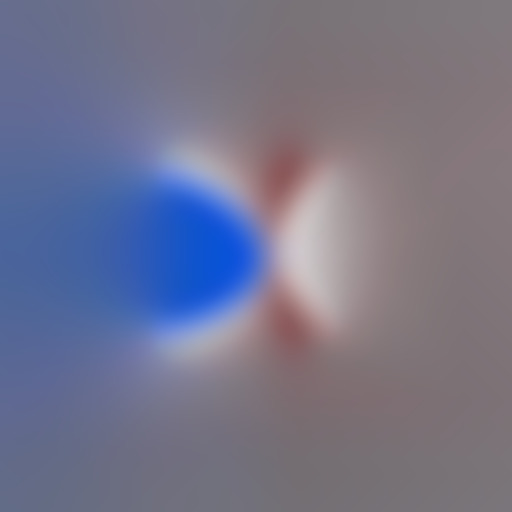}
  \includegraphics[width=1.86in]{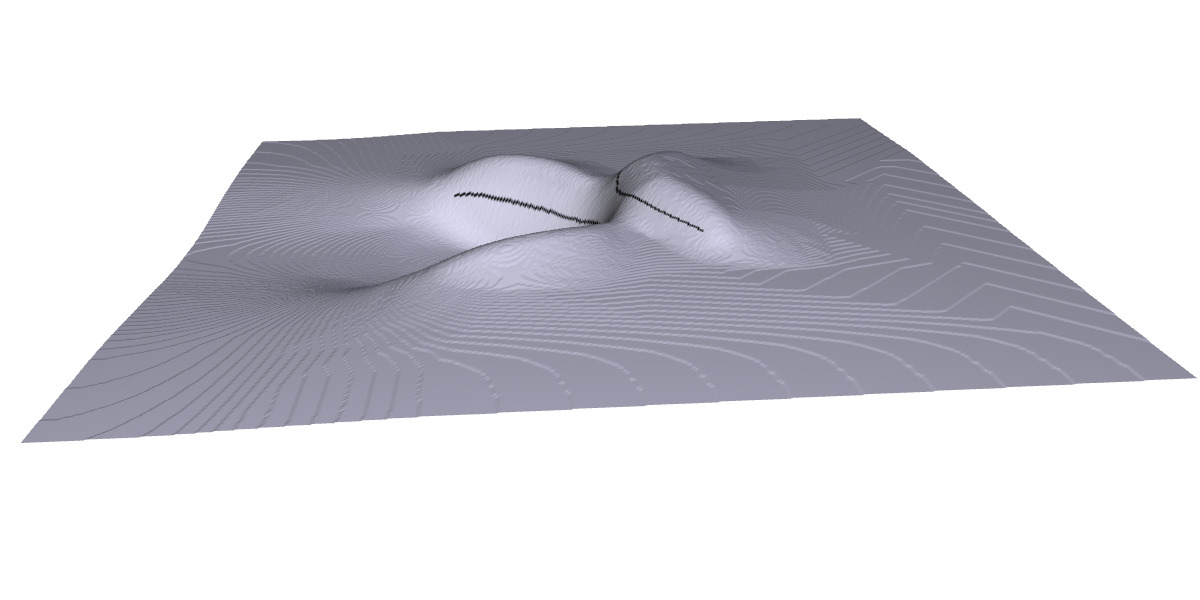}
  \includegraphics[width=1.86in]{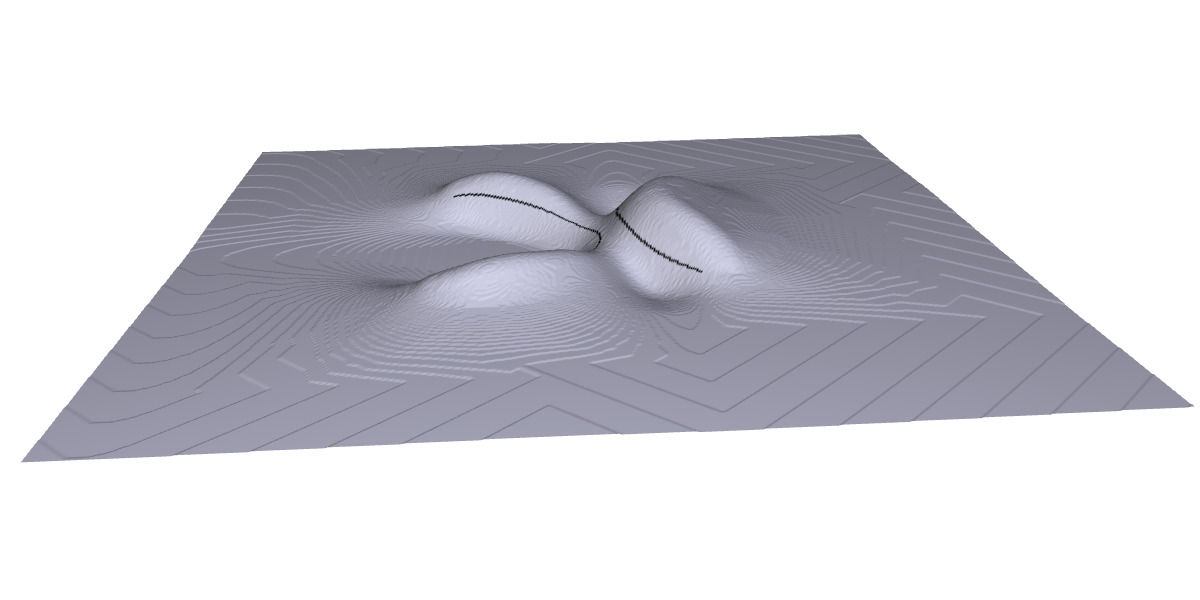}
  \includegraphics[width=1.86in]{DC_G_blur}\\
  \makebox[0.6in]{DCI}\makebox[1.8in]{Red}\makebox[1.8in]{Green}\makebox[1.8in]{Blue}
  \caption{Extrema control in DCI.
  Without blurring (row 1), the extrema are exactly on the diffusion curves, since the points off the curves have zero Laplacians.
  However, since the two diffusion curves are close to each other, their color diffusion competes each other.
  As row 2 shows, the blurring postprocess is helpful to reduce the artifacts.
  Unfortunately, it also introduces non-zero Laplacians to the nearby regions.
  As a result, the extrema are not on the diffusion curves (in black) any longer.
  }
  \label{fig:DCIartifact}
  \end{figure}

  \textbf{Comparison with thin-plate splines.}
  Finch et al.~\cite{Finch2011} showed that the solution to the bi-Laplace's equation $\Delta^2 u = 0$ provides for gradients that closely mimic smooth shading.
  They developed 5 types of basic curves, namely, tear, crease, slope, contour and value curves,
  for controlling color values and directional derivatives.
  Since the basic curves, in general, cannot be used alone,
  users need to combine a few types of curves to produce desired effects.
  Finch et al.'s system consists of more than 50 combinations of the basic curves.
  Such a large number of drawing tools is a double-edged sword.
  On one hand, it provides users more flexility and advanced controls in the form of value and gradient constraints;
  on the other hand, it increases the complexity of the system, hence deepening the learning curve.
  As Figure~\ref{fig:comparewithhoppe} shows, there are 13 types of primitives in a typical TPS image, some of them have very subtle differences and effects.
  In contrast, PVG provides only three types of primitives, each of which has a clear definition and purpose.
  A preliminary user study in Section~\ref{sec:results} shows that PVG is more intuitive and easy to use than TPS.
  Moreover, as pointed out in~\cite{Jacobson2012}\cite{Lieng2015},
  the biharmonic functions can be negative and have prevalent local extrema, leading to unexpected results (see Fig.~\ref{fig:artifactshoppe}).
  As mentioned in Section~\ref{subsec:features}, PVG allows users to control the extremum easily,
  i.e., the extrema are either on DCs and PCs or inside PRs, which have non-zero Laplacian.
  Readers can try the accompanying software to experience PVG.

  \begin{figure}
  \centering
  \includegraphics[width=1.5in]{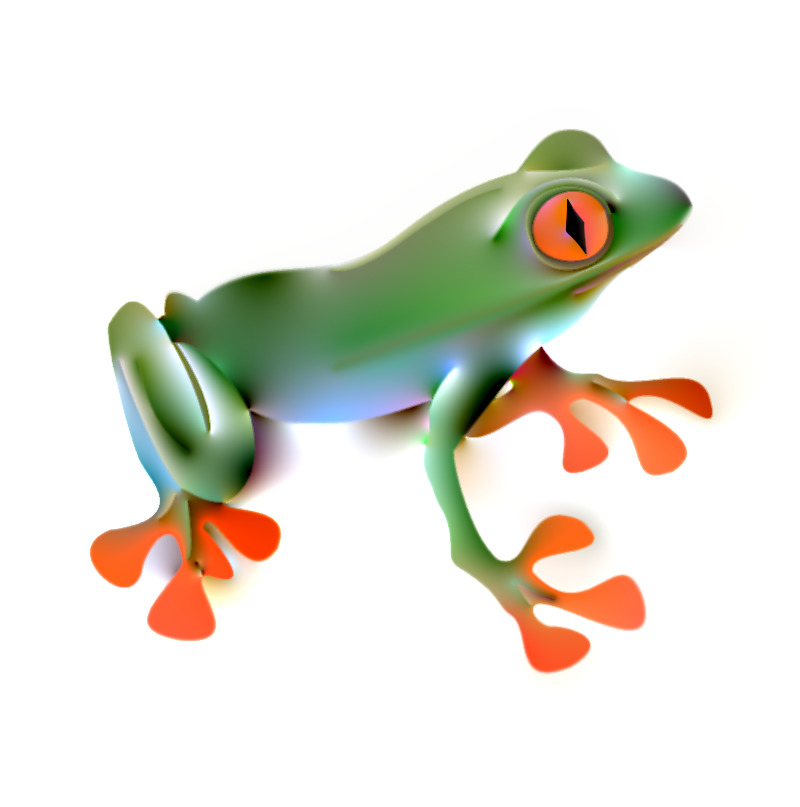}
  \includegraphics[width=1.5in]{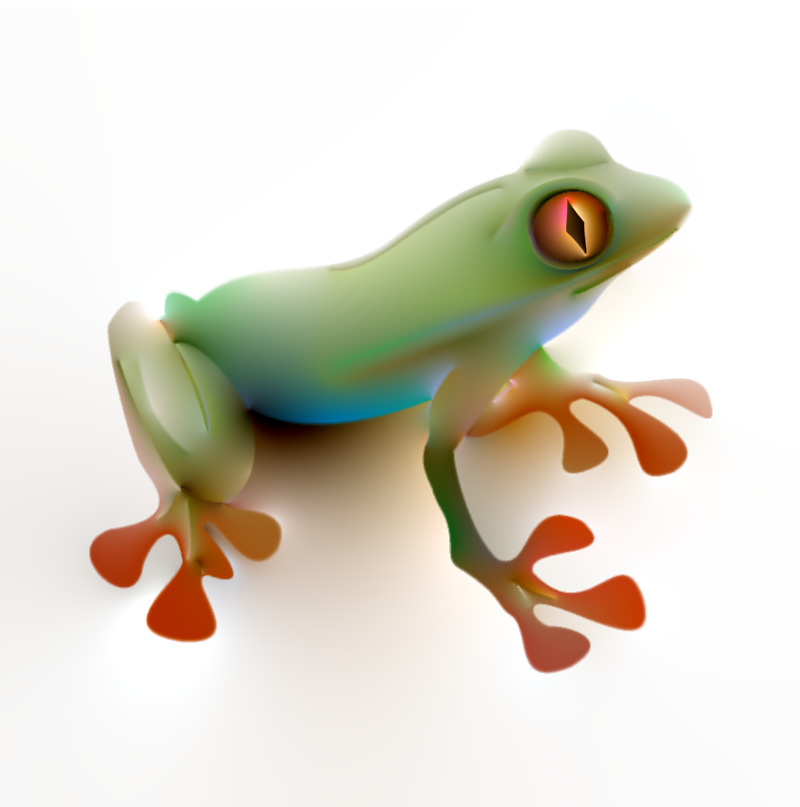}
  \includegraphics[width=1.5in]{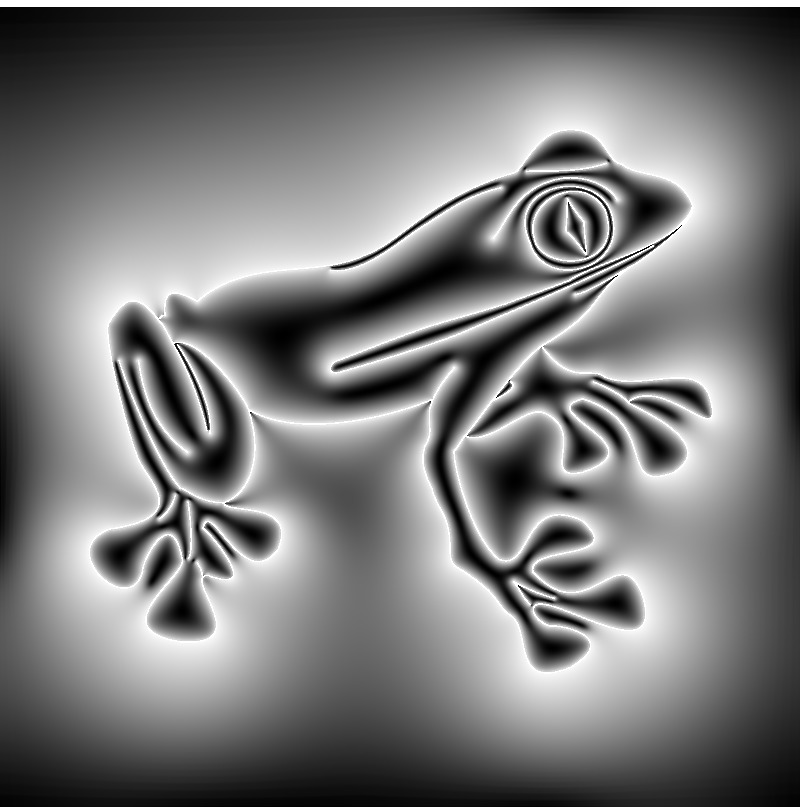}
  \includegraphics[width=1.5in]{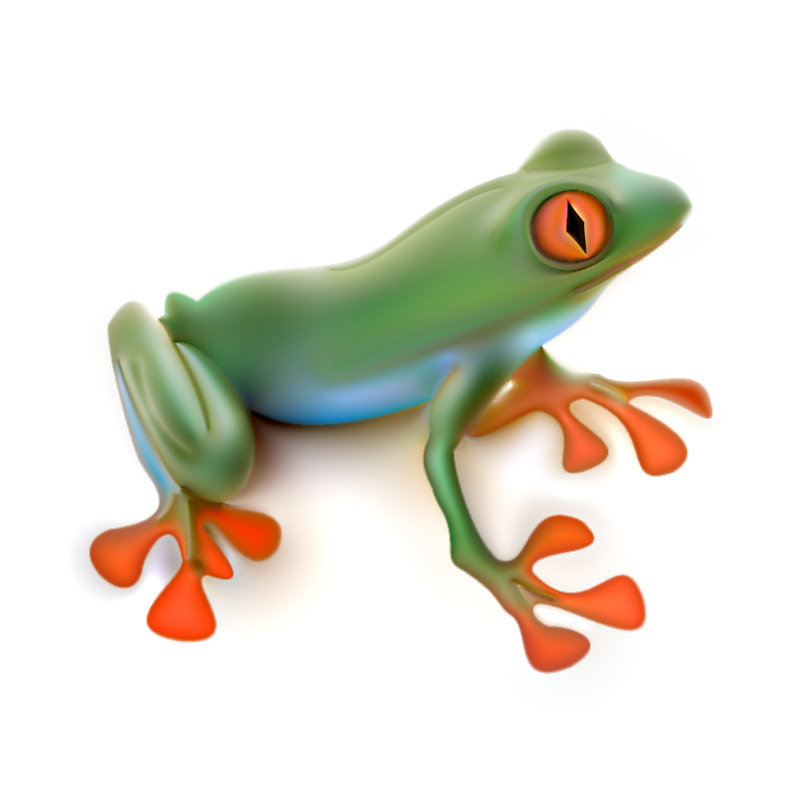}\\
  \makebox[1.5in]{DCI 1}\makebox[1.5in]{DCI 2}\makebox[1.5in]{Blending function}\makebox[1.5in]{GDCI}\\
  \includegraphics[width=1.5in]{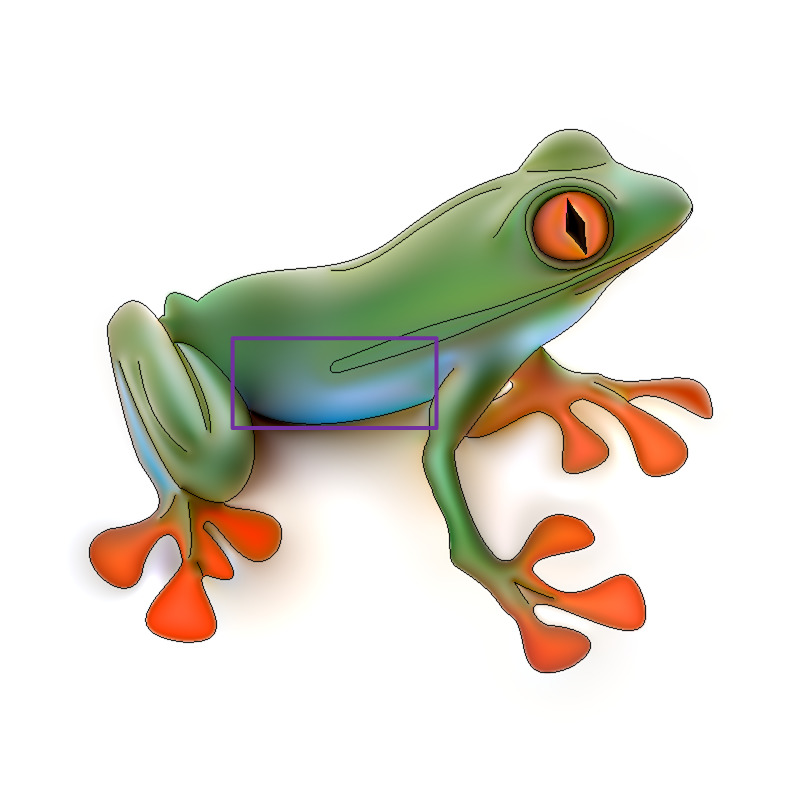}
  \includegraphics[width=1.5in]{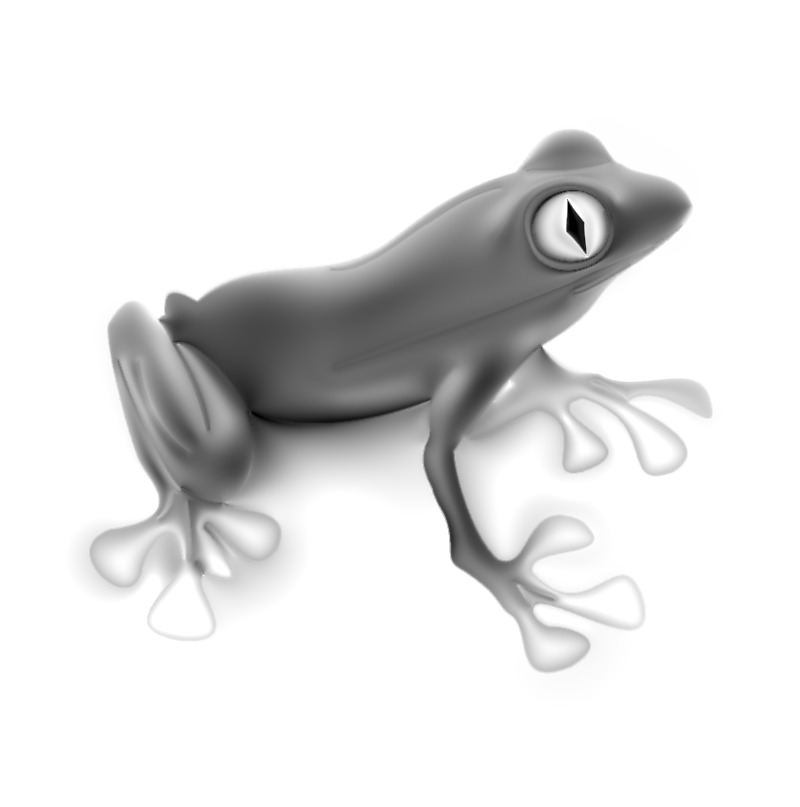}
  \includegraphics[width=1.5in]{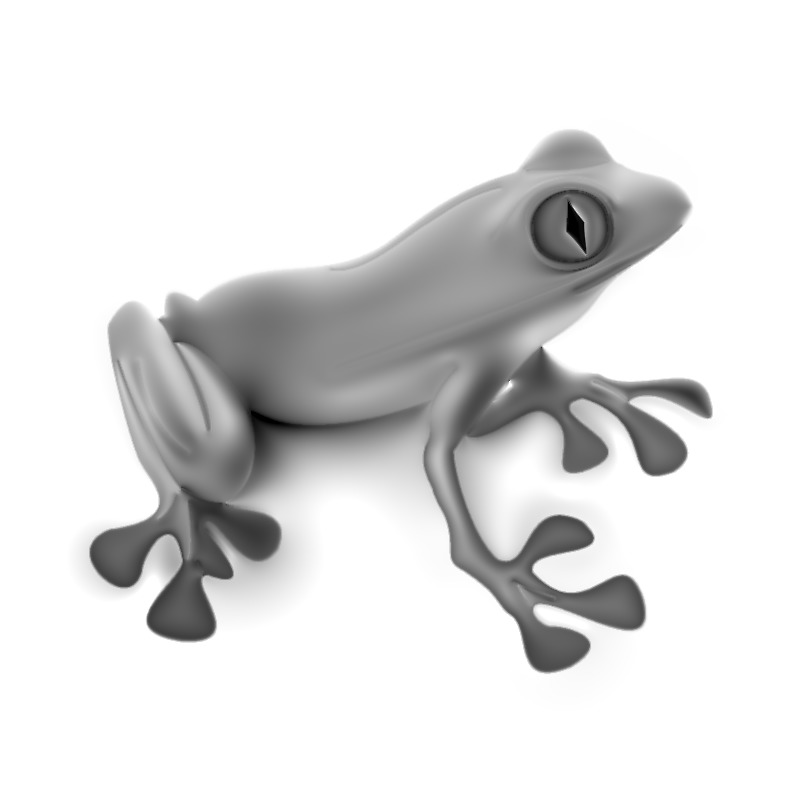}
  \includegraphics[width=1.5in]{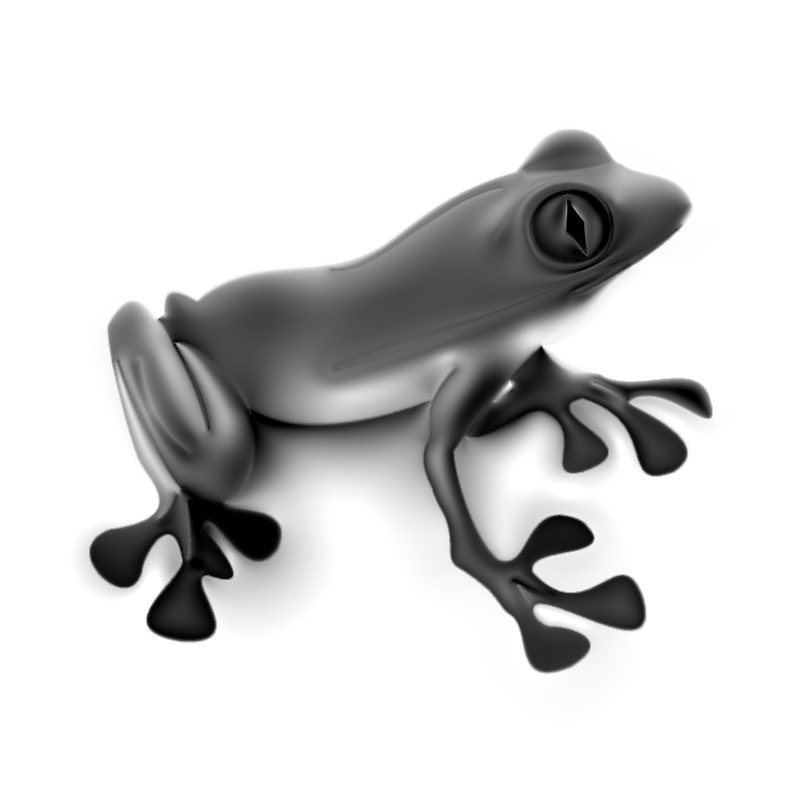}\\
  \includegraphics[width=1.5in]{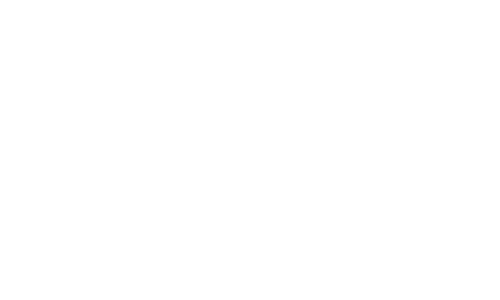}
  \includegraphics[width=1.5in]{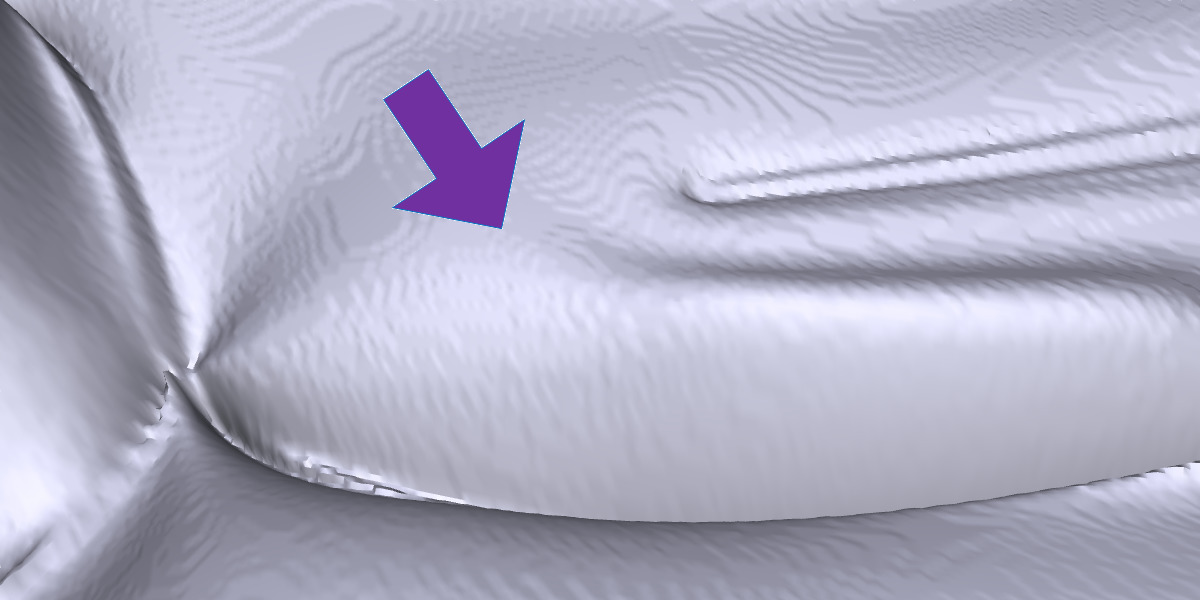}
  \includegraphics[width=1.5in]{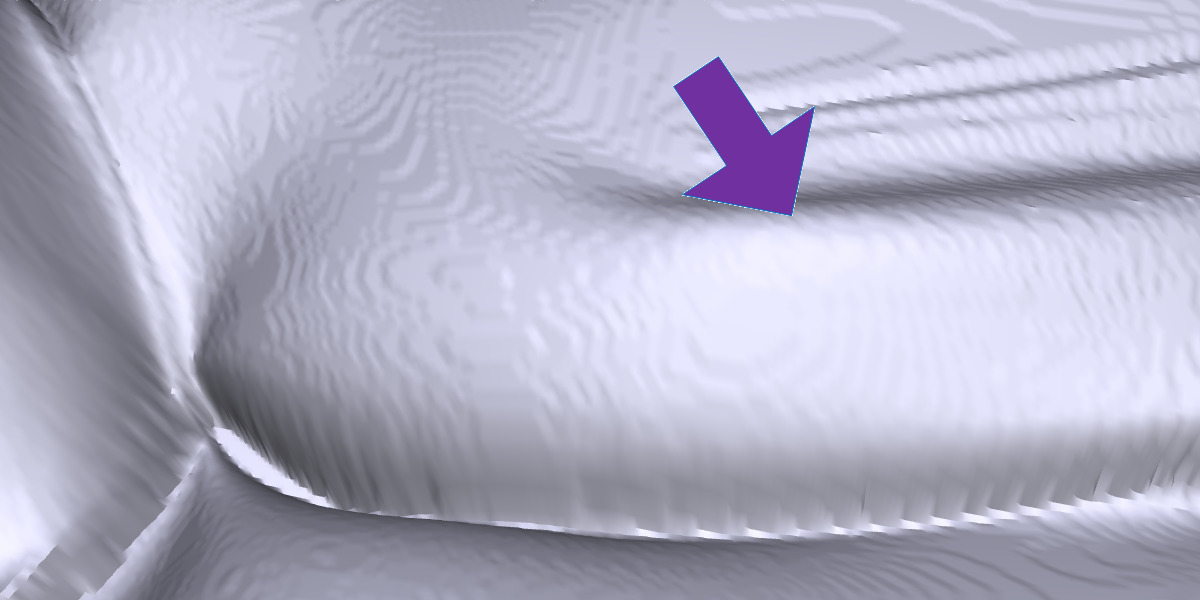}
  \includegraphics[width=1.5in]{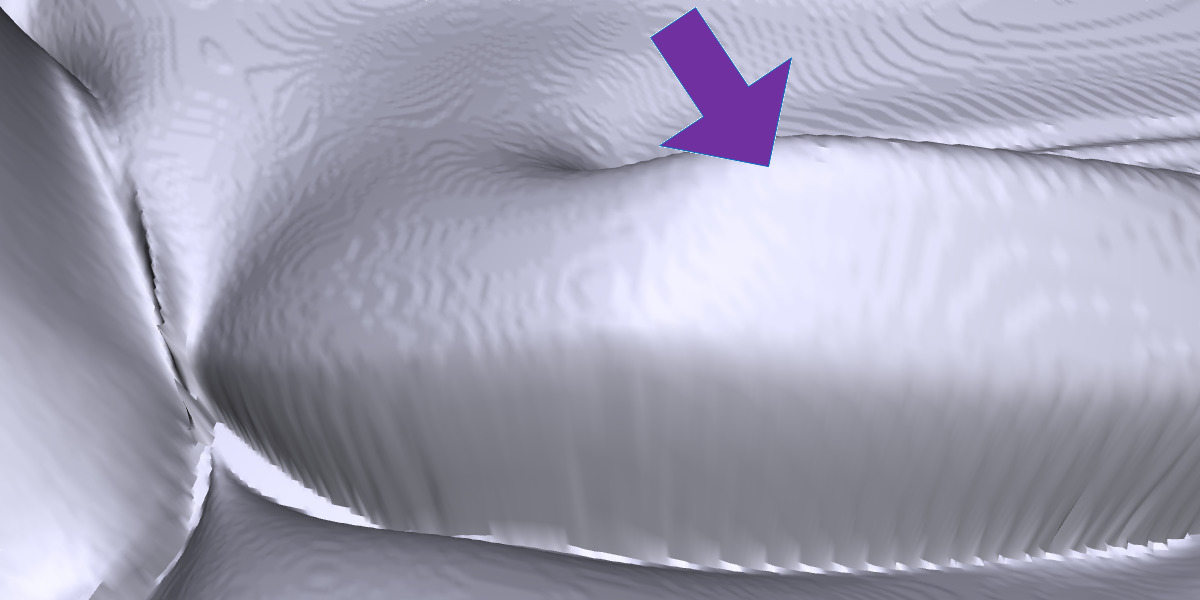}\\
  \makebox[1.5in]{}\makebox[1.5in]{Red}\makebox[1.5in]{Green}\makebox[1.5in]{Blue}\\
  \caption{Extrema of generalized DCI.
  Row 1 shows the two DCIs, the blending function and the GDCI (image courtesy of S. Jeschke~\cite{Jeschke2016}).
  Row 2 highlights extrema (indicated by the arrows), which are not on the diffusion curves in either DCI.
  }
  \label{fig:frog}
  \end{figure}

  \textbf{Comparison with generalized diffusion curves.}
  The generalized diffusion curves~\cite{Jeschke2016} spatially blend multiple conventional DCIs and are able to provide a similar expressive power of color control as TPS.
  Its solver is also highly efficient and numerically stable.
  However, since the blending functions are highly non-linear, it is non-trivial and non-intuitive to design them manually and control the extrema (see Figure~\ref{fig:frog}).
  Therefore, GDCI is often used in reverse engineering (e.g., image vectorization) and simple editing (e.g., changing the colors or modifying the curves),
  rather than authoring.

\begin{figure}[htbp]
  \centering
  \includegraphics[width=5in]{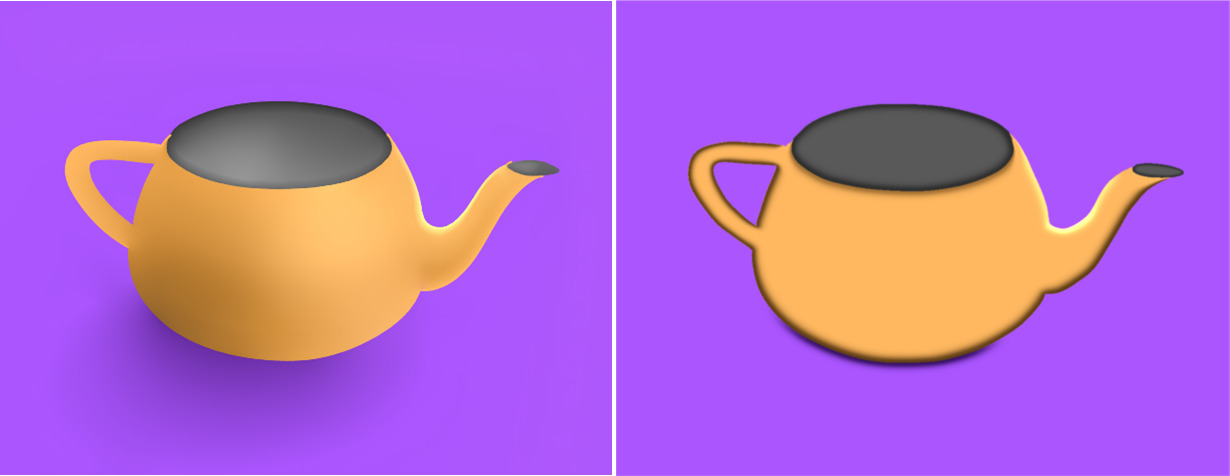}\\
  \makebox[2.5in]{PVG}\makebox[2.5in]{Shading curves}
  \caption{The colors produced by shading curves are not as vivid as those of PVG and diffusion curves.}
  \label{fig:compareshadingcurves}
  \end{figure}

  \textbf{Comparison with shading curves.}
  Lieng et al.~\cite{Lieng2015} proposed shading curves to simulate chiaroscuro drawing, providing strong contrasts between light and dark.
  Users first draw areas of constant tone with curves, fill in each individual area with constant color and specify the influence of that color to adjacent areas.
  Colors are then smoothed out with shading profiles, which are associated with each side of the curve.
  Finally, shading curves are converted to 3D control meshes and rendered as Catmull-Clark subdivision surfaces.
  Shading curves allow explicit control over color gradients, however, the produced colors are not as vivid as those of DC and PVG
  (see Fig.~\ref{fig:compareshadingcurves}).
  Moreover, due to the limitation of their region growing algorithm,
  shading curves are not able to handle curves with high curvatures and/or intersecting curves.
  It is also not clear whether shading curves support zooming-in of arbitrary resolution.

  \begin{table*}
  \centering
  \caption{Comparison of existing methods.}
  \setlength\tabcolsep{2pt}
  \begin{scriptsize}
  \begin{tabular}{|c||c|c|c|c|c|c|c|c|}
  \hline
  Method & Primitives & Color   & Domain         & Solver & Random   & Closed  & Gradient & Application \\
         &            & function& discretization &        & access  & form&  control &  \\
  \hline
  \hline
  Ours & DC+PC+PR & Non-harmonic & Voronoi diagrams & Harmonic B-splines & Yes & Yes & Easy & Authoring \\
  \hline
  \cite{Orzan2008} & DC & Harmonic & Hierarchical regular grids & Multigrid & No & No & No & \tabincell{c}{Vectorization \& \\ authoring}\\
  \hline
  \cite{Jeschke2009}& DC & Harmonic &  Regular grids  & Jacobi iteration & No & No & No & \tabincell{c}{Vectorization \&\\ authoring}\\
  \hline
  \cite{Bowers2011} & DC & Harmonic & Regular grids & Ray tracing & No & No & No& \tabincell{c}{Vectorization \& \\ authoring}\\
  \hline
  \cite{Finch2011} & DC variants & Biharmonic &Hierarchical regular grids & Multigrid & No & No & Difficult & Authoring\\
  \hline
  \cite{Pang2012} & DC & Harmonic & Mesh & Mean value coordinates & Yes & Yes & No & \tabincell{c}{Vectorization \&\\ authoring}\\
  \hline
  \cite{Boye2012} & DC & Biharmonic &Triangle meshes & Finite element method & Yes & Yes & Difficult & Authoring\\
  \hline
  \cite{Sun2012} & DC & Harmonic & Curved regions & Boundary element method & Yes & Yes & No & Texture mapping\\
  \hline
  \cite{Ilbery2013} & DC & Biharmonic &Regular grids&  Boundary element method & Yes & Yes & Difficult & \tabincell{c}{Vectorization \&\\ authoring}\\
  \hline
  \cite{Sun2014} & DC & Harmonic &Hierarchical regular grids & Boundary element method & Yes & Yes & No & Vectorization\\
  \hline
  \cite{Xie2014} & DC & (Bi-)harmonic &Curved regions & Boundary element method & Yes & Yes & Difficult  & Vectoirzation\\
  \hline
  \cite{Prevost2015} & DC & Harmonic &Triangle meshes & Ray tracing & No & No & Easy & Authoring\\
  \hline
  \cite{Lieng2015} & Shading curves& Piecewise polynomials & Subdivision surfaces& Subdivision & No & No & Easy & Authoring\\
  \hline
  \cite{Jeschke2016} & Generalized DC & Non-harmonic & Regular grids & Jacobi iteration & No & No & Difficult & \tabincell{c}{Vectorization \&\\ editing}\\
  \hline
  \end{tabular}
  \end{scriptsize}
  \label{tab:comparison}
  \end{table*}

  \textbf{Comparison with the existing DC solvers.}
  Sun et al.~\cite{Sun2012} developed a boundary element method for rendering diffusion curve textures.
  Their method is also based on Green's third identity,
  and can compute the texture value of any rectangular region in closed form.
  As their target is Laplace's equation, the double integral vanishes,
  hence they only needed to discretize the line integral in Eqn.~(\ref{eqn:green_identity}).
  Moreover, their method solves a dense linear system to obtain the normal derivatives of colors on diffusion curves.
  Due to its high computational cost, this step has to be done in a pre-processing stage.
  Therefore, their method applies to applications with fixed primitives (such as texture mapping),
  but it does not work for interactive applications, such as authoring.
  Our method solves Poisson's equation and deals with area integrals, which is more challenging than their problem.
  Moreover, our solver does not require pre-computation.
  Although we focus on authoring in this paper, our solver can be trivially extended to texture mapping.

  The multigrid solvers proposed in \cite{Orzan2008} and \cite{Finch2011} are efficient,
  however, they are not accurate and may suffer from artifacts, such as aliasing and flicking.
  With robust curve rasterization, Jeschke et al.~\cite{Jeschke2009} proposed a GPU based Jacobi iteration algorithm
  that can effectively eliminate most of these artifacts.
  However, this solver does not support random-access evaluation.
  Moreover, it produces visual artifacts on the image boundary during zooming-in,
  since the boundary condition obtained from the low-resolution DCI is not accurate enough to provide stable solution for high-resolution DCI.
  Such artifacts also occur when user pans the camera.
  Our method does not re-assign the boundary conditions, hereby it works well for camera panning and zooming-in.

  The FEM-based vectorial solver~\cite{Boye2012} provides closed-form solutions and allows random-access evaluation.
  Since this solver is designed for solving (bi-)Laplacian equations, it is unclear how to extend it for Poisson equations.
  The BEM-based solvers~\cite{Sun2012}\cite{Ilbery2013}\cite{Xie2014} can also provide closed form solutions,
  but they have to pre-compute normal derivatives on control curves,
  hence they can render DCIs with fixed primitives.
  The fast multipole method~\cite{Sun2014} is highly efficient, but it solves only Laplace equations.

  In contrast to the existing DC solvers,
  our method solves Poisson's equation with piecewise constant Laplacians,
  and it supports random-access evaluation, zooming-in of arbitrary resolution, and anti-aliasing.

  \textbf{Comparison with the quad-tree based Poisson solver~\cite{Agarwala2007}.}
  Both our solver and Agarwala's method adopt the quad-tree for domain discretization.
  Their method is highly efficient, but it produces \textit{numerical} solutions for quad-tree nodes only
  and adopts linear interpolation for points in a quad-tree cell.
  Our solver is designed for Poisson equations with piecewise constant Laplacians
  and it provides closed-form solution for any point in the domain.

  \textbf{Limitations.}
  In our current implementation, we adopt the harmonic B-spline based solver, which provides a closed-form solution.
  However, the price to pay is that we have to discretize all geometric primitives, even though they are represented using B-splines.
  Therefore, it is highly desired to develop efficient PVG solver that can work directly with \textit{continuous} boundary conditions.

\section{Conclusion}
\label{sec:conclusion}

  We presented Poisson vector graphics, an extension of the popular first-order diffusion curves, for generating smooth-shaded images.
  Armed with two new types of primitives, namely Poisson curves and Poisson regions,
  PVG can easily produce photorealistic effects such as specular highlights, core shadows, translucency and halos.
  PVG distinguishes itself from the existing drawing tools by separating color and tone,
  which brings three unique features, i.e., local hue change, ease of extrema control, and permit of intersection among geometric primitives.
  Our preliminary user study confirms that PVG is more intuitive and easy to use than diffusion curve and its biharmonic extension.
  We also developed a harmonic B-spline based PVG solver that supports random access evaluation, zooming-in of arbitrary resolution and anti-aliasing.
  Although the solution is approximate, computational results show that the relative mean error is less than 0.3\%, which is too small to be distinguished by naked eyes.

  \begin{figure}[ht]
  \centering
  \includegraphics[width=2.5in]{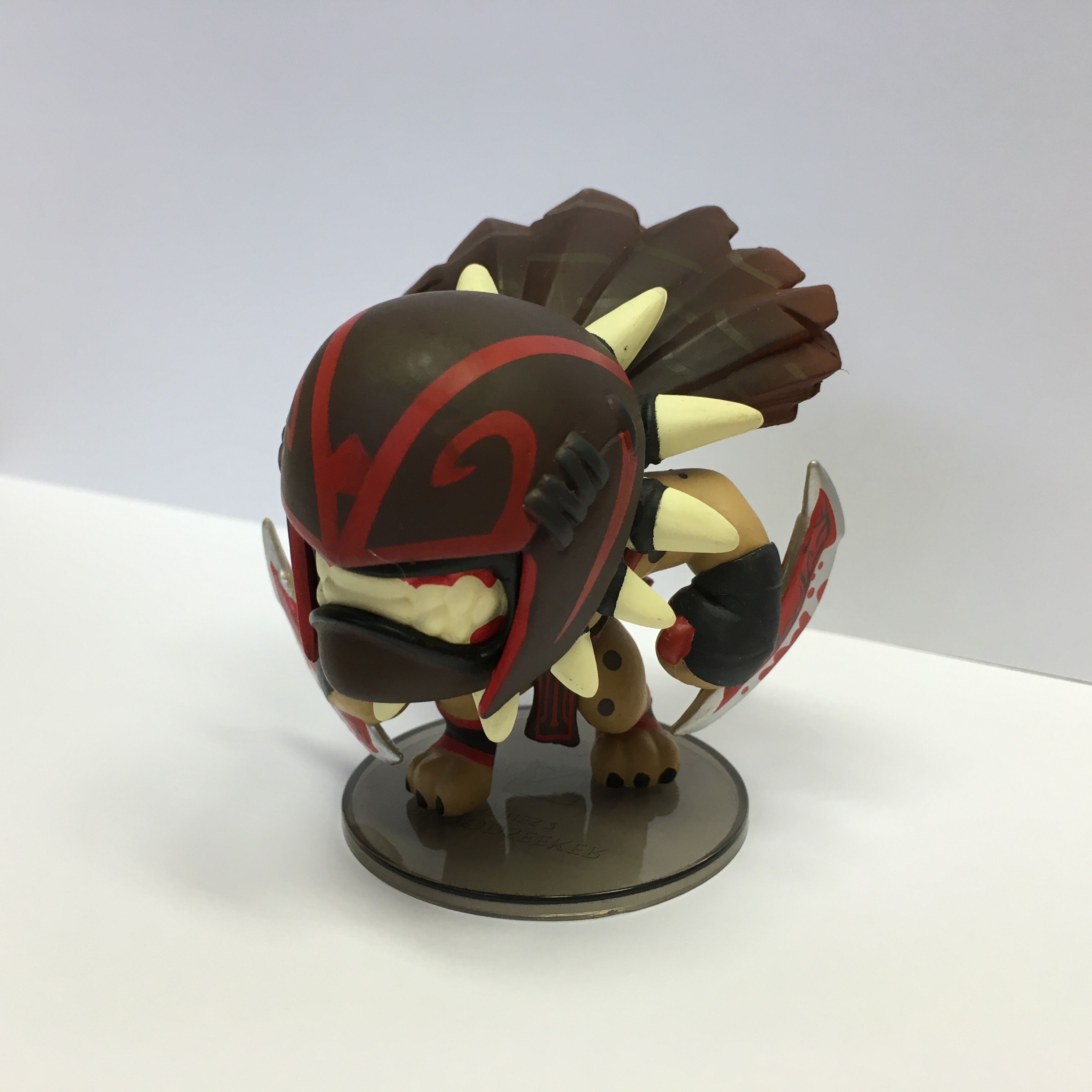}
  \includegraphics[width=2.5in]{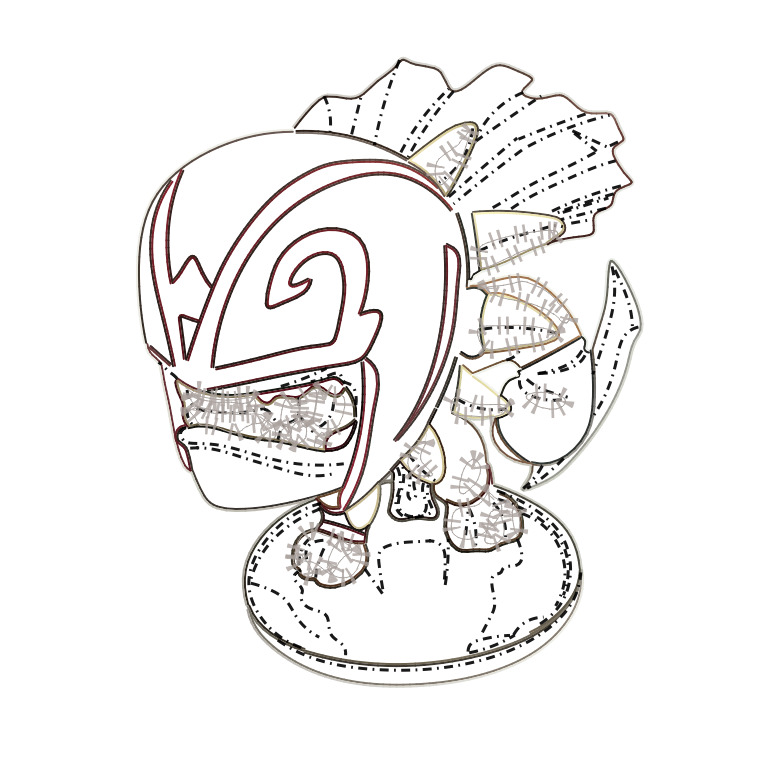}\\
  \makebox[2.50in]{Photo}\makebox[2.50in]{PVG}\\
  \includegraphics[width=5in]{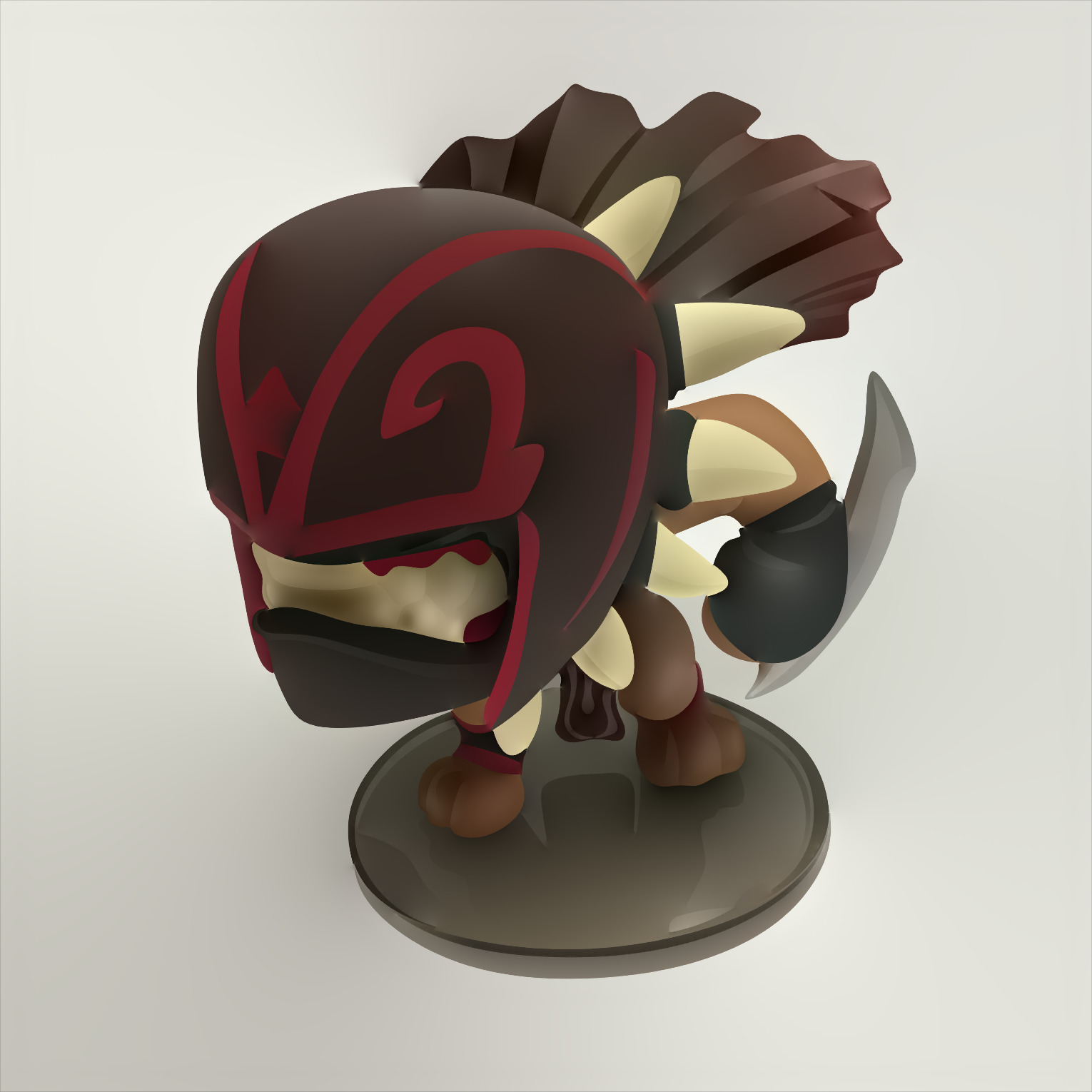}\\
  \makebox[5in]{PVG image}
  \caption{PVG is able to produce photo-realistic images with sparse geometric primitives.
  This PVG consists of 85 diffusion curves (solid lines), 71 Poisson curves (dashed lines) and 24 Poisson regions (loops with hatches).
  }
  \label{fig:realistic}
  \end{figure}

\bibliographystyle{abbrv}
\bibliography{reference}

\appendix

  To make the paper self-contained, we review some basis properties of harmonic B-splines.
  For details, we refer readers to \cite{Feng2012}.

  Let $\Omega\subset\mathbb{R}^2$ be a compact domain and $\mathcal{T}=\{t_i|t_i\in \Omega\}_{i=1}^{m}$ a set of knots.
  Taking $\{t_i\}$ as the generators, we construct a Voronoi diagram $\Omega=\bigcup_{i=1}^{m}\mathcal{V}_i$,
  where $\mathcal{V}_i$ is the Voronoi cell of knot $t_i$.

  For arbitrary points $x,y\in\Omega$, Green's function of the Laplace operator $\Delta$ satisfies
  \begin{equation}
  \label{eqn:green}
  \Delta \phi_y(x)=\delta_y(x).
  \end{equation}
  where $\delta_{y}(x)$ is the Dirac delta function centered at $y$.
  One symmetric solution to Equation~(\ref{eqn:green}) is $\phi_y(x)=\frac{1}{2\pi}\log(|x-y|)$.

  For a Voronoi cell $\mathcal{V}_j$, applying Green's theorem to $(\ref{eqn:green})$ yields
  \begin{equation}
  \label{eqn:divergence_theorem}
  \int_{\mathcal{V}_j}\Delta \phi_{y}(x)\:\mathrm{d}\sigma = \int_{\partial\mathcal{V}_j}\frac{\partial\Delta\phi_{y}(x)}{\partial \mathbf{n}}\:\mathrm{d}s,
  \end{equation}
  where $\bf n$ is the outward unit normal to the boundary $\partial\mathcal{V}_j$, $\mathrm{d}\sigma$ and $\mathrm{d}s$ are the area and line integral elements, respectively.

  Then define a function $\psi_j$ for each Voronoi cell $\mathcal{V}_j$ as
  \begin{equation}
  \label{eqn:basis}
  \psi_j(x)=\sum_i w_{ij}\phi_{t_i}(x),
  \end{equation}
  where $w_{ij}$ is the discrete Laplacian weight and $\sum_i w_{ij}\phi_{t_i}(x)$ is a boundary sum that approximates the line integral on the right hand side of Equation~(\ref{eqn:divergence_theorem}).

  Feng and Warren~\cite{Feng2012} showed that these functions $\psi_j$ share many properties of B-spline's basis functions,
  hence they called the linear combination $\sum_j\lambda_j\psi_j$, $\lambda_j\in\mathbb{R}^d$, a \textit{harmonic} B-spline.
  It is worth noting that the knots of a harmonic B-spline are completely free without any additional constraint.
  As a result, harmonic B-splines can be constructed on a set of fully irregular knots.
  Taking advantage of this feature, we express the solution of Eqn. (\ref{eqn:poissoneqn}) using harmonic B-spline,
  whose knots are the nodes of the quad-tree that discretizes the domain $\Omega$.

\end{document}